\documentclass[12pt,letterpaper]{article}
\usepackage[utf8]{inputenc}

\usepackage{graphicx,array}
\usepackage{url}
\usepackage{color}
\usepackage{latexsym}
\usepackage{amsthm}
\usepackage{amsmath}
\usepackage{amssymb}
\usepackage{amsfonts}
\usepackage[numbers,sort&compress]{natbib}
\usepackage{bm}
\usepackage{bbm}
\usepackage{slashed}
\usepackage{cancel}
\usepackage{mathrsfs}
\usepackage{enumerate}
\usepackage{tikz}
\usepackage{siunitx}
\usepackage{relsize}
\ExplSyntaxOn
\msg_redirect_name:nnn { siunitx } { physics-pkg } { none }
\ExplSyntaxOff
\usepackage{mdframed}
\usepackage{setspace}  
\usepackage{esvect}
\usepackage{physics}
\usepackage{enumitem}
\usepackage{booktabs}
\usepackage{tcolorbox}%

\numberwithin{equation}{section}

\usepackage{hyperref} 
\hypersetup{
    colorlinks=true,       
    linkcolor=red,          
    citecolor=blue,        
    filecolor=magenta,      
    urlcolor=blue           
}
\usepackage[all]{hypcap} 
\usepackage{multirow}
\usepackage{multicol}

\newcolumntype{M}[1]{>{\centering\arraybackslash}p{#1}}

\usepackage{natbib}
\setlist[description]{leftmargin=\parindent,labelindent=\parindent}

\usepackage[normalem]{ulem} 

\newcommand{\nc}{\newcommand}

\nc{\beq}{\begin{equation}}
\nc{\eeq}{\end{equation}}
\nc{\beqa}{\begin{eqnarray}}  
\nc{\eeqa}{\end{eqnarray}}  
\nc{\bit}{\begin{itemize}}  
\nc{\eit}{\end{itemize}}  

\DeclareMathOperator*{\argmin}{arg\,min}

\newcommand{\eg}{{\it e.g.}}
\newcommand{\ie}{{\it i.e.}}

\usepackage{floatrow}
\newfloatcommand{capbtabbox}{table}[][\FBwidth]

\usepackage{blindtext}

\allowdisplaybreaks

\title{ 
{\bf Baryoid Dark Matter from $\mathbb{Z}_N$ Domain Walls:} \\
{\bf \large The $(N-1):1$ origin of the dark matter-baryon coincidence}
}

\author{Yang Bai$^{\,a,b}$ and Ting-Kuo Chen$^{a}$}
\date{\small \it 
$^a$Department of Physics, University of Wisconsin-Madison, Madison, WI 53706, USA \\
$^b$HEP Division, Argonne National Laboratory, 
Argonne, IL 60439, USA
}

\begin{document}

\maketitle

\setlength{\parskip}{0.2ex}

\begin{abstract}
We propose an explanation for the dark matter-baryon coincidence based on collapsing $\mathbb{Z}_N$ domain walls, which form a novel compact baryonic state: the baryoid. A baryoid has an asteroid-scale mass and up-to-nuclear-scale energy density, and can serve as a dark matter candidate. Starting from equal baryon numbers in the domains formed in the early universe, the collapse of the domain walls after the QCD phase transition leads to a baryon-number ratio of $(N-1):1$ between the false- and true-vacuum domains. Since baryons are slightly lighter in the false-vacuum domains than in the true-vacuum domain, the resulting dark matter-to-baryon energy-density ratio is naturally close to, but slightly smaller than, $(N-1):1$, or $6:1$ for $N=7$. We calculate the domain-wall dynamics and the efficiency of baryon-number trapping, derive the resulting baryoid properties, and discuss a broad set of phenomenological probes.
\end{abstract}

\thispagestyle{empty}  
\newpage    
\setcounter{page}{1}  

\begingroup
\hypersetup{linkcolor=black,linktocpage}
\tableofcontents
\endgroup

\newpage

\section{Introduction}\label{sec:intro}

The presence of dark matter in our universe is firmly established by numerous pieces of observational evidence, one of which is the measurement of the cosmic microwave background (CMB). According to Planck data~\cite{Planck:2018vyg}, the ratio between the abundance of dark matter $\Omega_{\rm DM}$ and that of ordinary matter--mostly baryons, and thus we denote it by $\Omega_{\rm B}$--is given by $\Omega_{\rm DM}/\Omega_{\rm B}\approx 5.364\pm 0.077$. Ordinary baryons have a tiny asymmetry quantified by $(n_{\rm B} - n_{\bar{\rm B}} )/(n_{\rm B} + n_{\bar{\rm B}}) \approx 10^{-10}$, $n_{\rm B(\bar{B})}$ being the number density of (anti-)baryons, and their energy density, which is proportional to the nucleon mass $m_{\rm B}$, has a nontrivial origin from the asymptotic freedom of quantum chromodynamics (QCD). The fact that $\Omega_{\rm DM}/\Omega_{\rm B} = (n_{\rm DM}\,m_{\rm DM})/(n_{\rm B}\,m_{\rm B})$, $n_{\rm DM}$ and $m_{\rm DM}$ being the number density and mass of dark matter, respectively, happens to be an $\mathcal{O}(1)$ number raises the question of whether it is a mere coincidence or is actually established by a physical connection between dark matter and baryons, which physicists call the ``dark matter-baryon coincidence'' problem. 

Historically, there have been two main classes of solutions to this problem. The first class relates $\Omega_{\rm DM}$ and $\Omega_{\rm B}$ through an indirect portal: while dark matter has a particle nature distinct from that of baryons, its abundance is correlated with the baryon abundance. Examples in this class include asymmetric dark matter (see Ref.~\cite{Petraki:2013wwa} for a review), which seeks to achieve $n_{\rm DM}\approx n_{\rm B}$ and $m_{\rm DM}\approx 5m_{\rm B}$, and mirror-world scenarios (see Ref.~\cite{Foot:2004pa} for a review), which instead aim to achieve $n_{\rm DM}\approx 5n_{\rm B}$ and $m_{\rm DM}\approx m_{\rm B}$; see also Refs.~\cite{Bai:2013xga,Ritter:2024sqv}, which use conformal dynamics to explain the comparable masses. These frameworks inevitably introduce a large number of new degrees of freedom and an extended parameter space, making the solutions less natural and more fine-tuned. More recently, attempts to correlate the dark matter and baryon energy densities directly have been developed using composite-axion particle dynamics in Refs.~\cite{Brzeminski:2023wza,Banerjee:2024xhn}. 
While this latter approach establishes a seemingly more natural connection between dark matter and baryonic matter, one may still question the need to introduce a new particle origin for dark matter: the fact that $\Omega_{\rm DM}/\Omega_{\rm B}\sim\mathcal{O}(1)$ might instead point to a common origin for both types of matter. Motivated by this possibility, we explore in this paper the second class of solutions which directly associate dark matter with baryonic matter, and thereby address the coincidence problem in a more definitive manner.

Guided by this possibility, the basic mechanism studied in this paper is the following. A spontaneously broken $\mathbb{Z}_N$ symmetry partitions the universe in a way that there are about or more than $N^3$ total domains of the $N$-fold degenerate vacua in a Hubble patch, depending on the correlation length of the order parameter field which is bounded by causality according to the Kibble-Zurek mechanism~\cite{Kibble:1976sj,Zurek:1985qw}. Consequently, each of the $N$-fold degenerate vacua occupies $\sim 1/N$ of the total volume of the universe so that once a bias potential $V_{\rm bias}$ selects one true vacuum, the subsequent collapse of the domain walls naturally establishes an $(N-1):1$ baryon-number ratio between the false- and true-vacuum domains. If the baryons in the false-vacuum domains are then trapped and stabilized into compact objects, the ``baryoids",\footnote{In this model, the predicted baryonic mass of this state lies in the asteroid-mass range, while its density is of order below around nuclear density and much higher than the atomic density; therefore, we name it a ``baryoid.''} the resulting dark matter-to-baryon energy-density ratio is approximately
\begin{align}\label{eq:intro-ratio}
\frac{\Omega_{\rm DM}}{\Omega_{\rm B}} \;=\;
\frac{\Omega_{\rm BD}}{\Omega_{\rm B}} \;\approx\;
(N-1)\,\frac{E_{\rm BD}/N_{\rm B}}{m_{\rm B}}
\approx(N-1)\,\frac{m_{\rm B, F}}{m_{\rm B}}\,,
\end{align}
where $m_{\rm B}$ is the ordinary baryon mass, 
$E_{\rm BD}$ is the energy of baryoid with a large baryon number $N_{\rm B}$, and
$m_{\rm B, F}$ is the baryon mass in the false-vacuum domains and is close to but below $m_{\rm B}$. In particular, for $N=7$, this mechanism predicts a characteristic $6:1$ structure, naturally close to the observed value $\Omega_{\rm DM}/\Omega_{\rm B}\approx 5.4$ up to an ${\cal O}(1)$ correction. Here we have assumed zero baryon leakage rate during the baryoid formation process, which, if non-zero, will further slightly lower the ratio $\Omega_{\rm BD}/\Omega_{\rm B}=\Omega_{\rm DM}/\Omega_{\rm B}$, given that it does not compromise the stability of the baryoid state (see Section~\ref{subsec:formation:entrapment} for more discussion).

Before continuing, we first compare the baryoid state to another possible baryonic dark matter candidate: the quark nugget. In the Standard Model (SM), baryons mostly exist in the form of hadronic nuclei in the normal QCD vacuum, except at environments of high temperature (such as the early universe or ion colliders) or high density (such as the core of a massive neutron star) where they can exist as quark-gluon plasma or quark matter. In Ref.~\cite{Witten:1984rs}, Witten proposed the possible existence of ``quark nuggets'', which, if more stable than ordinary nuclei, can serve as dark matter candidates. Nevertheless, as the formation of quark nuggets requires a first-order QCD phase transition, which turns out to be a mere crossover in the SM~\cite{Aoki:2006we,Bhattacharya:2014ara}, the introduction of physics beyond the SM (BSM) is inevitable for this purpose. Moreover, although there exist some caveats, quark nuggets predicted by the SM QCD are unlikely to be more stable than ordinary nuclei~\cite{Bai:2024amm,Bai:2025zpm} (see Ref.~\cite{Holdom:2017gdc} for an alternative opinion). Some BSM quark nugget models include axion quark nuggets~\cite{Zhitnitsky:2002qa} and six-flavor quark nuggets~\cite{Bai:2018vik}. While these frameworks provide viable ways to realize the dark matter-baryon coincidence with nugget dark matter, their associated parameter spaces remain rather broad. This motivates us to explore a more predictive possibility, in which the observed ratio is tied directly to the discrete vacuum structure of a spontaneously broken $\mathbb{Z}_N$ symmetry, as discussed above.

\begin{figure}[ht!]
    \centering
    \includegraphics[width=0.98\linewidth]{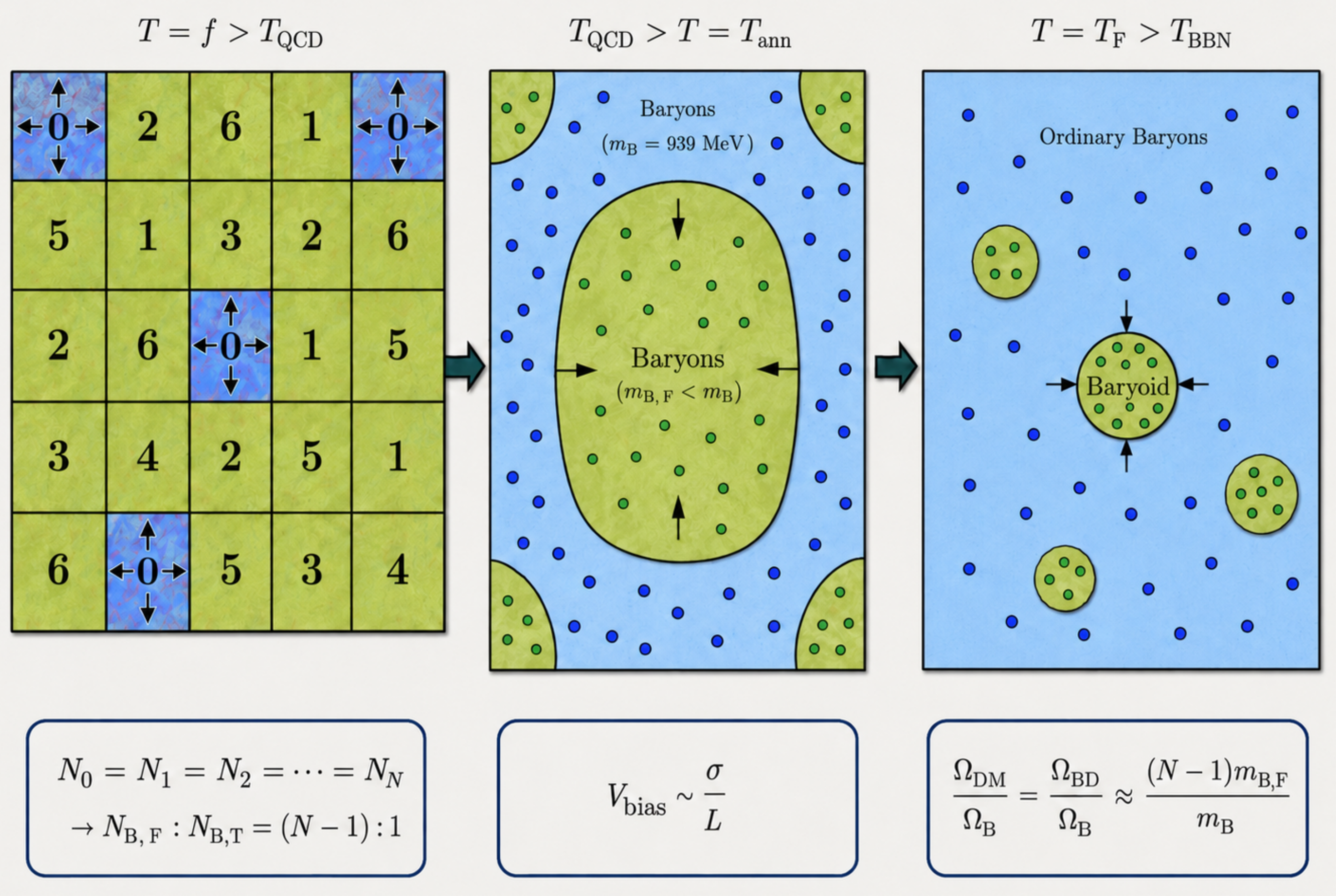}
    \caption{A schematic illustration of the baryoid explanation for the dark matter-baryon coincidence. 
    {\it Left:} Starting with a spontaneously broken $\mathbb{Z}_N$ symmetry at $T=f>T_{\rm QCD}\approx150~{\rm MeV}$, the universe is divided by the domain wall network into roughly equal numbers of domains occupying one of the $N$-fold degenerate vacua, thus establishing a baryon-number ratio between the false- and true-vacuum domains as $N_{{\rm B,F}}:N_{{\rm B,T}}=(N-1):1$. At this time, baryonic matter exists in the form of relativistic quark-gluon plasma.
    {\it Middle:} At $T=T_{\rm ann}<T_{\rm QCD}$ when the QCD-triggered bias potential $V_{\rm bias}\sim\sigma/L$, the tension force of the domain walls, the walls will begin to collapse and aggregate the non-relativistic baryons inside the false-vacuum regions as they sweep by. Because of the QCD-triggered bias, the baryon mass in the false-vacuum domains $m_{\rm B,F}<m_{\rm B}=939$~MeV, the average nucleon mass in the ordinary vacuum.
    {\it Right:} Once the collapsing domain walls are stabilized by the Fermi degeneracy pressure of the matter inside the false-vacuum domains at $T=T_{\rm F}>T_{\rm BBN}$ when the domain sizes are sufficiently small, the resulting compact baryonic objects are then referred to as the ``baryoids'', which will establish an abundance ratio of $\Omega_{\rm DM}/\Omega_{\rm B}=\Omega_{\rm BD}/\Omega_{\rm B}\approx(N-1)m_{\rm B,F}/m_{\rm B}$.
    }
    \label{fig:schematic}
\end{figure}

More specifically, we propose an explanation for the dark matter-baryon coincidence based on the novel domain-wall-bounded baryoid state, which is formed from collapsed $\mathbb{Z}_N$ domain walls that successfully drive baryons into the consequent compact object. As we have demonstrated in Eq.~\eqref{eq:intro-ratio}, the dark matter-to-baryon energy-density ratio naturally inherits the characteristic $(N-1):1$ structure in this framework, up to an ${\cal O}(1)$ correction. For consistency with cosmological observations, the domain walls must collapse, or annihilate, before Big Bang Nucleosynthesis (BBN) begins, at $T=T_{\rm BBN}\approx3~{\rm MeV}$~\cite{Bai:2021ibt,Bringmann:2023opz}; otherwise, they would either dominate the energy density of the universe or spoil the successful predictions of BBN.

In certain $\mathbb{Z}_N$ domain-wall models, the baryon mass in the false-vacuum domains is expected to be lower than the average nucleon mass $m_{\rm B}\approx939$~MeV. This can make the baryoids more stable than ordinary nuclei, and we also do not need to worry about the decay of ordinary nuclei into baryoids because of the challenging tunneling from the true-vacuum to the false-vacuum state. For example, in the QCD-anomalous $\mathbb{Z}_N$ symmetry model, which relates the strong CP phase to the order-parameter field $\theta$~\cite{Bai:2023cqj}, baryons in the domains with $\theta>0$ are expected to be lighter than ordinary nucleons~\cite{Lee:2020tmi}. Consequently, these stable baryoids can serve as dark matter candidates, with $\Omega_{\rm BD}=\Omega_{\rm DM}$. We illustrate this mechanism schematically in Figure~\ref{fig:schematic}. We note here that although the main purpose of this study is to explore the phenomenological consequences of the baryoid solution to the dark matter-baryon coincidence problem, our analysis is mainly motivated by the model presented in Ref.~\cite{Bai:2023cqj} (see Section~\ref{subsec:formation:model}), while one can certainly apply the same phenomenological framework to other models.

In the vanilla version of the $\mathbb{Z}_N$ domain-wall model, in which the domain walls do not interact with plasma particles, the walls undergo frictionless expansion and eventually enter the ``scaling regime'', where their characteristic length scale satisfies $L\propto t$ and they move with a constant root-mean-square velocity $v$. This behavior is both predicted by the velocity-dependent one-scale (VOS) model~\cite{Martins:2016ois} and confirmed by simulations~\cite{Hiramatsu:2012sc,Kawasaki:2014sqa}.  For the collapsing walls to successfully collect baryons, these baryons must exist in the form of non-relativistic hadrons. This implies that the walls should begin to collapse only at $T=T_{\rm ann}<T_{\rm QCD}\approx150$~MeV, where $T_{\rm QCD}$ is the QCD phase transition temperature. Assuming a radiation-dominated universe, we can write $L(T)\sim H^{-1}(T)/N=2t/N$. We also have the baryon number density $n_{\rm B}(T)=\eta_{\rm B}\,n_\gamma(T)$, where $\eta_{\rm B}=(1+\Omega_{\rm DM}/\Omega_{\rm B})\times\eta$, with $\eta\approx 6\times10^{-10}$ denoting the ordinary baryon-to-photon ratio. Consequently, assuming zero baryon leakage, the total number of baryons inside a single baryoid is $N_{\rm B}\sim n_{\rm B}(T_{\rm ann})L^3(T_{\rm ann})\sim10^{50}\times(50\,{\rm MeV}/T_{\rm ann})^3/N^3$,
which is on the verge of being excluded by microlensing constraints if we assume an energy per baryon number of $E/N_{\rm B}\approx1$~GeV; see Section~\ref{sec:pheno}.

Moreover, there are three additional reasons why this frictionless domain-wall model cannot generate stable baryoids. First, the wall velocity can easily become relativistic due to the combined acceleration from the curvature force and $V_{\rm bias}$, which would reduce the baryon trapping rate inside the false-vacuum pocket. Second, for the baryoids to stabilize, the walls must remain non-relativistic when the Fermi degeneracy pressure from the baryons and electrons, with the latter bound to the protons by the Coulomb force, becomes strong enough to counteract the collapsing forces as the domain volume shrinks. Third, to avoid the Deuterium-abundance constraint, which is sensitive to the length scale of baryon inhomogeneities generated by the collapsing domain-wall network, domain expansion must remain frustrated by friction; see Section~\ref{sec:pheno}. 

As a result, we follow the framework presented in Ref.~\cite{Hook:2026grn} to discuss domain-wall dynamics in the presence of significant friction, using a generic parametrization of the temperature dependence of the domain-wall tension $\sigma(T)$ and friction length $\ell_f(T)$. As we demonstrate in this study, combining the above conditions leads to a highly predictive framework that can explain the dark matter-baryon coincidence through domain-wall-bounded baryoids.

Our paper is organized as follows. 
In Section~\ref{sec:DW_summary}, we summarize the formulas for domain wall dynamics with the detailed derivation of the VOS model with a $T$-dependent wall tension $\sigma(T)$ given in Appendix~\ref{sec:gVOS} and details of the frustrated expansion and collapse dynamics discussed in Appendix~\ref{sec:DW}.
Next, we introduce the formation mechanism of baryoids and analyze their properties in Section~\ref{sec:formation}, in which we also demonstrate the success of the framework with a few benchmarks and propose a potential model realization.
After that, we discuss various phenomenological implications of baryoids and the associated constraints in Section~\ref{sec:pheno}, with the detailed derivation of the gravitational wave spectrum emitted by the collapsing frustrated domain walls given in Appendix~\ref{sec:GW_details}.
Finally, we discuss and conclude our study in Section~\ref{sec:conclusions}.

\section{Domain wall dynamics}\label{sec:DW_summary}

In this section, we provide a summary of our review and analysis on domain wall dynamics detailed in Appendices~\ref{sec:gVOS} and \ref{sec:DW}. We begin with a qualitative overview of the topic and then present the technical details used in this study.

Domain walls are the topological defects naturally produced from the spontaneous breaking of certain symmetries. In the cosmological setting, the dynamics of domain walls produced in the early universe is mainly governed by three types of forces: (1) The surface tension force, which tends to minimize the total surface energy of the walls by increasing their curvature radii through wall straightening and expansion. (2) The damping force consisting of the Hubble friction (from the universal expansion of the spacetime), the plasma friction (from the interaction between the walls and the plasma particles), and, if the surface tension varies with time (and thus temperature), the relaxation of the tension (which actually subtracts from the damping force rather than adds to it). (3) The bias pressure, which lifts the degeneracy of the vacua and drives the domain walls to collapse in the direction of the metastable vacua, leaving only the stable-vacuum domains to survive. As we briefly discussed in Section~\ref{sec:intro}, the collapse of domain walls must finish before the BBN begins in order to be consistent with cosmological observations.

In a scenario without plasma friction, the surface tension force will play the dominant role upon the formation of the domain walls, driving them to straighten and expand roughly at the Hubble rate; consequently, the domain wall sizes will scale roughly with the Hubble length scale. At the same time, the surface tension force itself will decrease along with the increasing curvature radii of the walls; once the bias pressure becomes comparable to the surface tension force, the domain walls will then collapse almost instantaneously. Nevertheless, if the plasma friction is sufficiently strong compared to the surface tension of the walls and the Hubble rate, it is likely for the walls to undergo ``frustrated'' expansion and collapse: they will first expand at a rate much slower than the Hubble rate, making their size significantly smaller than the Hubble length scale (and thus populating much more domains than in the scenario without the plasma friction), and then collapse at a non-relativistic velocity rather than instantaneously. As we will discuss more in Section~\ref{sec:formation}, the frustrated dynamics is crucial for the successful trapping of non-relativistic baryons and thus the formation of baryoids. To model the dynamics of domain walls with all these factors taken into consideration, we adopt the VOS model~\cite{Martins:2016ois} and phenomenologically parametrize all the mentioned forces, while we will discuss the potential model realization in Section~\ref{subsec:formation:model}.

The VOS model for domain walls is derived from the Nambu-Goto action for a worldsheet by treating a domain wall effectively as a 2D object. Depending on the specifics of the domain wall network, there are two ways to parametrize it. The first way, which describes an unbiased domain wall network that can be parametrized by its average energy density $\rho_w$ and root mean square velocity $v$, is used to model the average behavior of an expanding domain wall network before the bias potential $V_{\rm bias}$ kicks in. In this scenario, the wall network is parametrized by its characteristic length scale $L\equiv \sigma/\rho_w$, $\sigma$ being the universal domain wall tension, and root mean square velocity $v$. The second way is to focus on an individual domain wall, which, assuming a simple geometry for the wall such as a spherical shell, is then parametrized by its radius $R$ and velocity $v$. This framework is used in this study to model the collapse dynamics of a biased domain wall since we have to treat the expanding and collapsing domain walls separately.

We now turn to the technical details summarized from Appendices~\ref{sec:gVOS} and \ref{sec:DW}. After the spontaneous breaking of a given discrete symmetry--in this study we only consider $\mathbb{Z}_N$--at the energy scale $f$, domain walls will form and divide the universe into domains of distinct vacuum expectation values. Choosing the $\mathcal{O}(1)$ factor to be exactly one, we denote the domain wall tension by $\sigma(T)= T^{x_\sigma}f^{3-x_\sigma}$, with $0\leq x_\sigma\leq3$, where the $T$-dependence could arise if the intrinsic scale of the theory is affected by the interaction between the wall and the plasma. For example, in $SU(N)$ Yang-Mills theories, if the $\mathbb{Z}_N$ center symmetry is spontaneously broken, the resulting domain walls will have $\sigma\propto T^3$, which is not associated with any particular scale in the Lagrangian, when $T>\Lambda_{SU(N)}$, the fundamental scale of the $SU(N)$ theories~\cite{Bhattacharya:1990hk}. We also consider the friction induced by the wall-plasma interaction, which we parametrize using the friction length $\ell_f=T^{x_f-4}f^{3-x_f}\beta^{-1}$ with $0\leq x_\sigma \leq x_f\leq 3$ ($x_f=x_\sigma+x_P$ with $x_P$ parametrizing the $T$-dependence of the pressure exerted on the wall by the plasma particles; see Appendix~\ref{subsec:DW:frustrated} for details), where we fix $\beta\sim\mathcal{O}(1)$ as a constant. We only consider integer values of $x_{\sigma}$ and $x_f$. For details and alternative considerations, see Appendix~\ref{subsec:DW:frustrated}. For convenience, we further define the domain-wall formation temperature $T_c=f$ and $t_c=\sqrt{90/(4\pi^2 g_*)}M_{\rm Pl}/T_c^2$, where $M_{\rm Pl}=2.43\times10^{18}$~GeV is the reduced Planck mass and $g_*(T)$ denotes the relativistic energy density degrees of freedom, which we fix as a constant number $g_*(T)=g_*(100~{\rm MeV})\approx 10$. We separate the discussion into two parts: the expansion, which includes scaling and frustrated expansions, and the collapse of the domain wall network.

For expansion, we model it with the VOS model that describes the average behavior of a domain wall network~\cite{Martins:2016ois}, which is a coupled differential equation system of the characteristic length scale $L$ and the root mean square velocity of the wall $v$,
\begin{equation}\label{eq:VOS:main}
\begin{aligned}
    \frac{dL}{dt} &= HL+v^2\frac{L}{\ell_d} + c_w v ~, \\
    \frac{dv}{dt} &= (1-v^2)\left(\frac{k_w}{L}-\frac{v}{\ell_d}\right) ~,
\end{aligned}
\hspace*{1.5cm}\text{[Domain wall network]}
\end{equation}
where $H=a^{-1}(da/dt)$ is the Hubble scale, $a$ being the scale factor and $t$ the cosmic time, $\ell_d^{-1}=3H+\ell_f^{-1}+\dot{\sigma}/\sigma$ is the damping factor, and $c_w$ and $k_w$ are the phenomenological parameters given by simulations that describe possible energy losses from the domain wall network. We briefly explain the meaning and/or implication of each term: (1) $HL$ describes the increase of the domain wall size induced by Hubble expansion. (2) $v^2 L/\ell_d$ in $dL/dt$ and $v/\ell_d$ in $dv/dt$ describe the damping effects on the wall motion, which tend to decrease the energy density for fixed wall tension $\sigma$ (and thus increase the wall size) and the velocity $v$ of the wall network. Note that this term is absent from the $dR/dt$ equation in Eq.~\eqref{eq:VOS:R:main} but present here because $L$ is defined from the average wall energy density $L=\sigma/\rho_w$ rather than denoting the actual physical domain wall size which $R$ stands for. (3) $c_wv$ describes the energy loss effect from inter-wall interactions~\cite{Martins:2016ois} on $dL/dt$ which tends to increase the wall size. (4) $k_w/L$ describes the effective curvature force under some potential phenomenological energy loss mechanism similar to that behind the $c_wv$ term. (5) Finally, the $(1-v^2)$ factor in $dv/dt$ simply comes from the Lorentz factor $\gamma(v)=1/\sqrt{1-v^2}$.

\begin{figure}[ht!]
    \centering
    \includegraphics[width=0.48\linewidth]{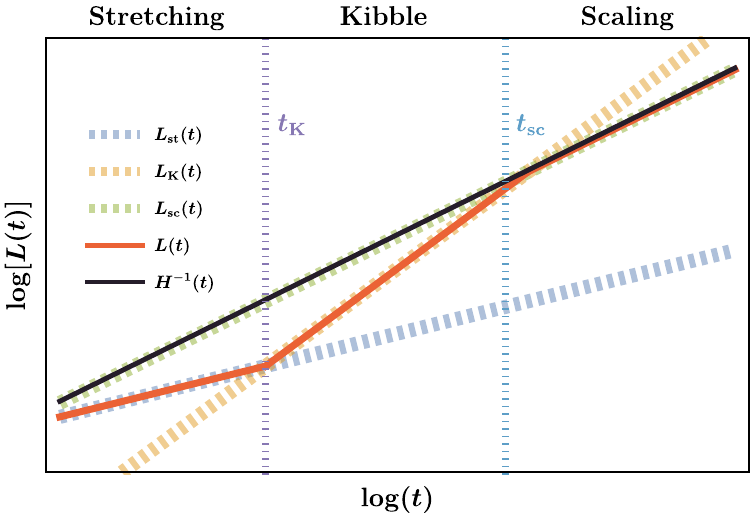}
    \hspace{3mm}\includegraphics[width=0.48\linewidth]{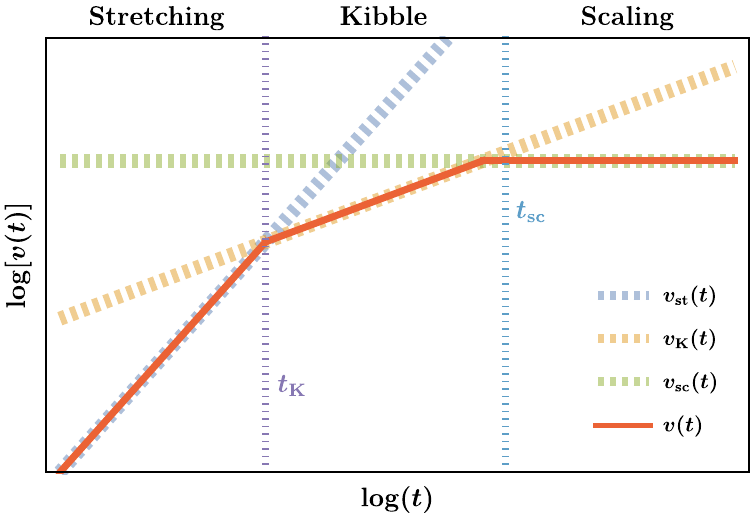}
    \caption{Schematic ({\it left}) $L(t)$ and ({\it right}) $v(t)$ profiles of an expanding frustrated domain wall network. The wall evolution begins in the stretching regime, transits to the Kibble regime at $t=t_{\rm K}$, and finally enters the scaling regime at $t\sim t_{\rm sc}$. We also show the Hubble length scale $H^{-1}(t)$ in the left panel for comparison.
    }
    \label{fig:Lt_vt}
\end{figure}

We describe the scaling and frustrated expansions below:
\begin{itemize}[leftmargin=*]
    \item {\bf Scaling expansion:} If $3H+\dot{\sigma}/\sigma\gg\ell_f^{-1}$, \ie, friction is negligible, Eq.~\eqref{eq:VOS:main} has the solution
    \beqa\label{eq:Lv_sc:main}
        L_{\rm sc}(t) = L_0\,t ,\qquad  v_{\rm sc}(t) = v_0 ~~~\quad [x_\sigma<3] ~,
    \eeqa
    where $L_0=2\sqrt{k_w(k_w+c_w)/(3-{x_\sigma})}$ and $v_0=\sqrt{k_w/[(3-{x_\sigma})(k_w+c_w)]}$. Note that the scaling solution only exists for $x_\sigma<2$, while for $x_\sigma\geq3$, since $3H+\dot{\sigma}/\sigma=(3-x_\sigma)/2t\leq0$ in the radiation-dominated universe, there is not only no effective damping term, but a ``boosting term'' might be induced, leading to unphysical evolution. 
    
    \item {\bf Frustrated expansion:} When friction is strong enough, we can asymptotically split the solution to  Eq.~\eqref{eq:VOS:main} into three regimes: the stretching regime, when $\ell_d^{-1}\simeq \ell_f^{-1}$ and $v\simeq0$; the Kibble regime~\cite{Kibble:1981gv}, when the $\mathcal{O}(v)$ term in $dL/dt$ becomes important but the $\mathcal{O}(v^2)$ terms are still negligible; and the scaling regime, when $v$ is no longer negligible and $\ell_d^{-1}\simeq 3H+\dot{\sigma}/\sigma$. While the scaling regime is identical to the scaling expansion discussed in the previous item, the stretching solution is given by
    \begin{equation}\label{eq:Lv_st:main}
    \begin{aligned}
        L_{\rm st}(t) &\simeq L(t_c)\left(\frac{t}{t_c}\right)^{1/2} ~, \\
        v_{\rm st}(t) &\simeq \frac{1}{L(t_c)M_{\rm Pl}^{1/2}}\frac{t^{(3-{x_f})/2}}{t_c^{(2-{x_f})/2}}\sqrt{\frac{2}{\beta^2}\left(\frac{\pi^2g_*}{90}\right)^{1/2}} ~,
    \end{aligned}
    \end{equation}
    where $L(t_c)\simeq H^{-1}(t_c)/N$ for $\mathbb{Z}_N$, and the Kibble solution is given by
    \begin{equation}\label{eq:Lv_K:main}
    \begin{aligned}
        L_{\rm K}(t) &\simeq \frac{t^{(6-{x_f})/4}}{M_{\rm Pl}^{1/4}t_c^{(3-{x_f})/4}}\sqrt{\frac{4}{4-{x_f}}\frac{c_wk_w}{\beta}\left(\frac{4\pi^2g_*}{90}\right)^{1/4}} ~, \\
        v_{\rm K}(t) &\simeq \frac{t^{(2-{x_f})/4}}{M_{\rm Pl}^{1/4}t_c^{(3-{x_f})/4}}\sqrt{\frac{4-{x_f}}{4}\frac{k_w}{\beta c_w}\left(\frac{4\pi^2g_*}{90}\right)^{1/4}} ~.
    \end{aligned}
    \end{equation}
    The asymptotic transition time $t_{\rm K}$ from the stretching to the Kibble regime is given by
    \begin{equation}\label{eq:t_K:main}
        t_{\rm K} = \frac{t_c^{({x_f}-5)/({x_f}-4)}}{M_{\rm Pl}^{1/({x_f}-4)}}\left[\frac{N}{2}\sqrt{\frac{4}{4-{x_f}}\frac{c_wk_w}{\beta}\left(\frac{4\pi^2g_*}{90}\right)^{1/4}}\right]^{4/({x_f}-4)} ~~~[{x_f}<2] ~,
    \end{equation}
    while the asymptotic transition time $t_{\rm sc}$ from the Kibble to the scaling regime is given by
    \begin{equation}\label{eq:t_sc:main}
        3H(t_{\rm sc})+\frac{\dot{\sigma}(t_{\rm sc})}{\sigma(t_{\rm sc})}\sim\ell_f^{-1}(t_{\rm sc}) \to t_{\rm sc} = \frac{t_c^{({x_f}-3)/({x_f}-2)}}{M_{\rm Pl}^{1/({x_f}-2)}}\left[\frac{(3-{x_\sigma})^2}{2\beta^2}\sqrt{\frac{\pi^2g_*}{90}}\right]^{1/({x_f}-2)} ~~~[{x_f}<2] ~.
    \end{equation}
    Note that for $x_f=3$, the solution will always stay in the stretching regime, while for ${x_f}=2$, there is only one transition from the stretching regime to the scaling regime at $t=t_{\rm K}=t_{\rm sc}$ as defined in Eq.~\eqref{eq:t_K:main} since there is no asymptotic Kibble regime in this case.
\end{itemize}
We show the schematic profiles of $L(t)$ and $v(t)$ for an expanding frustrated domain wall network in Figure~\ref{fig:Lt_vt}, in the left panel of which we also show the Hubble length scale $H^{-1}(t)$ for comparison. Note that $t_{\rm sc}$, which is defined as the time when $3H+\dot{\sigma}/\sigma\sim\ell_f^{-1}$, is not exactly at but very close to the actual transition time from the Kibble regime to the scaling regime in both $L(t)$ and $v(t)$.

Next, to avoid contradicting astrophysical and cosmological observations such as the BBN measurements, we introduce a bias potential $V_{\rm bias}$ to collapse the domain wall at the temperature $T_{\rm ann}\gtrsim T_{\rm BBN}\approx3$~MeV~\cite{Bai:2021ibt,Bringmann:2023opz}, which is determined by $V_{\rm bias}\approx0.4\,\sigma(T_{\rm ann})/L(T_{\rm ann})$ for $\mathbb{Z}_N$ domain walls~\cite{Hiramatsu:2012sc,Kawasaki:2014sqa}. To model the collapse of a single domain wall, we use the worldsheet equation of motion derived from the Nambu-Goto action (see Appendix~\ref{sec:gVOS}), which is a coupled differential equation system of the wall radius $R$ and velocity $v=-dq/d\tau=-d(R/a)/d(t/a)$, $\tau$ being the conformal time and $q$ the comoving radius,
\begin{equation}\label{eq:VOS:R:main}
\begin{aligned}
    \frac{dR}{dt} &= HR - v ~, \\
    \frac{dv}{dt} &= (1-v^2)\left(\frac{s}{R}+\frac{V_{\rm bias}}{\sigma}-\frac{v}{\ell_d}\right) ~,
\end{aligned}
\hspace*{1.5cm}\text{[Single domain wall]}
\end{equation}
where $s$ is the curvature parameter. Eq.~\eqref{eq:VOS:R:main} is similar to Eq.~\eqref{eq:VOS:main} except for three major differences: the parametrization of an individual domain wall by its radius $R$ instead of by the characteristic length scale $L$ for an unbiased domain wall network, the appearance of $V_{\rm bias}$ that drives the collapse, and that $v$ stands for the velocity of an individual wall in Eq.~\eqref{eq:VOS:R:main} while it denotes the root mean square velocity of a wall network in Eq.~\eqref{eq:VOS:main}.

In the following, we only consider spherical domain walls and fix $s=2$ as a constant.
We are only interested in the friction-dominated collapse scenario where $\ell_d^{-1}\approx\ell_f^{-1}$ (this becomes exact when $x_\sigma=3$). Assuming that the terminal velocity $v_t(t)$\footnote{
Note that the terminal velocity in this context is $t$-dependent (see Appendix~\ref{subsec:DW:frustrated} for details), which is different from the usual constant terminal velocity defined in other mechanical systems.
} is reached immediately at $t=t_{\rm ann}$, we can derive 
\begin{equation}\label{eq:v_t_friction:main}
    v_t(t) = \left(\frac{4\pi^2g_*}{90}\right)^{1/2}\frac{T_c}{\beta M_{\rm Pl}}\left[\frac{s}{R}+\frac{0.2\,N\,t^{x_\sigma/2}}{t_c^{1/2}t_{\rm ann}^{(1+x_\sigma)/2}}\right]\frac{t^{(4-x_f)/2}}{t_c^{(2-x_f)/2}} ~,
\end{equation}
and plug it into the first line of Eq.~\eqref{eq:VOS:R:main} to get
\begin{equation}
    \frac{dR}{dt} \approx HR - v_t(t) = \frac{R}{2t} - \left(\frac{4\pi^2g_*}{90}\right)^{1/2}\frac{T_c}{\beta M_{\rm Pl}}\left[\frac{s}{R}+\frac{0.2\,N\,t^{x_\sigma/2}}{t_c^{1/2}t_{\rm ann}^{(1+x_\sigma)/2}}\right]\frac{t^{(4-x_f)/2}}{t_c^{(2-x_f)/2}} ~,
\end{equation}
subject to the initial condition $R(t_{\rm ann})=(2/N)t_{\rm ann}^{1/2}t_c^{1/2}$. Notice that when $HR-v_t(t) > 0$, even though the wall velocity is pointing inward, the wall size is still increasing along with Hubble expansion. To ensure that $t_{\rm ann}<t_{\rm K}$, we can derive an upper bound on $V_{\rm bias}$ for $x_f\leq2$
\begin{equation}
    V_{\rm bias} \lesssim \frac{N}{5}\left[\frac{4-x_f}{N^2}\frac{\beta}{c_wk_w}\right]^{\frac{x_\sigma-4}{x_f-5}}\left(\sqrt{\frac{90}{4\pi^2g_*}}M_{\rm Pl}\right)^{\frac{x_\sigma-x_f+1}{x_f-5}}T_{\rm ann}^{-\frac{x_\sigma-5x_f+21}{x_f-5}} ~~~[{x_f}\leq 2] ~,
\end{equation}
while we are in principle free of this concern for $x_f=3$. However, since this bound also roughly implies the bound for which the wall collapse velocity can easily become relativistic, we still consider it for the case of $x_f=3$ in what follows. Finally, given $T_{\rm ann}$ and $V_{\rm bias}$, one has
\begin{equation}
    f = \left(\frac{5}{N}\sqrt{\frac{90}{4\pi^2g_*}}\frac{M_{\rm Pl}V_{\rm bias}}{T_{\rm ann}^{1+x_\sigma}}\right)^{1/(4-x_\sigma)} ~.
\end{equation}
For these solutions to hold, we need to ensure $v_t(t)^2\ll1$ throughout the time interval of our interest, which we will discuss in the next section when studying baryon entrapment within the collapsing domains. Considering the parameter space of our interest and potential realistic models, we focus on a few $(x_\sigma,x_f)$ choices and derive the corresponding scaling of the upper bounds on $V_{\rm bias}$ as
\begin{equation}\label{eq:V_bias_bounds:main}
    V_{\rm bias} \lesssim
    \begin{cases}
        (3.54~{\rm MeV})^4\left(\frac{7}{N}\right)^{3/5}\left(\frac{T_{\rm ann}}{50\,{\rm MeV}}\right)^{21/5}\left(\frac{\beta}{c_wk_w}\right)^{4/5} &,~\quad \left[(x_\sigma,x_f)=(0,0)\right] \\[2ex]
        (770~{\rm MeV})^4\left(\frac{7}{N}\right)^{5/12}\left(\frac{T_{\rm ann}}{50\,{\rm MeV}}\right)^{11/12}\left(\frac{\beta}{c_wk_w}\right)^{1/3} &,~\quad \left[(x_\sigma,x_f)=(0,2)\right] \\[2ex]
        (0.127~{\rm MeV})^4\left(\frac{7}{N}\right)^{0}\left(\frac{T_{\rm ann}}{50\,{\rm MeV}}\right)^{9/2}\left(\frac{\beta}{c_wk_w}\right)^{1/2} &,~\quad \left[(x_\sigma,x_f)=(3,3)\right]
    \end{cases} ~,
\end{equation}
which correspond to
\begin{equation}\label{eq:f_bounds:main}
    f \lesssim
    \begin{cases}
        2.26\times10^2~{\rm GeV} \left(\frac{T_{\rm ann}}{50\,{\rm MeV}}\right)^{4/5}\left(\frac{7}{N}\right)^{2/5}\left(\frac{\beta}{c_wk_w}\right)^{1/5} &,~\quad \left[(x_\sigma,x_f)=(0,0)\right] \\[2ex]
        4.91\times10^4~{\rm GeV} \left(\frac{T_{\rm ann}}{50\,{\rm MeV}}\right)^{2/3}\left(\frac{7}{N}\right)^{2/3}\left(\frac{\beta}{c_wk_w}\right)^{1/3} &,~\quad \left[(x_\sigma,x_f)=(0,2)\right] \\[2ex]
        3.44\times10^7~{\rm GeV} \left(\frac{T_{\rm ann}}{50\,{\rm MeV}}\right)^{1/2}\left(\frac{7}{N}\right)^{2/2}\left(\frac{\beta}{c_wk_w}\right)^{1/2} &,~\quad \left[(x_\sigma,x_f)=(3,3)\right]
    \end{cases} ~.
\end{equation}

In Sections~\ref{subsec:formation:entrapment} and \ref{subsec:formation:properties}, we will present benchmarks of the $(x_\sigma,x_f)=(0,0)$ and $(3,3)$ cases, while the $(x_\sigma,x_f)=(0,2)$ case will be used as an example of one potential realistic model to be discussed in Section~\ref{subsec:formation:model}, in which $\beta\ll1$ due to its scaling $\sim m_e^2$, $m_e$ being the electron mass, instead of $f^2$. 

Before closing, we also comment on the possible overclosure of domain wall energy, which can be determined by estimating
\begin{equation}
    \frac{\rho_w}{\rho_c} = \frac{T^{x_\sigma}f^{3-x_\sigma}}{2\,t_c^{1/2}t^{1/2}/N}\times\frac{1}{3H^2M_{\rm Pl}^2} = \sqrt{\frac{10}{\pi^2g_*}}\frac{N\,T^{x_\sigma-3}f^{4-x_\sigma}}{M_{\rm Pl}} ~,
\end{equation}
where $\rho_c$ is the critical energy density. For $x_\sigma=0$, $\rho_w$ increase along with decreasing temperature until $T=T_{\rm ann}$, which sets a bound on $f\lesssim10^{3-4}~{\rm GeV}$; as for $x_\sigma=3$, the maximum of $\rho_w$ happens at $T=f$, and the corresponding bound is $f\lesssim10^{18}~{\rm GeV}$.

\section{Baryoids and dark matter-baryon coincidence}\label{sec:formation}

In this section, we outline the road map to realizing the dark matter-baryon coincidence using domain walls and baryoid dark matter. Suppose that the domain wall network has a uniform surface tension $\sigma(t)$, then, given the details of wall evolution and $V_{\rm bias}$, one can determine the annihilation temperature $T_{\rm ann}$ and the baryoid stabilization/formation time $t_{\rm F}$. For a fixed $V_{\rm bias}$, $T_{\rm ann}$ will decrease with increasing $\sigma(t_{\rm ann})$. It can be expected that the wall velocity $v(t)$ will increase during the collapse until it is successfully countered at $t_{\rm F}$ by the gradually increasing Fermi degeneracy pressure of the baryons and electrons inside the false-vacuum domains, \ie, $v(t)\leq v(t_{\rm F})$. Moreover, the thermal pressure generated by relativistic particles produced through proton-anti-proton annihilation before this process freezes out at $T\approx22$~MeV~\cite{Kolb:1990vq} could also help decelerate the collapse in the early stage, but since it does not influence the later collapse dynamics, the main conclusions are not dramatically affected by ignoring it. We also assume that the average baryon mass in the false-vacuum domains is given by $m_{\rm B, F} =  m_{\rm B}-\Delta m_{\rm B}$ with $m_{\rm B}\approx939$~MeV as the average nucleon mass in the ordinary QCD vacuum and $\Delta m_{\rm B} > 0$ (expected in, for example, models predicting domains with a non-zero strong CP phase~\cite{Lee:2020tmi,Bai:2023cqj}) which we explain below, while the baryon number densities at $t=t_{\rm F}$ contained within ordinary baryons and within all false-vacuum domains (which eventually become baryoids) are given by $n_{\rm B}$ and $n_{\rm BD}$, respectively. For an $\Omega_{\rm BD}/\Omega_{\rm B} \approx 5$ to be established, the following conditions must be satisfied:

\begin{figure}[ht!]
    \centering
    \includegraphics[width=0.7\linewidth]{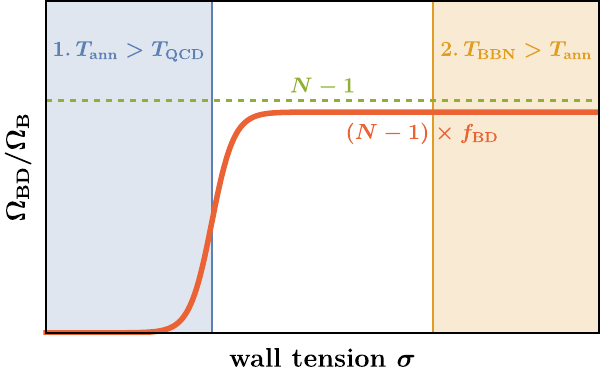}
    \caption{A schematic plot of the relation between the wall tension $\sigma$ and $\Omega_{\rm BD}/\Omega_{\rm B}$ (red line). For a given $V_{\rm bias}$, when $\sigma$ is small and thus $T_{\rm ann}>T_{\rm QCD}$, the baryons, which exist in the form of relativistic quark-gluon plasma, cannot be effectively aggregated by the collapsing domain walls, and thus $\Omega_{\rm BD}/\Omega_{\rm B}\to 0$. As temperature reaches below around $T_{\rm QCD}$, the baryons will gradually become massive hadrons that can be swept along by the moving walls, and thus $\Omega_{\rm BD}/\Omega_{\rm B}$ will eventually reach an asymptotic constant value of $(N-1)\times f_{\rm BD}$ for a large enough $\sigma$ (which will give a low enough $T_{\rm ann}$), with $f_{\rm BD}\approx m_{\rm B,F}/m_{\rm B}\lesssim1.0$ and subject to further $\mathcal{O}(1)$ corrections coming from factors including surface tension energy, vacuum energy, baryon leakage rate, etc.. We also overlay the first and second required conditions listed in Section~\ref{sec:formation} for baryoids to form and explain the dark matter-baryon coincidence on the plot, with the third and fourth conditions being $v(t)\leq v(t_{\rm F})\ll1$ and $\Delta m_{\rm B}>0$, respectively. The actual values of the relevant quantities depend on the details of domain wall evolution (see Section~\ref{sec:DW_summary} and Appendix~\ref{sec:DW}). 
    }
    \label{fig:roadmap}
\end{figure}

\begin{description}
    \item[1. $\boldsymbol{T_{\rm ann}<T_{\rm QCD}\approx 150}$~MeV:] In order for the collapsing domain walls to successfully aggregate baryons inside the false-vacuum domains, baryons must exist in the non-relativistic hadron state for wall-baryon scattering to be effective. Note that since QCD phase transition is a crossover in the normal QCD vacuum~\cite{Aoki:2006we,Bhattacharya:2014ara}, both $m_{\rm B}$ and $\Delta m_{\rm B}$ could be gradually increasing as temperature decreases if , in addition to the former, the latter is also induced by QCD phase transition. As a result, we expect $\Omega_{\rm BD}/\Omega_{\rm B}\to0$ when $T_{\rm ann}\gtrsim T_{\rm QCD}$ (since baryons still exist in the form of relativistic quark-gluon plasma), and will saturate to a constant value as $\sigma$ increases (which leads to a lower $T_{\rm ann}$) if all other conditions to be discussed below are satisfied, as we demonstrate with the red line in the schematic plot shown in Figure~\ref{fig:roadmap}. For the purpose of examining baryoid properties, we assume that $m_{\rm B}$ and $\Delta m_{\rm B}$ will reach their respective asymptotic constant values soon after $T\lesssim T_{\rm QCD}$ and use them for our analysis below.
    
    \item[2. $\boldsymbol{T_{\rm ann}>T_{\rm BBN}\approx 3}$~MeV:] If the walls do not collapse before BBN, the compact objects eventually formed will be bags of heavy elements and thus cannot serve as dark matter candidates. Also, they will influence the process of BBN and contradict the current observations~\cite{Bai:2021ibt,Bringmann:2023opz}.
    
    \item[3. $\boldsymbol{v(t)\leq v(t_{\rm F})\ll1}$ during the collapse of domain walls:] If the wall velocity $v(t)$ becomes relativistic, it will result in two outcomes: the leakage of baryons due to their enhanced scattering with the wall and the complete annihilation of the false-vacuum domains that cannot be stopped by the Fermi degeneracy pressure of the matter inside. Therefore, $v(t)$ must remain non-relativistic until $t_{\rm F}$, which is satisfied as long as $v(t_{\rm F})\ll1$. As we have shown in Section~\ref{sec:DW_summary} and Appendix~\ref{sec:DW}, this may require additional interactions between the domain walls and plasma particles.
    
    \item[4. $\boldsymbol{\Delta m_{\rm B}>0}$:] If $\Delta m_{\rm B}\leq 0$, meaning that the average baryon mass in the false-vacuum domains is equal to or larger than the average nucleon mass in the normal QCD vacuum, these baryoids would be less stable than ordinary nuclei (after including finite-size effects) and could decay to the latter states. Although such decays are expected to take place only near the baryoid surface and thus the baryoids can potentially have a lifetime longer than the age of the universe, we only focus on the case of stable baryoids and leave the studies of long-lived baryoids to the future.

\end{description}
We overlay a summary of these conditions on the schematic plot shown in Figure~\ref{fig:roadmap}. If all of the conditions mentioned above are satisfied, then, for a spontaneously broken $\mathbb{Z}_N$ symmetry (see Section~\ref{subsec:formation:separation}), one can infer $\Omega_{\rm DM}/\Omega_{\rm B} = \Omega_{\rm BD}/\Omega_{\rm B}=(N-1)\times f_{\rm BD}$, where $f_{\rm BD}$ depends on $n_{\rm B}$, $n_{\rm BD}$, $m_{\rm B}$, and $\Delta m_{\rm B}$, as well as other energy components such as surface tension and $V_{\rm bias}$, which are expected to be sub-dominant. In the ideal scenario where most baryons inside the false-vacuum regions are successfully entrapped by the domain walls and $\Delta m_{\rm B}\lesssim m_{\rm B}$, one should expect $f_{\rm BD}\approx m_{\rm B,F}/m_{\rm B}\lesssim1.0$ [see Eq.~\eqref{eq:intro-ratio}], while an $\mathcal{O}(1)$ correction to it is expected from a potential small baryon leakage rate that does not compromise the stability of the baryoid (see Section~\ref{subsec:formation:entrapment} for more discussion). For $f_{\rm BD}$ close to unity, the dark matter--baryon coincidence is naturally explained by the discrete model integer $N-1$.

    In the following, we discuss in detail the formation mechanism and properties of baryoids. The formation mechanism includes two stages: baryon separation, the key stage that establishes the observed $\Omega_{\rm DM}/\Omega_{\rm B}=\Omega_{\rm BD}/\Omega_{\rm B}$, and baryon entrapment, the enclosing and stabilization of these domain-wall-bounded solitons which later become baryoids. Once these baryoids survive to $T=0$, we can analyze their properties, including their size, mass, and stability compared to ordinary nuclei, which will be used to study their search prospects in Section~\ref{sec:pheno}.

\subsection{Baryon separation}\label{subsec:formation:separation}

If there exists a spontaneously broken $\mathbb{Z}_N$ symmetry in the early universe, the resulting domain walls will naturally divide the universe in a way that each of the $N$-fold degenerate vacua will occupy $\sim 1/N$ of the total volume of the universe with roughly the same baryon number density given that $V_{\rm bias}$ only becomes important in later time (at around $t_{\rm ann}$), thus establishing a baryon-number ratio of $N-1:1$ between the $N-1$ false- and one true-vacuum domains per group. Since the Planck data~\cite{Planck:2018vyg} indicates $\Omega_{\rm DM}/\Omega_{\rm B}\approx 5.4$, we expect that a natural interpretation can be obtained with $N=7$ or $8$, depending on the baryon leakage rate during the baryon entrapment process and the baryon mass in the false-vacuum domains, $m_{\rm B,F}=m_{\rm B}-\Delta m_{\rm B}$. Although one can also consider $N\geq9$ with a higher baryon leakage rate or a smaller baryon mass to establish the observed ratio, we do not consider such scenarios given the inconceivability of such large values of $N$ and the concern for baryoid stability that depends on the balance between the Fermi degeneracy pressure of the baryons and electrons inside the baryoid and the collapsing force, the latter including both $V_{\rm bias}$ and the curvature force.

We assume that the domain walls form at $T=T_c=f$,\footnote{
For $x_\sigma=3$, although the domain wall tension $\sigma\propto T^3$ is not associated with the scale $f$, we still assume that it quantifies the correlation length of the order parameter field after spontaneous symmetry breaking.}
resulting in an initial domain wall length scale of $L(t_c)\simeq H^{-1}(t_c)/N=2\,t_c/N$. Assume that friction is sufficiently strong so that the walls are frustrated in the beginning of the expansion, they will evolve according to the stretching behavior [see Eq.~\eqref{eq:Lv_st:main} and Figure~\ref{fig:Lt_vt}] until the wall velocity $v$ grows large enough to affect the overall dynamics, which marks the beginning of the transient Kibble regime at $t=t_{\rm K}$ [see Eq.~\eqref{eq:t_K:main}]. For the purpose of entrapping baryons, allowing baryoids to stabilize, and avoiding the baryon inhomogeneity constraint (see Section~\ref{subsec:pheno:inhomogeneity}), we require that walls begin to collapse at $t=t_{\rm ann}<t_{\rm K}$, which is realized by bounding $V_{\rm bias}$ according to Eq.~\eqref{eq:V_bias_bounds:main}.

\subsection{Baryon entrapment}\label{subsec:formation:entrapment}

When the domain walls begin to collapse at $t_{\rm ann}$, they will potentially sweep the baryons along to the interior of the false-vacuum domains, eventually forming baryoids. This is possible as long as either the wall velocity remains non-relativistic or the potential barrier exerted by the wall is sufficiently hard to tunnel through so that the baryons inside remain trapped until the baryoids stabilize. To see this, we consider the scattering of baryons in the false-vacuum domain against an incoming planar domain wall with an average ensemble velocity $v\hat{z}$ in the rest frame of the wall. Assuming that the baryons are thermalized and thus follow the Fermi-Dirac distribution, we have
\beqa
    f_{\rm B}(v) = \frac{g_{\rm B}}{\exp[\gamma(v)(E-p_zv)/T]+1} ~,
\eeqa
where $\gamma(v)=(1-v^2)^{-1/2}$, $E=\sqrt{(m_{\rm B}-\Delta m_{\rm B})^2+p^2}$, and $g_{\rm B}=2\times2=4$ denotes the spin and isospin degrees of freedom of baryons. Here, we have assumed that the baryon chemical potential $\mu_{\rm B}\to 0$ due to the low baryon number density. Suppose that the domain wall acts as a potential barrier such that baryons with $p_z>p_{z,{\rm th}}$ can transmit through the wall~\cite{Blasi:2022ayo} and that the wall velocity $v$ is constant, we can define the average  baryon leakage rate as
\beqa\label{eq:leakage}
    1-S(v) \equiv \frac{\int d^2p \int_{p_{z,{\rm th}}}^\infty dp_z f_{\rm B}(v)}{\int d^3p f_{\rm B}(v)} ~,
\eeqa
with $S(v)$ as the trapping rate of baryons inside the false-vacuum domain.
Note that this estimation is only valid for a thick domain wall. For thin-wall scenarios, one needs to consider the correlation between the height and width of the potential barrier, which does not necessarily lead to a higher baryon leakage rate.

\begin{figure}[ht!]
    \centering
    \includegraphics[width=0.48\linewidth]{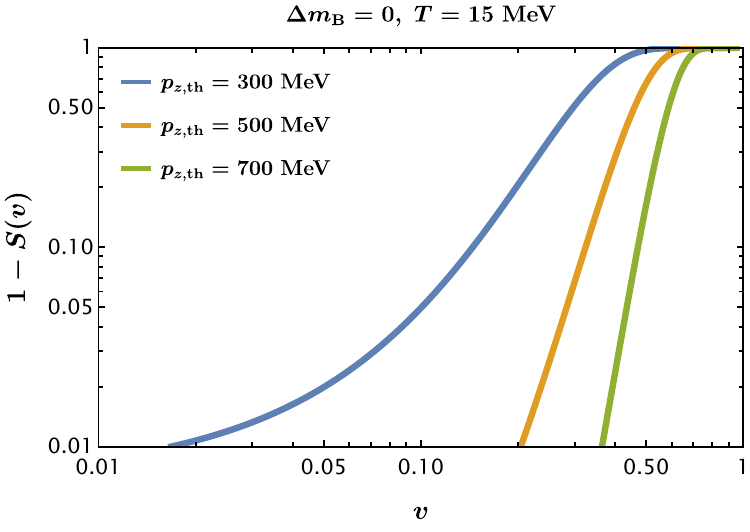}
    \includegraphics[width=0.48\linewidth]{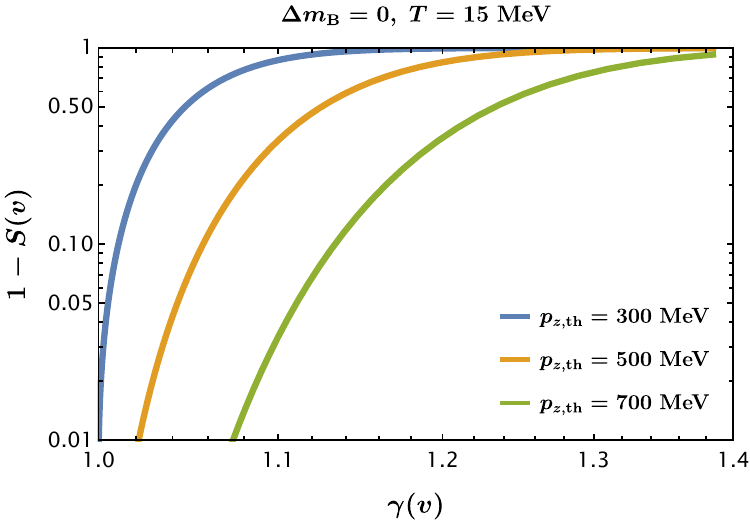}
    \caption{The $1-S(v)$ profiles against $v$ and $\gamma(v)$ with the parameter choices $\Delta m_{\rm B}=0$, $T=15$~MeV, and $p_{z,{\rm th}}=300,500,700$~MeV [see Eq.~\eqref{eq:leakage}]. Note that $S(v)$ is the baryon trapping rate in the false-vacuum pockets and correspondingly $1-S(v)$ the leakage rate.
    }
    \label{fig:leakage}
\end{figure}

We take $\Delta m_{\rm B}=0$ as a conservative estimation (lighter baryons are more likely to transmit through the wall because of the less suppressed Boltzmann factor). As we will discuss below, we want the baryoids to form at $T=T_{\rm F}\lesssim15$~MeV to avoid the evaporation constraint, and thus we choose $T=15$~MeV for the estimation. While the scale of $p_{z,{\rm th}}$ depends on the microscopic details of the baryon-wall interaction, we choose $p_{z, {\rm th}}=300,500,700$~MeV as examples and show the corresponding $1-S(v)$ profiles against $v$ and $\gamma(v)$ in Figure~\ref{fig:leakage}. One can see that the leakage rate increases very rapidly with increasing $v$ unless $p_{z,{\rm th}}$ is high enough (for higher $p_{z,{\rm th}}$, one can expect the leakage rate to be even more suppressed). If we further consider the acceleration caused by the collapsing force, which can only be eventually countered by the Fermi degeneracy pressure of the matter inside the false-vacuum domains, the actual leakage rate might be even higher, making the formation of baryoids even more challenging. Despite the fact that this is not completely impossible, since the purpose of this study is to demonstrate how baryoid dark matter can explain the dark matter-baryon coincidence, we only focus on the scenarios with $v(t)\ll1$, for which one needs to introduce a way to efficiently slow down the wall motion. As we have summarized in Section~\ref{sec:DW_summary} and analyzed in Appendix~\ref{subsec:DW:collapse}, this is possible when friction is dominant, \ie, $\ell_f^{-1}\gg 3H+\dot{\sigma}/\sigma$. This also provides suitable conditions for the walls to remain in the stretching regime (see Section~\ref{sec:DW_summary} and Appendix~\ref{subsec:DW:frustrated}) so that $a^{-1}(t_{\rm ann})L(t_{\rm ann})\lesssim d_n(t_{\rm ann})\approx d_p(t_{\rm ann})$ assuming $T_{\rm ann}>T_\nu$, which allows us to avoid the baryon inhomogeneity constraint~\cite{Bagherian:2025puf} while also evading the microlensing constraint as each baryoid will contain less baryons and have smaller masses (see Section~\ref{sec:pheno}).

We now examine the pressure balance conditions. We first denote the numbers and chemical potentials of electrons, protons, and neutrons inside a baryoid by $N_{e,p,n}$ and $\mu_{e,p,n}$, respectively. For a charge-neutral baryoid in $\beta$-equilibrium, two conditions must be satisfied: $N_e=N_p$ and $\mu_n=\mu_p+\mu_e$. In the relativistic regime, $n_e\approx\mu_eT^2/3\approx \eta_{\rm B}n_\gamma(T)$, where $\eta_{\rm B}\approx(1+\Omega_{\rm DM}/\Omega_{\rm B})\times(6\times10^{-10})$ includes both the baryon and dark matter abundance and $n_\gamma(T)=2\,\zeta(3)\,T^3/\pi^2$, $\zeta$ denoting the Riemann Zeta function. Consequently, this gives $\mu_e\approx3\eta_{\rm B}n_\gamma(T)/T^2=(2.81\times10^{-9}) \times T\ll m_{n,p}\lesssim \mu_{n,p}$ for $T<T_{\rm QCD}\approx150$~MeV, which implies $\mu_n\approx\mu_p$ and correspondingly $N_e=N_p\approx N_n\approx N_{\rm B}/2$ at $T\gg Q\approx1.3$~MeV. As a result, $P_F=P_{\rm B}+P_e\approx P_e$ is dominated by the electron pressure. As a conservative estimate, we only consider the $T$-independent degeneracy pressure $P_e^0$ (the pure thermal pressure, which only takes effect through a non-zero $v$ because of the homogeneity in the electron plasma, has already been included as a part of friction, while we do not consider the potential slight enhancement in the degeneracy pressure through thermal corrections), which is given by
\begin{equation}
    P_e^0(T) = \frac{m_e^4}{24\pi^2}\left[x(x^2-3)\sqrt{x^2+1}+3\sinh^{-1}(x)\right] ,~ x=\frac{[3\pi^2 N_e/V(T)]^{1/3}}{m_e} ~,
\end{equation}
where $N_e \sim n_{\rm B,F}(t_{\rm ann})R(t_{\rm ann})^3/2$ with $R(t_{\rm ann})=2\,t_{\rm ann}^{1/2}\,t_c^{1/2}/N$ in the stretching regime, $V(T) \sim R(T)^3$, and we have assumed zero baryon leakage (which should be valid as long as the wall velocity remains non-relativistic). When $P_e^0(T_{\rm F})=s\,\sigma(T_{\rm F})/R(T_{\rm F})+V_{\rm bias}$ with $s=2$ at the formation temperature $T_{\rm F}$, if $v^2(T_{\rm F})\ll1$, then the baryoid will stabilize and survive until the present time. Note that one can further take into account the non-zero leakage rate associated with a considerable value of $v(T)$ that allows tunneling of the baryons across the wall, which can help with the precise establishment of $\Omega_{\rm DM}/\Omega_{\rm B}\approx 5.4$. Since the purpose of this study is to introduce and justify the formation mechanisms, we leave the detailed numerical construction of these scenarios to future studies.

\begin{figure}[ht!]
    \centering
    \includegraphics[width=0.48\linewidth]{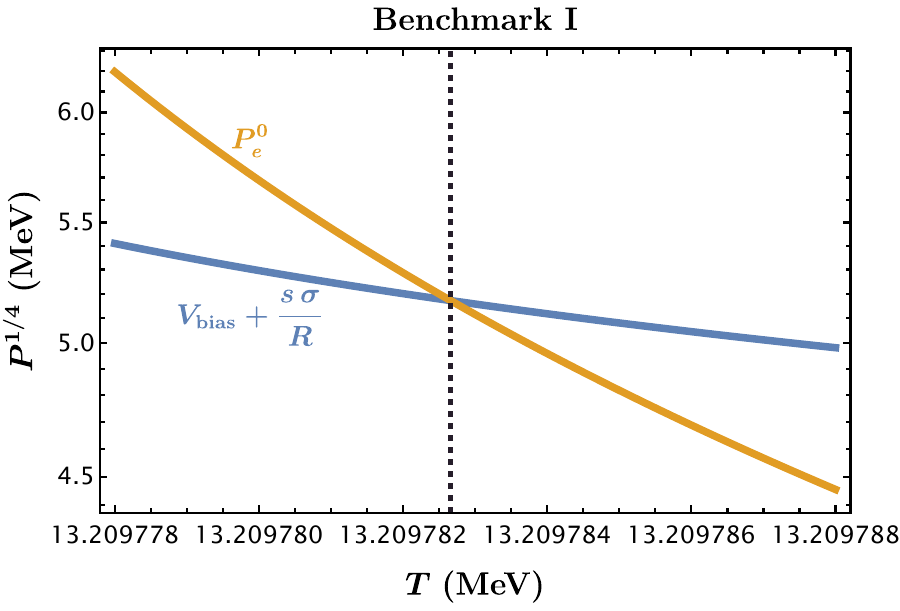}
    \includegraphics[width=0.48\linewidth]{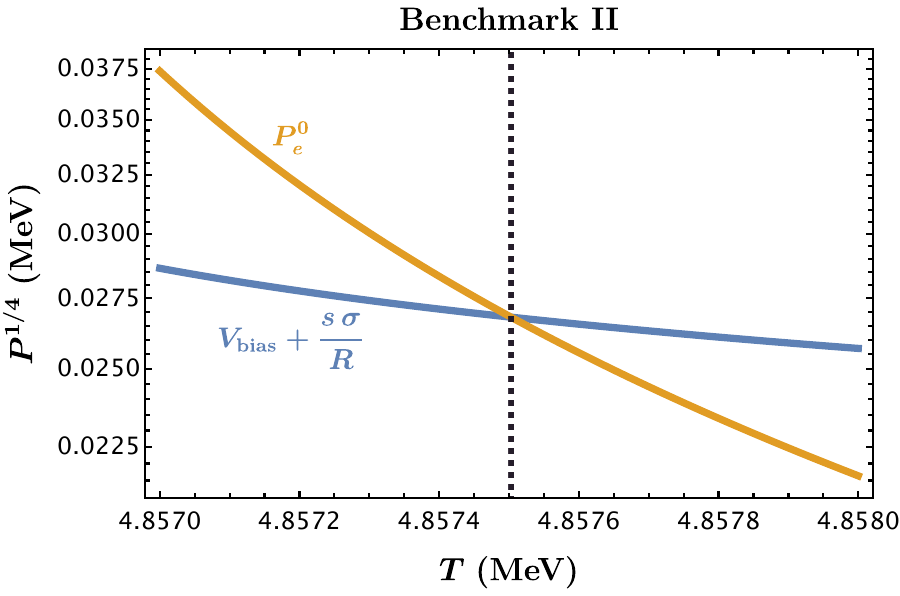}
    \includegraphics[width=0.48\linewidth]{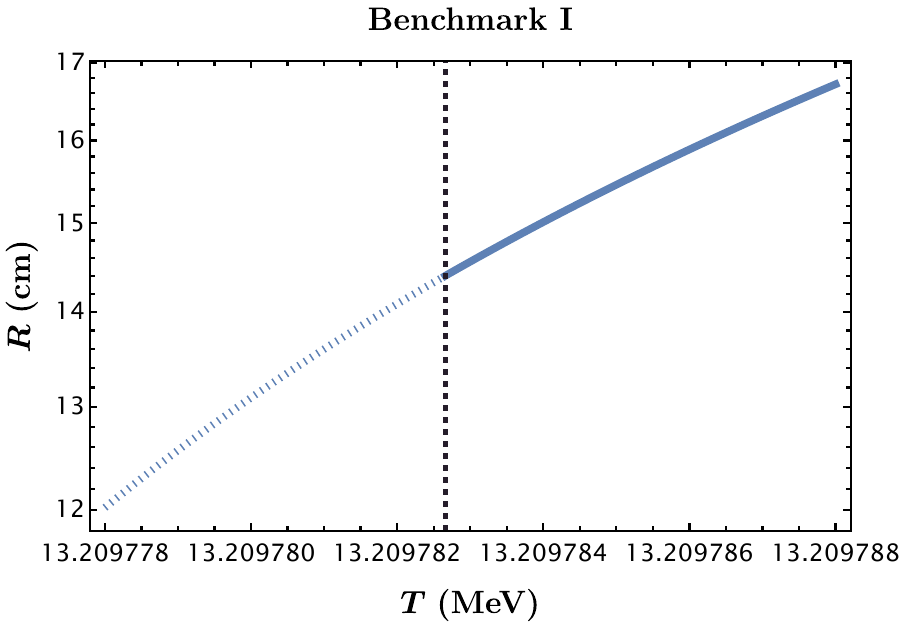}
    \includegraphics[width=0.48\linewidth]{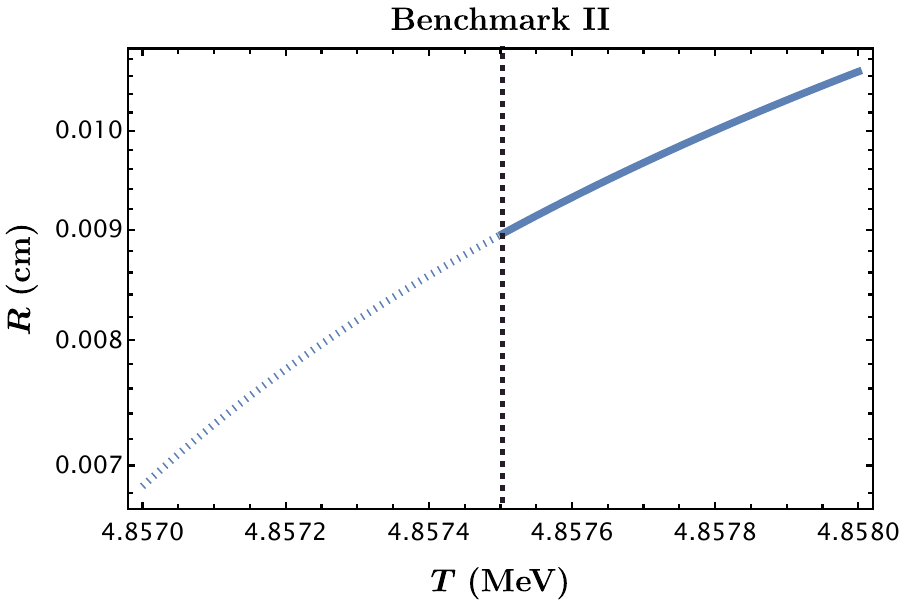}
    \includegraphics[width=0.48\linewidth]{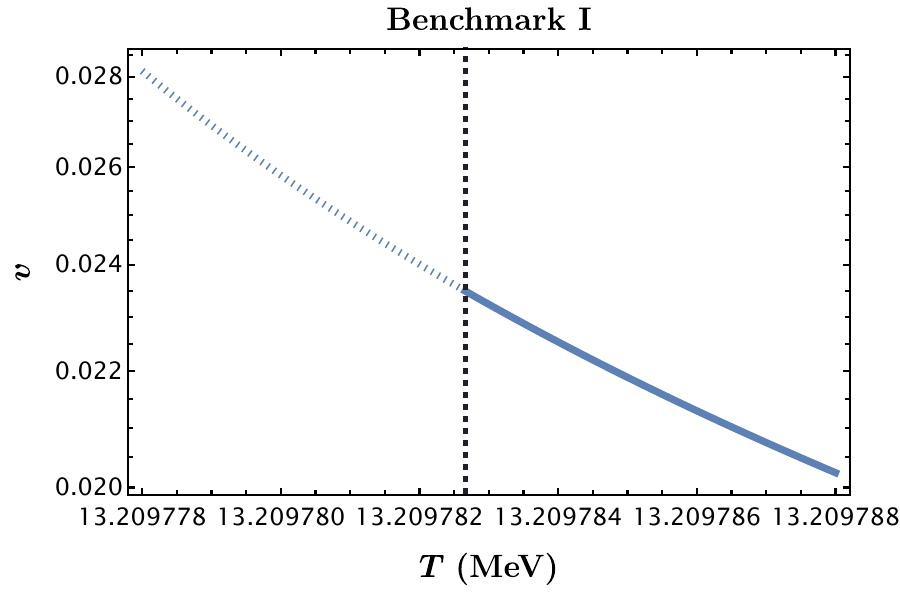}
    \includegraphics[width=0.48\linewidth]{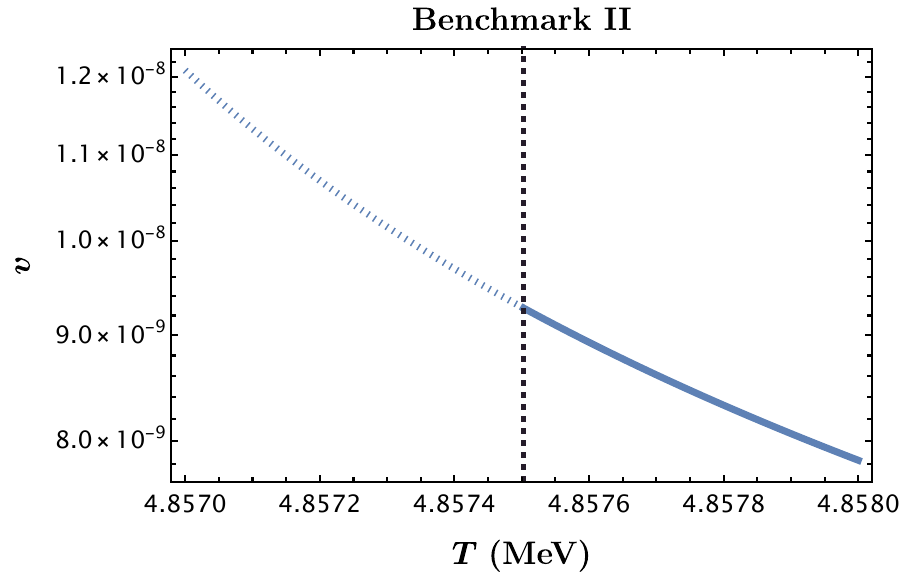}
    \caption{The $P_e^0(T)$ and $V_{\rm bias}+s\,\sigma(T)/R(T)$ (top row), $R(T)$ (middle row), and $v(T)$ (bottom row) profiles in the late stage of the collapse for benchmarks I (left column) and II (right column), respectively. The vertical dashed lines denote $T=T_{\rm F}$. For details of the benchmarks, see main text and Eqs.~\eqref{eq:benchmark_I&II_1} and \eqref{eq:benchmark_I&II_2}. }
    \label{fig:benchmarks}
\end{figure}

We present two benchmarks in the following, whose parameters are given by (all energy scales are given in MeV)
\begin{equation}\label{eq:benchmark_I&II_1}
\begin{aligned}
    \mathbf{\rm I}&:~(T_{\rm ann},V_{\rm bias}^{1/4},f,x_\sigma,x_f,N,\beta,c_wk_w) = (50,\,1,\,6.38\times10^{4},\,0,\,0,\,7,\,1,\,1) ~, \\
    \mathbf{\rm II}&:~(T_{\rm ann},V_{\rm bias}^{1/4},f,x_\sigma,x_f,N,\beta,c_wk_w) = (5,\,0.01,\,1.33\times10^{10},\,3,\,3,\,7,\,0.1,\,1) ~,
\end{aligned}
\end{equation}
with the $P_e^0(T)$ and $V_{\rm bias}+s\,\sigma(T)/R(T)$, $R(T)$, and $v(T)$ profiles in the late stage of the collapse shown in Figure~\ref{fig:benchmarks}. One can see that $P_e^0$ balances the collapse force at $T_{\rm F}\lesssim15$~MeV with $v(T)\leq v(T_{\rm F})\ll1$ always satisfied for both benchmarks. We emphasize that the rapid variations in these profiles do not imply fine-tuning but are a natural result of the $s\,\sigma/R$-acceleration in Eq.~\eqref{eq:VOS:R:main} which significantly boosts both $P_e^0$ and the collapse force in the late stage of the collapse. Asymptotically, we approximate $dv/dt\approx s/R$ and consequently $d^2R/dt^2\approx -2/R$ in Eq.~\eqref{eq:VOS:R:main}, leading to $R(t)\sim R(t_{\rm ann})\exp\{-{\rm ierf}^2[-2(t-t_{\rm ann})/\sqrt{\pi}R(t_{\rm ann})]\}$ with ${\rm ierf}$ denoting the inverse error function, which explains why $R$ is decreasing at an exponential pace. Therefore, we conclude that these baryoids can stabilize and survive to $T=0$ with zero baryon leakage. In this limit, we can derive
\begin{equation}\label{eq:NB:main}
\begin{aligned}
    N_{\rm B} &\sim n_{\rm B,F}(t_{\rm ann})R(t_{\rm ann})^3 \approx \eta_{\rm B}\, n_\gamma(T_{\rm ann})\left(\frac{2\,t_{\rm ann}^{1/2}\,t_c^{1/2}}{N}\right)^3 =\frac{8\,\eta_{\rm B}}{N^3}n_\gamma(T_{\rm ann})\left(\sqrt{\frac{90}{4\pi^2g_*}}\frac{M_{\rm Pl}}{T_{\rm ann}T_c}\right)^3 \\
    &=  3.35\times10^{37}\left(\frac{7}{N}\right)^3\left(\frac{100\,{\rm GeV}}{f}\right)^3 ~.
\end{aligned}
\end{equation}
Consequently, this gives $N_{\rm B}=1.29\times10^{38}$ and $N_{\rm B}=1.44\times10^{22}$ for the two benchmarks, respectively. The average baryon number stored in the baryoids formed in the stretching wall regime is smaller by a factor of $\sim (T_{\rm ann}/f)^3 = 1.25\times 10^{-10}\times (T_{\rm ann}/50\,\mbox{MeV})^3 (100\,\mbox{GeV}/f)^3$, compared to the ones formed during the scaling region with $N_{\rm B}^{\rm sc} \sim 3.41\times10^{46}\times (7/N)^3 (50\,\mbox{MeV}/T_{\rm ann})^3$.

We can use Eq.~\eqref{eq:V_bias_bounds:main} and Eq.~\eqref{eq:NB:main} to estimate the scaling of the lower bounds on $N_{\rm B}$ as
\begin{equation}\label{eq:NB_lower}
    N_{\rm B} \gtrsim \begin{cases}
        1.91\times10^{36}\left(\frac{7}{N}\right)^{9/5}\left(\frac{50\,{\rm MeV}}{T_{\rm ann}}\right)^{12/5}\left(\frac{c_wk_w}{\beta}\right)^{3/5} &,~\quad \left[(x_\sigma,x_f)=(0,0)\right] \\[2ex]
        1.42\times10^{29}\left(\frac{7}{N}\right)^{1}\left(\frac{50\,{\rm MeV}}{T_{\rm ann}}\right)^{2}\left(\frac{c_wk_w}{\beta}\right)^{1} &,~\quad \left[(x_\sigma,x_f)=(0,2)\right] \\[2ex]
        2.91\times10^{20}\left(\frac{7}{N}\right)^{0}\left(\frac{50\,{\rm MeV}}{T_{\rm ann}}\right)^{3/2}\left(\frac{c_wk_w}{\beta}\right)^{3/2} &,~\quad \left[(x_\sigma,x_f)=(3,3)\right]
    \end{cases} ~,
\end{equation}
and, requiring $f\gtrsim T_{\rm QCD}\approx150$~MeV, we have $N_{\rm B}\lesssim 1.0 \times 10^{46}$. Note that the actual lower bounds on $N_{\rm B}$ should be more stringent than those listed in Eq.~\eqref{eq:NB_lower} since saturating the upper bound on $V_{\rm bias}$ (and thus $f$) will in principle lead to a relativistic $v(t_{\rm F})$, which is also why even though there is no such upper bound for the case of $x_f=3$, we still consider it as a rough bound for the velocity to remain non-relativistic (see Section~\ref{sec:DW_summary}). Moreover, for the $(x_\sigma,x_f)=(0,2)$ case that we will study in the Section~\ref{subsec:formation:model}, $\beta\ll1$ and thus the lower bound on $N_{\rm B}$ will be much higher than the scale presented above. On the other hand, if $f$ is low enough to be in the experimentally reachable range, it might be constrained by some existing searches depending on the particle nature of the $\mathbb{Z}_N$ order parameter field, which we do not consider in this paper and leave to future studies.

Before concluding this section, we comment on a few potential issues. The first one is the possible evaporation of baryoids induced by the flux of energetic neutrinos. Ref.~\cite{Alcock:1985vc} has studied the continuous evaporation of strange quark nuggets (with similar properties to baryoids) from $T=150$~MeV to $T=1$~MeV, assuming that they are formed at $T=150$~MeV due to a fictitious first-order QCD phase transition and the evaporation lasts until the neutrinos become too cold to effectively impact the nuggets. They came to the conclusion that only strange quark nuggets with $N_{\rm B}\gtrsim10^{42}$ can survive until the present time. In contrast, in the scenario of baryoids formed from collapsing domain walls, since the formation temperature is no longer restricted to the QCD phase transition temperature and can be as low as $T_{\rm F}\sim 5-15$~MeV, these baryoids will in principle be safe from evaporation, and thus we do not consider its influence on the baryoid formation process. Second, one might concern the population of strange baryons and their effects on baryoid properties. Consider the lightest strange baryon $\Lambda^0$ with $m_{\Lambda^0}\approx1.116$~GeV~\cite{ParticleDataGroup:2024cfk} at $T=T_{\rm F}=15$~MeV, one can derive $n_{\Lambda^0}/n_{\rm B}\approx\exp[-(m_{\rm \Lambda^0}-m_{\rm B})/T_{\rm F}]\sim\mathcal{O}(10^{-5})$, and thus strange baryons can be safely ignored in this regime. Finally, the proton-anti-proton annihilation process freezes out at $T\lesssim 22$~MeV~\cite{Kolb:1990vq}, and thus we should also not be concerned with it.

\subsection{Baryoid properties}\label{subsec:formation:properties}

Once these baryoids survive and stabilize until the present time, we can calculate their physical properties. The quantities of the most interest are their masses and sizes, which are determined by minimizing the total energy of the baryoid system that comprises the matter energy, the surface tension energy, and the bias potential energy. Unlike the finite-temperature scenario studied in Section~\ref{subsec:formation:entrapment} where the excited electron gas plays the main role, the protons and electrons inside the baryoids at zero temperature might combine to form neutrons to change the overall energy configuration. As a result, we introduce the parameter ``proton fraction'' $r_p\equiv N_p/N_{\rm B}=N_e/N_{\rm B}$ [and thus $N_n=(1-r_p)N_{\rm B}$], with which we can write down the total energy of a baryoid with $N_{\rm B}$ total baryons at $T=0$ in the thin-wall limit as
\begin{equation}
    E(r_{p,c},R_c) = \min_{r_p,R}E(r_p,R) = \frac{4\pi R^3}{3}\left(\sum_{f=n,p,e}\rho_f + V_{\rm bias}\right) + 4\pi R^2\sigma(T=0) ~,
\end{equation}
where
\begin{equation}
    \rho_f = \frac{m_f^4}{4\pi^2}\left[
        x_f\sqrt{1+x_f^2}\left(\frac{1}{2}+x_f^2\right) - \frac{1}{2}\sinh^{-1}\left(x_f\right)
    \right] ,~ x_f = \frac{\left(9\pi N_f/4\right)^{1/3}}{m_fR} ~,
\end{equation} 
$N_f$ is the number of the fermion $f$ in the baryoid, and we have chosen $m_n=m_p=m_{\rm B,F}=m_{\rm B}-\Delta m_{\rm B}$ with $\Delta m_{\rm B}=100$~MeV. Note that for $x_\sigma>0$ with $\sigma=T^{x_\sigma}f^{3-x_\sigma}$, one might naively assume $\sigma=0$ at $T=0$; however, it is imaginable that during the cooling process, the $T$-dependence of the wall tension, which implies a direct interaction between the plasma and the domain wall, eventually becomes replaced by a fixed scale related to the wall-baryon interaction. Despite this, we expect the surface tension energy in this case to be much smaller than the other two energy components, and thus we set it to zero in the following analysis. Consequently, for the two benchmarks [see Eq.~\eqref{eq:benchmark_I&II_1}], we have 
\begin{equation}\label{eq:benchmark_I&II_2}
\begin{aligned}
    \mathbf{\rm I}:&~ (E/N_{\rm B},R_c,r_{p,c}) = (841~{\rm MeV},\, 3.78~{\rm cm},\, 2.32\times10^{-5}) ~, \\
    \mathbf{\rm II}:&~ (E/N_{\rm B},R_c,r_{p,c}) = (839~{\rm MeV},\, 3.61\times10^{-3}~{\rm cm},\, 2.66\times10^{-12}) ~,
\end{aligned}
\end{equation}
which give $\Omega_{\rm BD}/\Omega_{\rm B}=(N-1)\times(E/N_{\rm B})/m_{\rm B}\approx5.4$, successfully realizing the dark matter-baryon coincidence with stable baryoids. Finally, their energy densities are $\rho_{\rm BD}=8.55\times10^{11}~{\rm g/cm^3}$ and $1.09\times10^5~{\rm g/cm^3}$, respectively, both of which are much lower than the nuclear-scale energy density $\sim10^{14}~{\rm g/cm^3}$. This is mostly due to the small $V_{\rm bias}$ in these scenarios, which is not strong enough to compress the baryoids to an even higher density. 

We further provide an analytical approximation below: treating neutrons and protons as non-relativistic gases and electrons as a relativistic gas, we have
\begin{equation}
    \rho_{n,p} = \frac{10\,m_{\rm B,F}^2\,x_{n,p}^3+3\,x_{n,p}^5}{30\,m_{\rm B,F}\,\pi^2} ,~\qquad  \rho_e = \frac{x_e^4}{4\pi^2} ~.
\end{equation}
Assuming $r_p\ll1$, we can determine
\begin{equation}
    r_{p,c} = \argmin_{r_p}\sum_{f=n,p,e}\rho_f \to r_{p,c} = \frac{9\,\pi N_{\rm B}}{32\,m_{\rm B,F}^3\,R^3} ~.
\end{equation}
Next, we consider two limit cases. The first one is when the $V_{\rm bias}$ contribution to the total energy is negligible compared to the $\sigma$ contribution, such as in benchmark I. In this case, we can derive
\begin{equation}
    R_c \approx \argmin_{R} \frac{4\pi R^3}{3}\left(\sum_{f=n,p,e}\rho_f + 4\pi R^2\sigma  \right) \to R_c^{\rm I}\approx 0.544 \frac{N_{\rm B}^{5/12}}{(m_{\rm B,F}\,\sigma)^{1/4}} \approx 3.78~{\rm cm} ~,
\end{equation}
and $r_{p,c}^{\rm I}\approx2.76\times10^{-5}$. The second case, on the contrary, is when we can neglect the $\sigma$ contribution compared to the $V_{\rm bias}$ contribution, such as in benchmark II. For this, we have
\begin{equation}
    R_c \approx \argmin_{R} \frac{4\pi R^3}{3}\left(\sum_{f=n,p,e}\rho_f + V_{\rm bias}\right) \to R_c^{\rm II} \approx 0.706\frac{N_{\rm B}^{1/3}}{(m_{\rm B,F}V_{\rm bias})^{1/5}} \approx 3.52\times10^{-3}~{\rm cm} ~,
\end{equation}
and $r_{p,c}^{\rm II}\approx3.81\times10^{-12}$. Both results agree very well with the exact results presented above.

Before closing, we note that these baryoids might further stabilize due to the nuclear forces between the entrapped baryons and thus possess an even higher energy density. In the case where $V_{\rm bias}$ can be much higher because of a sufficiently strong friction [see Eq.~\eqref{eq:V_bias_bounds:main} and the discussion in Section~\ref{subsec:formation:model}], these baryoids might turn out to be closer to the nuclear-scale density. We will assume that all baryoids will eventually reach the nuclear-scale energy density when estimating the phenomenological constraints in Section~\ref{sec:pheno} and note in passing that these constraints could shift around for different baryoid configurations.

\subsection{Potential realization with QCD-anomalous \texorpdfstring{$\mathbb{Z}_N$}{} symmetry}\label{subsec:formation:model}

So far in this study, we have proposed a bottom-up framework to explain the dark matter-baryon coincidence using collapsing $\mathbb{Z}_N$ domain walls. In this section, we suggest a potential top-down realization of the baryoid formation mechanism using a QCD-anomalous $\mathbb{Z}_N$ symmetry, which we studied in Ref.~\cite{Bai:2023cqj}. Introducing a complex scalar field $S$ that transforms as $S\to \exp(i\,2\pi/N)S$ under $\mathbb{Z}_N$, we can write down a $\mathbb{Z}_N$-symmetric potential
\begin{equation}
    V(S) = - m^2 SS^\dagger + \lambda (SS^\dagger)^2 - \mu (S^N + S^{\dagger N}) ~,
\end{equation}
whose symmetry will be spontaneously broken at $T\sim f=T_c$, leading to the formation of a domain wall network with tension $\sigma\sim f^3$ that divides the universe into roughly equal numbers of domains of vacua $\langle S\rangle=f\exp(i\,\theta)=f\exp(i\,2\pi j/N),\,j=0,1,\cdots,N-1$. For $N>4$, as we have also pointed out in Ref.~\cite{Bai:2023cqj}, $V(S)$ contains non-renormalizable terms. Moreover, for odd $N$'s, $V(S)$ is apparently unbounded from below, while for even $N$'s this can also be the case within specific parameter space. Both of these imply the need for a UV-completion of the model. For the purpose of this study, we assume that the UV-completed model exists and does not affect the phenomenological study here, especially for our benchmark studies in which we choose $N=7$. Finally, because of the previous reasons, $\mu$ can either be of the same scale as $f$ or be suppressed by the UV cutoff scale. In the latter case, $\theta$ could gain a relatively light mass as in the scenario of an axion (or axion-like-particle) while the domain wall tension could also be much smaller than the expected scale of $\sim f^3$, which could lead to the generation of additional light relics in the universe. While such scenarios are in principle worth exploring, we do not consider them in this study and simply assume $\mu\sim f$ where $f$ is not much smaller than the UV cutoff scale.

If $\theta$ couples to $n_f$ pairs of heavy quarks $\psi$ that are chiral under $\mathbb{Z}_N$ but vector-like under QCD with $\mathbb{Z}_N$ charge $q_\psi$ in the fundamental representation of color $SU(3)$ through the Yukawa interaction
\begin{equation}
    \mathcal{L}_{\rm Yuk} = -\sum_{n_f}y_SS\bar{\psi}_L\psi_R + {\rm H.c.} ~,
\end{equation}
the QCD instanton effects will accordingly generate an effective chiral potential. At $T=0$ and the leading order in chiral expansion, the two-flavor potential is given by~\cite{DiVecchia:1980yfw}
\begin{equation}\label{eq:V_chiral}
    V(\theta) = -m_\pi^2 f_\pi^2\sqrt{1-\frac{4m_um_d}{(m_u+m_d)^2}\sin^2\left(\frac{n_f\,q_\psi\,\theta}{2}\right)} ~,
\end{equation}
where $m_\pi=135$~MeV, $f_\pi=92$~MeV, and $m_u/m_d=0.49$~\cite{Fodor:2016bgu} in the $\overline{\rm MS}$ scheme at 2~GeV. If ${\rm gcd}(n_fq_\psi,N)=1$ (gcd standing for ``greatest common divisor''), then $V(\theta)$ will break the $N$-fold degeneracy and play the role of $V_{\rm bias}$.\footnote{For the strong CP problem, we have assumed that CP is an exact symmetry in UV realized through some Nelson-Barr-like models. Moreover, for $n_f$ flavors of $\psi$ and an integer-valued $q_\psi$, the $\mathbb{Z}_N$ degeneracy can only be broken to $\mathbb{Z}_{{\rm gcd}(n_fq_\psi,N)}$. For more discussion on these two problems, see Ref.~\cite{Bai:2023cqj}.
} As suggested by Ref.~\cite{Lee:2020tmi} using the chiral perturbation theory, one can derive
\begin{equation}\label{eq:m_B_theta}
    m_{\rm B}(\theta) = m_0 - 4c_1m_\pi^2(\theta) - \frac{3g_A^2m_\pi^3(\theta)}{32\pi f_\pi^2} ~,
\end{equation}
where $m_0\simeq865$~MeV~\cite{Hoferichter:2015hva}, $g_A=1.27$, $c_1=-1.1~{\rm GeV}^{-1}$~\cite{Hoferichter:2015tha}, and
\begin{equation}\label{eq:m_pi_theta}
    m_\pi^2(\theta) = m_\pi^2\sqrt{1-\frac{4m_um_d}{(m_u+m_d)^2}\sin^2\left(\frac{n_f\,q_\psi\,\theta}{2}\right)} = -\frac{V(\theta)}{f_\pi^2} ~.
\end{equation}
Consequently, $\Delta m_{\rm B}\sim\mathcal{O}(10-100)~{\rm MeV}$ in domains of $\vert\theta\vert\lesssim\pi$, which naturally provides a QCD-triggered mechanism to make baryoids more stable than ordinary nuclei while also correlating $\Delta m_{\rm B}(\theta)$ to the chiral potential $V(\theta)$. Nevertheless, as $V(\theta)\sim[\mathcal{O}(100)~{\rm MeV}]^4$, the requirement for the wall velocity $v(t)\ll1$ until $t=t_{\rm F}$ may not be satisfied without additional elements. We discuss two possible directions for future explorations.

The first one is to introduce a sufficiently strong coupling between the walls and plasma so that the friction parameter $\beta$ can be of multiple orders of magnitude large to significantly slow down the wall expansion and collapse, allowing a higher scale for $V_{\rm bias}$ as reflected in Eq.~\eqref{eq:V_bias_bounds:main}. While this is challenging for a perturbative theory, we expect there to be some non-trivial (perhaps also nonperturbative) mechanism that could realize this in the early universe. For instance, Ref.~\cite{Bodeker:2017cim} proposes the possibility of $\gamma$-enhanced soft processes, which limits the bubble wall velocity to $\gamma\sim1/\alpha$, $\alpha$ being the fine structure constant, in the context of first-order electroweak phase transition. In this case, the baryoid energy density would be closer to the nuclear-scale density, in contrast to the cases when $V_{\rm bias}$ is much suppressed below the QCD scale.

The second direction is to set $n_fq_\psi=N$ so that $V(\theta)$ does not break any degeneracy; instead, there might exist some other QCD-triggered $V_{\rm bias}$ (\eg, $\mu\cos\theta\,(\bar{u}\gamma^5u+\bar{d}\gamma^5d)\to\mu\cos\theta\,\langle\bar{u}u+\bar{d}d\rangle$ with a scale $\mu \langle \bar{u}u + \bar{d}d\rangle\ll[\mathcal{O}(100)~{\rm MeV}]^4$) which, because of its supposed correlation to $\Delta m_{\rm B}$ through QCD [see Eqs.~\eqref{eq:V_chiral}, \eqref{eq:m_B_theta}, and \eqref{eq:m_pi_theta}], will suppress the baryon mass difference between different $\theta$ domains. As an estimate, we take $\Delta m_{\rm B}=5$~MeV in the following benchmark study. In this case, we can introduce a coupling between $\theta$ and electrons in a similar fashion to the axion-fermion coupling studied in Ref.~\cite{Blasi:2022ayo} to induce the friction. Accordingly, we have $x_\sigma=0$, $x_f=2$, and $\beta\sim(m_e/f)^2\ll1$. To demonstrate that this can still lead to successful formation of baryoids, we present a third benchmark (all energy scales are given in MeV)
\begin{equation}\label{eq:benchmark_III_1}
    \mathbf{\rm III}:~(T_{\rm ann},V_{\rm bias}^{1/4},f,x_\sigma,x_f,N,\beta,c_wk_w) = (50,\,0.157,\,1.00\times10^{4},\,0,\,2,\,7,\,2.61\times10^{-9},\,1) ~,
\end{equation}
which stabilizes at $T=T_{\rm F}\approx5.2803$~MeV with $v(T_{\rm F})\ll1$ and gives $N_{\rm B}\approx3.34\times10^{40}$. Consequently, we get
\begin{equation}\label{eq:benchmark_III_2}
    \mathbf{\rm III}:~ (E/N_{\rm B},R_c,r_{p,c}) = (934~{\rm MeV},\, 1.53\times10^2~{\rm cm},\, 2.57\times10^{-7}) ~,
\end{equation}
with $\rho_{\rm BD}\approx3.29\times10^{9}~{\rm g/cm^3}$, which, if the overall baryon trapping rate is $S\approx0.90$, also produces $\Omega_{\rm BD}/\Omega_{\rm B}=(N-1)\times(E/N_{\rm B})/m_{\rm B}\approx5.4$. The $P_e^0(T)$ and $V_{\rm bias}+s\,\sigma(T)/R(T)$, $R(T)$, and $v(T)$ profiles in the late stage of the collapse for this benchmark are similar to those presented in Figure~\ref{fig:benchmarks} for benchmarks I and II. While one might worry about the constraints from some existing searches, it is possible that this scenario turns out to have $f\gtrsim 100$~GeV [see Eq.~\eqref{eq:f_bounds:main}] and is thus free of such concerns. Either of these directions can offer a framework that naturally triggers the generation of $V_{\rm bias}$ and $\Delta m_{\rm B}$ below the QCD scale, providing even more predictability for this framework; nevertheless, we expect the first scenario to be realized in a more economic and natural manner, while there is a larger uncertainty in the wall dynamics associated with the second scenario due to the potentially required non-zero baryon leakage rate.

\section{Phenomenology of baryoids}\label{sec:pheno}

\begin{figure}[ht!]
    \centering
    \includegraphics[width=0.8\linewidth]{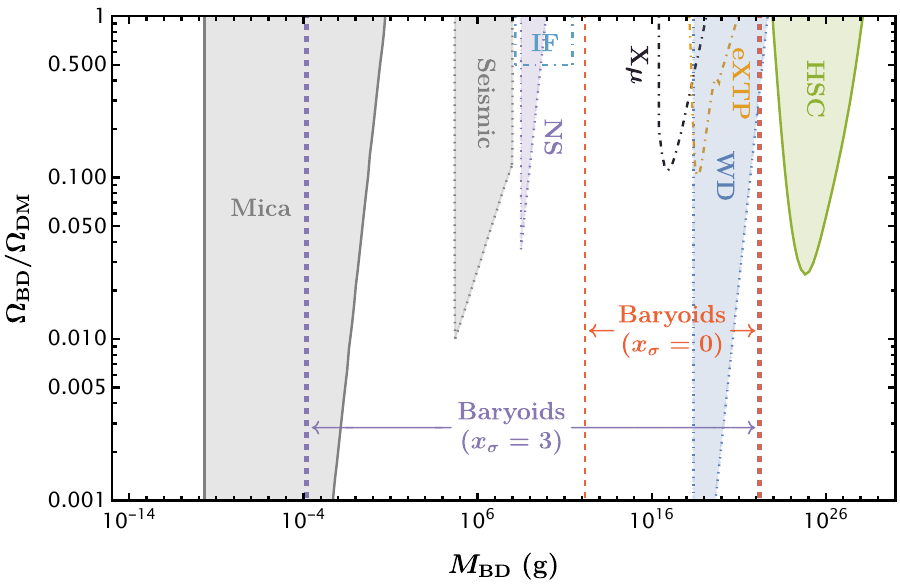}
    \caption{The phenomenological constraints on and future sensitivities to $M_{\rm BD}$ and $\Omega_{\rm BD}/\Omega_{\rm DM}$ for nuclear-density ($10^{14}~{\rm g/cm^3}$) baryoids, as well as the approximate predicted ranges of $M_{\rm BD}$ for $x_\sigma=0,3$, assuming $m_{\rm B, F}=m_{\rm B}-\Delta m_{\rm B}=850$~MeV and taking $N_{\rm B}\sim10^{36}-10^{46}$ and $N_{\rm B}\sim10^{20}-10^{46}$ for the two cases, respectively. The Mica constraint is taken from Ref.~\cite{DeRujula:1984axn}, the seismic constraint from Ref.~\cite{Herrin:2005kb}, the neutron star (NS) superbursts and white dwarf (WD) supernovae constraints from Ref.~\cite{SinghSidhu:2019tbr}, and the Subaru/HSC microlensing (HSC) constraint from Ref.~\cite{Sugiyama:2026kpv}. The X-ray microlensing sensitivity of eXTP (300 days) is taken from Ref.~\cite{Bai:2018bej} and that of X$\mu$ (30 days) from Ref.~\cite{Tamta:2024pow}, and the direct-detection mass range with future GW interferometers (IF) from Ref.~\cite{Jiang:2025xln}. Note that we show the Seismic, NS, and WD constraints in dotted lines as there are still large uncertainties, and we do not show the sensitivity to $\Omega_{\rm BD}/\Omega_{\rm DM}$ for IF. These constraints and sensitivities can also shift if $V_{\rm bias}$ is too weak to maintain a nuclear-scale baryoid density, as discussed in Section~\ref{sec:formation}. In the presented $m_{\rm BD}$ range, only the WD constraint has density-dependent sensitivities, and it becomes ineffective once $\rho_{\rm BD}\lesssim10^{10}~{\rm g/cm^3}$.
    }
    \label{fig:pheno}
\end{figure}

In this section, we discuss the phenomenology of baryoids. Like quark nuggets~\cite{Witten:1984rs,DiClemente:2024lzi}, their kinetic nature as macroscopic compact objects implies various potential astrophysical consequences. We discuss them below and summarize their implied constraints on and future sensitivities to the baryoid mass $M_{\rm BD}$ and the dark matter fraction accounted for by baryoids $\Omega_{\rm BD}/\Omega_{\rm DM}$ in Figure~\ref{fig:pheno}, assuming that the energy density of the baryoids is close to the nuclear-scale density~\cite{Bai:2018vik}, \ie, $\sim10^{14}~{\rm g/cm^3}$. Note that these constraints could shift around if $V_{\rm bias}$ is not strong enough and thus the baryoids have an energy density lower than the nuclear-scale density, as discussed in Section~\ref{sec:formation}. In particular, for the mass range of our interest, only the white dwarf (WD) constraint is sensitive to baryoid energy densities,\footnote{Other searches mostly concern long-range gravitational interactions, whose typical length scale (\ie, the Einstein radius for microlensing) is too large to be sensitive to the baryoid density range of our interest} and it will eventually become completely insensitive once $\rho_{\rm BD}\lesssim10^{10}~{\rm g/cm^3}$. We also overlay the approximate predicted ranges of $M_{\rm BD}$ for the $x_\sigma=0$ and $x_\sigma=3$ cases on the plot, assuming $m_{\rm B, F}=m_{\rm B}-\Delta m_{\rm B}=850$~MeV and taking $N_{\rm B}\sim10^{36}-10^{46}$ and $N_{\rm B}\sim10^{20}-10^{46}$ for the two cases, respectively. Since both cases (in fact, all cases) cover the predicted $M_{\rm BD}$ range of the $x_\sigma=0$ case, which is $\sim10^{12}-10^{22}~{\rm g}$, we will treat it as the primary $M_{\rm BD}$ range of our interest.
For even lighter baryoids, one can consider detection methods such as underground direct detection and neutrino or large-volume detection experiments (see Ref.~\cite{Bramante:2026wzh} for a review).
On the other hand, because of the collapse of the domain walls during the formation process, stochastic gravitational wave background (SGWB) might also be generated; however, as we show below and detail in Appendix~\ref{sec:GW_details}, these signals will either be exactly absent or highly suppressed due to the $t$-dependence of their quadrupolar nature in the stretching regime.
Finally, the inhomogeneity in baryon distribution generated by the domain wall motions is also constrained by BBN observations, of which we focus on the measurement of Deuterium abundance.
We briefly discuss each of these in the following.

\subsection{Gravitational microlensing}\label{subsec:pheno:lensing}

When a compact object passes close to our line of sight to a star, its gravitational field will temporarily amplify the apparent brightness of this background star. This phenomenon is known as gravitational microlensing~\cite{Paczynski:1985jf,Griest:1990vu} and has long been used to search for macroscopic dark matter. The first experiments were conducted by the MACHO and EROS collaborations~\cite{EROS-2:2006ryy}, with subsequent long-timescale surveys carried out by OGLE~\cite{Wyrzykowski:2010bh,Wyrzykowski:2015ppa,2015AcA....65....1U,Mroz:2024wia}, MOA~\cite{Bond:2001kt}, and Subaru Telescope's Hyper Suprime-Cam (Subaru/HSC)~\cite{2018PASJ...70S...1M,2018PASJ...70S...2K,2018PASJ...70...66K,2018PASJ...70S...3F}. Recently, Ref.~\cite{Sugiyama:2026kpv} performed an updated analysis on the Subaru/HSC data and presented the new microlensing constraints on primordial black holes, which can also be used to constrain the properties of baryoids given their high energy density. To see this, we choose $M\sim10^{23}~{\rm g}$, the lower bound on the lens mass sensitivity of the Subaru/HSC experiment, and $D_S\approx765~{\rm kpc}$, the distance from the earth to M31, to estimate the minimum Einstein radius $R_E\sim\sqrt{4G_NMD_S}\sim10^9~{\rm cm}$, $G_N$ denoting the Newtonian constant, which is much larger than the typical size of a baryoid $\lesssim\mathcal{O}(1)~{\rm cm}$ and therefore cannot distinguish between a primordial black hole and a baryoid. However, given the limit of optical microlensing and the upper bound on $M_{\rm BD}$ imposed by the requirement of $f\gtrsim T_{\rm QCD}$ (see Section~\ref{subsec:formation:entrapment}), the Subaru/HSC constraints turn out to be irrelevant for baryoids (see Figure~\ref{fig:pheno}). In addition to optical microlensing, another promising probe is the microlensing of X-rays~\cite{Bai:2018bej,Tamta:2024pow}, from which we quote the sensitivities of eXTP (300 days, targeting the SMC X-1 pulsar source)~\cite{eXTP:2016rzs,Bai:2018bej} and X$\mu$ (30 days, targeting a to-be-discovered high-flux X-ray source)~\cite{Tamta:2024pow,STROBE-X}.

\subsection{Thermonuclear runaway and proto-neutron star conversion}\label{subsec:pheno:runaway}

If a baryoid with a sufficiently large cross section impacts a white dwarf or a neutron star, it could lead to a thermonuclear runaway which then leads to a supernova explosion or an X-ray superburst~\cite{Graham:2015apa,Graham:2018efk,SinghSidhu:2019tbr}. Alternatively, if, with enough mass, it penetrates a newly formed neutron star (proto-neutron star) before its crust forms, it could convert the proto-neutron star into a quark star, which has not yet been observed to date~\cite{Madsen:1988zgf,Watts:2006hk}. These will impose constraints on the baryoid properties given the observational data of these compact stars. Nevertheless, the latter constraint is superseded by other astrophysical constraints~\cite{Madsen:1988zgf}, and thus we do not show it in Figure~\ref{fig:pheno}. Note that, as mentioned in Ref.~\cite{SinghSidhu:2019tbr}, there are still large uncertainties in these constraints, and thus we show the NS and WD constraints in dotted lines.

\subsection{Direct detection with future GW interferometers}\label{subsec:pheno:direct}

When a baryoid passes near the interferometers for GW detection, its gravitational influence will cause differential acceleration among the test masses of the detectors and produce a detectable Doppler signal. Based on this, we can project the sensitivities to the baryoid mass of future experiments~\cite{Seto:2004zu,Adams:2004pk,Jiang:2025xln}, including LISA~\cite{LISA:2017pwj}, TianQin~\cite{TianQin:2015yph}, and Taiji~\cite{Hu:2017mde}. Ref.~\cite{Jiang:2025xln} also studied the BBO experiment~\cite{Crowder:2005nr} and found no relevant sensitivity to probe $\Omega_{\rm BD}/\Omega_{\rm DM} \leq 1$ since it targets a higher GW frequency band than the three discussed experiments and is thus more sensitive to lighter $m_{\rm BD}$ below the primary range of our interest. Following this logic, we expect some nHz GW experiments such as SKA~\cite{Dewdney:2009tmd}, IPTA~\cite{Manchester:2013ndt}, EPTA~\cite{Kramer:2013kea}, and NANOGrav~\cite{Brazier:2019mmu} to be the most promising probes for baryoids in the future.

\subsection{Seismic events}\label{subsec:pheno:seismic}

For lighter baryoids, one can consider their potential impacts on earth and on the moon, which could induce seismic events with more of a ``shockwave'' nature in contrast to those of a tectonic nature~\cite{Herrin:2005kb,Burdin:2014xma}. Based on the old data, Ref.~\cite{Herrin:2005kb} performed an analysis on the implied constraints on macroscopic dark matter given the absence of observation of such seismic events, while Ref.~\cite{Cyncynates:2016rij} later performed an updated analysis and found no such limits on nuclear-density macroscopic dark matter. Therefore, there are still large uncertainties in these constraints and we show them in dotted lines. Interestingly, Ref.~\cite{Marsquake} has reported the recent detection of the largest Marsquake to date. Unlike earth and the moon, tectonic events are very rare on Mars, and thus it is very likely that these events are induced by external factors, including the impact of compact objects. As a result, the study of Marsquakes in the future could be a promising probe for baryoids.

\subsection{Traces in natural minerals}\label{subsec:pheno:mica}

If the baryoids are sufficiently light, they could have a high enough number density to leave an imprint into the natural minerals as they penetrate into earth's crust. This technique, called ``paleo-detection'', has also been used to study other types of dark matter, such as weakly interacting massive particles (see Ref.~\cite{Theodosopoulos:2026ehn} for a recent study). An early study of paleo-detection of nuclear-dense objects using ancient Muscovite mica was conducted in Ref.~\cite{DeRujula:1984axn} (see Ref.~\cite{Jacobs:2014yca} for a recent review), which found no such traces in the sample (also considered in Ref.~\cite{Price:1986ky} for magnetic monopole searches), and thus could place a constraint on the properties of these compact objects.

\subsection{Stochastic gravitational wave background}\label{subsec:pheno:GW}

We leave the detailed derivation of the SGWB spectrum generated by the collapse of stretching domain walls to Appendix~\ref{sec:GW_details} and only conclude here that for $x_\sigma=0,2,4$, SGWB will not be generated due to the particular $t$-dependence of the domain wall quadrupole moment, while for $x_\sigma=1,3$, the peak amplitudes are on the orders of $10^{-54}$ and $10^{-112}$, respectively, in the parameter range of our interest, both of which are too small to be probed by existing and future GW experiments. This results from the negligible domain wall energy density and nonrelativistic wall velocities during the collapse.

\subsection{Baryon inhomogeneity and Deuterium abundance}\label{subsec:pheno:inhomogeneity}

In principle, due to the static nature of the $\mathbb{Z}_N$-symmetry separation mechanism, the size of the domain walls does not play a major role in the first half of the baryoid formation process. Nevertheless, as the walls collapse later and the baryons are swept into the shrinking baryoids, inhomogeneity in the baryon distributions within the true-vacuum domains will be induced, which could potentially violate the BBN constraints. Ref.~\cite{Bagherian:2025puf} has shown that for an inhomogeneity of a length scale $\gg d_n(T)\geq d_p(T)$, $d_{n,p}$ being the comoving neutron and proton diffusion lengths, respectively, the observed Deuterium-to-Helium abundance ratio ($D/H$) can put a constraint on the root mean square inhomogeneity amplitude as~\cite{Burns:2023sgx}
\begin{equation}
    \epsilon_{\rm RMS} \equiv \sqrt{\langle\epsilon(x)\rangle^2}\leq 0.28 ~,
\end{equation}
where
\begin{equation}
    \eta(x) = \eta_{\rm CMB}[1+\epsilon(x)] ~,
\end{equation}
$\eta_{\rm CMB}\sim10^{-10}$ being the CMB-inferred value of the baryon-to-photon ratio.

To see how this constraint favors the stretching domain wall scenario for baryoid formation, we perform a simple analysis below. After the domain walls collapse, we treat the resulting baryoids as zero-volume point-like objects compared to the surviving true-vacuum domain. Denoting the average baryon number density in the final true-vacuum domain by $\bar{n}_{\rm B}$ and the baryon leakage rate from the false-vacuum domains by $1-S$, one gets $\bar{n}_{\rm B}=[(N-1)(1-S)n_0+n_0]/N=\eta_{\rm CMB}n_\gamma$, where $n_0$ is the initial baryon number density in each domain. It can be shown that 
\begin{equation}
    \epsilon_{\rm RMS} = \sqrt{\frac{\Delta\bar{n}_{\rm B,T}^2+(N-1)\Delta \bar{n}_{\rm B,F}^2}{N\bar{n}_{\rm B}^2}} = \frac{\sqrt{N-1}\,S}{N + S - N S} ~,
\end{equation}
and that
\begin{equation}
    \frac{\Omega_{\rm DM}}{\Omega_{\rm B}} = \frac{(N-1)S}{1+(N-1)(1-S)} ~,
\end{equation}
where
\begin{equation}
    \Delta\bar{n}_{\rm B,T} = \frac{(N-1)Sn_0}{N} ,~ \Delta \bar{n}_{\rm B,F} = -\frac{Sn_0}{N}
\end{equation}
are the baryon number density variations with respect to $\bar{n}_{\rm B}$ in the true and false-vacuum domains, respectively. Assuming $\Delta m_{\rm B}=0$ and plugging in $\Omega_{\rm DM}/\Omega_{\rm B}=5.4$, we get $\epsilon_{\rm RMS}=2.20$ and $2.04$ for $N=7$ and $8$, respectively, both clearly exceeding the upper limit of $0.28$. However, if the domain walls couple strongly enough to the plasma, they can be sufficiently frustrated so that their comoving size $L(T_\nu)/a(T_\nu)\lesssim d_p(T_\nu)\approx d_n(T_\nu)\approx 10^5$~cm and can further induce plasma flows to redistribute the baryons. As an estimation, we consider the Hubble length scale at $T=T_\nu$, which is roughly $2\,t_\nu\approx10^{11}$~cm, while, choosing $a(T=1~{\rm MeV})=1$~\cite{Bagherian:2025puf}, we have $a(T_\nu)d_n(T_\nu)\approx10^5$~cm. Consequently, we must require that one Hubble patch be divided into roughly $\sim(10^6)^3$ domains to satisfy the above-mentioned condition, which is unachievable if the domain walls enter the scaling regime. On the other hand, for stretching domain walls, we have $L_{\rm st}(t_\nu)/H^{-1}(t_\nu)\approx (T_\nu/f)/N=1.14\times10^{-6}\times(100\,{\rm GeV}/f)(7/N)$ [see Eq.~\eqref{eq:Lv_st:main}], which can satisfy the constraint as long as $f\gtrsim100$~GeV, and thus it is achievable even for the case of $x_\sigma=x_f=0$ [see Eq.~\eqref{eq:f_bounds:main}]. Since we only consider scenarios with strong friction in this study, we assume that this constraint is always satisfied. 

Before closing this section, we note that although this separation mechanism does not require a non-zero $\Delta m_{\rm B}$, after the baryoids stabilize at $T=0$, it is still preferred for $\Delta m_{\rm B}>0$ for the baryoids to be more energetically stable than ordinary nuclei.

\section{Discussion and conclusions}\label{sec:conclusions}

In this study, we have assumed that all false-vacuum domains will eventually form individual baryoids. However, it is not impossible to imagine a scenario where multiple false-vacuum domains clump together to form a ``structured'' baryoid, in which there still exist several different domains that are nearly degenerate, while the more unstable domains are eventually merged into other more stable domains. On the other hand, as we have pointed out in Ref.~\cite{Bai:2023cqj}, inside the false-vacuum domains with $\vert\theta\vert\lesssim\pi$, QCD phase transition might be first-order, which allows the formation of quark nuggets. Since these quark nuggets can potentially be observed independently from the baryoids, the dark matter candidates could be a combination of both states. Both possibilities will lead to a more diverse spectrum of macroscopic dark matter sizes and masses, which could have very interesting phenomenological implications.

In addition to the static $\mathbb{Z}_N$ separation mechanism, a second, conceptually distinct thermal-selection mechanism can also generate an $\mathcal{O}(1)$ dark matter-to-baryon ratio in this framework, analogous to the neutron-to-proton number density ratio, $n_n/n_p \simeq 1/6$--$1/7$~\cite{ParticleDataGroup:2024cfk}, at the onset of BBN. A non-zero $\Delta m_{\rm B}$ can dynamically generate a hierarchy between the baryon number densities in the true- and false-vacuum domains (see Ref.~\cite{Atreya:2014sca} for an example in a different context). If the average comoving domain-wall size satisfies $L/a(T)\lesssim d_p(T)$, the baryons inside individual domains can thermalize and establish thermal and chemical equilibrium across the domain wall, which acts as a potential barrier between the domains. For $\Delta m_{\rm B}>0$, this equilibrium naturally favors a higher baryon number density in the false-vacuum domains. Once the temperature drops below an out-of-chemical-equilibrium temperature $T_{\rm OCE}$ such that the Hubble rate exceeds the baryon exchange rate, $H(T_{\rm OCE})\gtrsim \Gamma_{\rm ex}(T_{\rm OCE})$, the baryon numbers in the individual domains freeze out. If all false-vacuum domains share the same baryon mass $m_{\rm B,F}\equiv m_{\rm B}-\Delta m_{\rm B}$, one then expects
\begin{align}
\frac{\Omega_{\rm BD}}{\Omega_{\rm B}}
\sim
(N-1)\,e^{\Delta m_{\rm B}/T_{\rm OCE}} \, .
\end{align}
Since $T_{\rm BBN}\approx 3~{\rm MeV}<T_{\rm ann}<T_{\rm OCE}<T_{\rm QCD}\approx 150~{\rm MeV}$, the exponential factor is naturally an $\mathcal{O}(1)$ number for $\Delta m_{\rm B}\sim \mathcal{O}(10\text{--}100)~{\rm MeV}$. This mechanism therefore provides an alternative route to the observed dark matter-baryon ratio and can relax the requirement that the discrete symmetry alone, through $N-1$, account for the full numerical value. In the minimal case with $N=2$, $\Delta m_{\rm B}/T_{\rm OCE}\sim \mathcal{O}(1)$ is the only parameter controlling the ratio. If this turns out to be realized, it would point to a natural target range for $\Delta m_{\rm B}$, directly analogous to the familiar relation $n_n/n_p\sim e^{-Q/T}$ for $T\lesssim T_\nu$, with $Q\simeq 1.3~{\rm MeV}$ being the neutron-proton mass difference.

Another phenomenological implication of baryoids (or macro dark matter in general) is the potential observation of meteors when they pass through the atmosphere. Ref.~\cite{SinghSidhu:2019cpq} has studied this using combined data from the retired U.S. Prairie, Canadian, and Eastern bolide networks~\cite{1986AJ.....92..595H} as well as those from the currently operating Desert Fireball Network (DFN) in Australia~\cite{2017ExA....43..237H}. Although the current sensitivity cannot yet reach nuclear-scale densities, it will be interesting to see if the planned future upgrades for DFN will be useful for the study of baryoids. We also emphasize again that the most promising probes in the near future will be X-ray microlensing~\cite{Bai:2018bej,Tamta:2024pow} and direct interferometer detection with future GW experiments~\cite{Seto:2004zu,Adams:2004pk,Jiang:2025xln}.

In conclusion, we have explored a novel explanation for the dark matter-baryon coincidence based on baryoid formation from $\mathbb{Z}_N$ domain walls. When the wall expansion is sufficiently frustrated by wall-plasma interactions, baryons can be efficiently trapped inside false-vacuum domains and later form stable baryoids without conflicting with astrophysical or cosmological bounds. Moreover, the number of domain walls also becomes exponentially greater than that in the scaling case due to the highly suppressed domain wall size. We have presented benchmark scenarios showing that this mechanism can naturally reproduce the observed ratio $\Omega_{\rm BD}/\Omega_{\rm B}=\Omega_{\rm DM}/\Omega_{\rm B}\approx5.4$, and we have derived the corresponding baryoid baryon-number range, $10^{20}\lesssim N_{\rm B}\lesssim10^{46}$. We have also discussed a possible realization in a QCD-anomalous $\mathbb{Z}_N$ domain wall model, along with future directions for completing the framework and the phenomenological probes of baryoid dark matter.

\vspace{1cm}
\subsubsection*{Acknowledgments}
We thank Nima Arkani-Hamed for useful discussion. This work is supported by the U.S. Department of Energy under the contract DE-SC-0017647 and DE-AC02-06CH11357 at Argonne National Laboratory. TKC is also supported by the Ministry of Education, Taiwan, under the Government Scholarship to Study Abroad. 

\begin{appendix}
\section{VOS model for \texorpdfstring{$T$-dependent}{} domain wall tension}\label{sec:gVOS}

In this appendix, we extend the derivation of the VOS model in Refs.~\cite{Martins:2016ois,Blasi:2022ayo} to discuss the situation with a $T$-dependent domain wall tension $\sigma=\sigma(T)=\sigma(t)$. In the thin-wall limit, a domain wall can be effectively described as a two-dimensional object with the Nambu-Goto action for a worldsheet
\begin{equation}
    S = -\int d^3\zeta\,\sigma\left(x^\mu(\zeta)\right)\sqrt{\gamma} ~,
\end{equation}
where $\gamma$ is the determinant of the induced metric on the three-dimensional worldvolume traced out by the worldsheet trajectory. This metric is defined by
\begin{equation}
    \gamma_{ab} = g_{\mu\nu}\frac{\partial x^\mu}{\partial\zeta^a}\frac{\partial x^\nu}{\partial\zeta^b} \equiv g_{\mu\nu}x^\mu_{,a}x^\nu_{,b} ~,
\end{equation}
with $g_{\mu\nu}$ being the metric of the physical spacetime. Varying the action, one has
\begin{equation}
    \delta S = -\int d^3\zeta\left(\sqrt{\gamma}\,\delta\sigma + \sigma\,\delta\sqrt{\gamma}\right) = 0 ~,
\end{equation}
where
\begin{equation}
    \delta\sigma = (\partial_\mu\sigma)\delta x^\mu ~,
\end{equation}
and
\begin{equation}
    \delta\sqrt{\gamma} = \frac{\delta\gamma}{2\sqrt{\gamma}} = \frac{1}{2}\,\sqrt{\gamma}\,\gamma^{ab}\delta\gamma_{ab} ~,
\end{equation}
with
\begin{equation}
    \delta\gamma_{ab} = \delta\left(g_{\mu\nu}x^\mu_{,a}x^\nu_{,b}\right) = (\partial_\lambda g_{\mu\nu}\delta x^\lambda)x^\mu_{,a}x^\nu_{,b} + 2g_{\mu\nu} x^\nu_{,b}\delta x^\mu_{,a} ~.
\end{equation}
Consequently,
\begin{equation}
    \int d^3\zeta\left\{\sqrt{\gamma}\,\partial_\mu\sigma\delta x^\mu + \frac{\sigma}{2}\sqrt{\gamma}\,\gamma^{ab}\left[(\partial_\lambda g_{\mu\nu}\delta x^\lambda)x^\mu_{,a}x^\nu_{,b} + 2g_{\mu\nu} x^\nu_{,b}\delta x^\mu_{,a}\right]\right\} = 0 ~.
\end{equation}
Performing integration by parts,
\begin{equation}
    \int d^3\zeta\, \sigma\sqrt{\gamma}\gamma^{ab}g_{\mu\nu} x^\nu_{,b}\delta x^\mu_{,a} = -\int d^3\zeta\partial_a\left( \sigma\sqrt{\gamma}\gamma^{ab}g_{\mu\nu} x^\nu_{,b}\right)\delta x^\mu ~,
\end{equation}
which leads to the equation of motion
\begin{equation}
    \partial_a\left( \sigma\sqrt{\gamma}\gamma^{ab}g_{\mu\nu} x^\nu_{,b}\right) - \frac{\sigma}{2}\sqrt{\gamma}\gamma^{ab}x^\nu_{,a}x^\rho_{,b}\partial_\mu g_{\nu\rho} = \sqrt{\gamma}\,\partial_\mu\sigma ~.
\end{equation}
Isolating the $\partial_a\sigma$ part from the first term on the left-hand side, we have
\begin{equation}\label{eq:GVOS:eom:prep}
    \frac{1}{\sqrt{\gamma}}\partial_a\left(\sqrt{\gamma}\gamma^{ab} x^\nu_{,b}\right)g_{\mu\nu} + \gamma^{ab} x^\nu_{,b}\partial_ag_{\mu\nu} - \frac{1}{2}\gamma^{ab}x^\nu_{,a}x^\rho_{,b}\partial_\mu g_{\nu\rho} = \frac{\partial_\mu\sigma}{\sigma} -  \gamma^{ab}g_{\mu\nu} x^\nu_{,b}\frac{\partial_a \sigma}{\sigma} ~.
\end{equation}
The second and third terms on the left-hand side can be arranged into
\begin{equation}
    \gamma^{ab} x^\nu_{,b}\partial_ag_{\mu\nu} - \frac{1}{2}\gamma^{ab}x^\nu_{,a}x^\rho_{,b}\partial_\mu g_{\nu\rho} = \gamma^{ab}x^\nu_{,a}x^\rho_{,b}\left(\partial_\rho g_{\mu\nu}-\frac{1}{2}\partial_\mu g_{\nu\rho}\right) = \Gamma_{\mu\nu\rho}\gamma^{ab}x^\nu_{,a}x^\rho_{,b} ~,
\end{equation}
where $\Gamma_{\mu\nu\rho}$ denotes the Christoffel symbols
\begin{equation}
    \Gamma_{\mu\nu\rho} = \frac{1}{2}\left(\partial_\rho g_{\mu\nu}+\partial_\nu g_{\mu\rho}-\partial_\mu g_{\nu\rho}\right) ~.
\end{equation}
Finally, by contracting both sides of Eq.~\eqref{eq:GVOS:eom:prep} with $g^{\lambda\mu}$, we get
\begin{equation}\label{eq:GVOS:eom:full}
    \frac{1}{\sqrt{\gamma}}\partial_a\left(\sqrt{\gamma}\gamma^{ab} x^\lambda_{,b}\right) + \Gamma^\lambda_{\nu\rho}\gamma^{ab}x^\nu_{,a}x^\rho_{,b} = \left(g^{\lambda\mu} -  \gamma^{ab} x^\lambda_{,a}x^\mu_{,b}\right)\frac{\partial_\mu \sigma}{\sigma} ~.
\end{equation}

Now, as we work with the Friedmann-Lema\^{i}tre-Robertson-Walker (FLRW) metric,
\begin{equation}
    ds^2 = a^2(\tau)(-d\tau^2+dx^2) ~,
\end{equation}
and choose the transverse gauge by identifying $\zeta^0=\tau$ while requiring $\partial_0 x\cdot\partial_ix\equiv\dot{x}\cdot x_{,i}=0$, $i=1,2$, to form an orthogonal system, we have
\begin{equation}
    \gamma_{ab} = a^2(\tau)\,{\rm diag}(1-\dot{x}^2,-x_{,1}^2,-x_{,2}^2) ~.
\end{equation}
In this study, we only consider a spatially uniform wall tension $\sigma=\sigma(t)=\sigma(\tau)$. Consequently, one can show that the time component of Eq.~\eqref{eq:GVOS:eom:full} reads
\begin{equation}
    \frac{\dot{\epsilon}}{\epsilon} + \left(3\mathcal{H}+\frac{\dot{\sigma}}{\sigma}\right)\dot{x}^2 = 0 ~,
\end{equation}
where $\mathcal{H}=\dot{a}/a$ is the conformal Hubble parameter and~\cite{Blasi:2022ayo}
\begin{equation}
    \epsilon = \sqrt{\frac{x^2_{,1}x^2_{,2}}{1-\dot{x}^2}} ~.
\end{equation}
Then, the spatial components of Eq.~\eqref{eq:GVOS:eom:full} can be simplified to
\begin{equation}\label{eq:GVOS:eom:final}
    \ddot{x}^i+\left(3\mathcal{H}+\frac{\dot{\sigma}}{\sigma}\right)\dot{x}^i(1-\dot{x}^2) = \frac{1}{\epsilon}\frac{\partial}{\partial\zeta^1}(x_{,2}^2x_{,1}/\epsilon) + \frac{1}{\epsilon}\frac{\partial}{\partial\zeta^2}(x_{,1}^2x_{,2}/\epsilon) ~.
\end{equation}
Defining $v^i=\dot{x}^i$, one can obtain a coupled first-order differential equation system of $x^i$ and $v^i$ analogous to the VOS model. The terms on the right-hand side of Eq.~\eqref{eq:GVOS:eom:final} are the curvature terms, which are analogous to the $(1-v^2)k_w/L$ term in the original VOS model [see Eq.~\eqref{eq:VOS}]. From Eq.~\eqref{eq:GVOS:eom:final}, one can clearly see that the net effect of having $\sigma=\sigma(\tau)$ is the inclusion of an additional friction term $\propto \dot{\sigma}/\sigma$. As a result, we get the overall damping factor expressed in terms of cosmic time $t$ as
\begin{equation}\label{eq:l_d:full}
    \ell_d^{-1} = 3H + \ell_f^{-1} + \frac{\dot{\sigma}(t)}{\sigma(t)} ~.
\end{equation}
We note that for $\sigma(T)\propto T^{x_\sigma}\propto t^{-{x_\sigma}/2}$ in the radiation-dominated universe, $\dot{\sigma}/\sigma=-{x_\sigma}/2t<0$ is not effectively a ``damping'' term. One physical interpretation of this is that the $T$-dependence of the domain wall tension is induced by the non-trivial interaction between the wall and plasma, which effectively induces inhomogeneity in the plasma pressure around the wall that acts as a net force countering the damping effects of the Hubble and friction terms.

\section{Details of domain wall dynamics}\label{sec:DW}

In this appendix, we discuss the formation and evolution of domain walls. Domain walls are a common product of spontaneously broken discrete symmetries. For simplicity, we illustrate this with an effective $\mathbb{Z}_2$ symmetry model. For a detailed discussion of more general $\mathbb{Z}_N$ symmetries, we refer to Ref.~\cite{Bai:2023cqj} and only summarize the results here.

Assuming that there exists a scalar field $S$ which transforms as $S\to -S$ under a $\mathbb{Z}_2$ symmetry, we can write down the corresponding renormalizable potential
\begin{equation}
    V(S) = \frac{\lambda}{4}(S^2-f^2)^2 ~~~ [N=2] ~,
\end{equation}
where $f$ denotes the breaking scale of the symmetry. After spontaneous symmetry breaking, $S$ can settle into either of the two degenerate vacuum expectation values, $\langle S\rangle=\pm f$. The topological structure of the $\mathbb{Z}_2$ vacuum manifold implies the existence of ``domain walls'' in the physical spacetime, which are a class of time-independent classical solutions to the theory that interpolate between the two degenerate vacua. The profile of one such domain wall is given by 
\begin{equation}
    S(z) = f\,\tanh\left(\sqrt{\frac{\lambda}{2}}fz\right) ~~~ [N=2] ~,
\end{equation}
which implies a wall thickness $\delta\sim(\sqrt{\lambda}f)^{-1}$ and wall tension 
\begin{equation}
    \sigma = \int_{-\infty}^{\infty}T_{00}dz = \frac{2\sqrt{2}}{3}\sqrt{\lambda} f^3 ~~~ [N=2] ~,
\end{equation}
where $T_{\mu\nu}$ denotes the energy-momentum tensor. 

For $N>2$, $S$ becomes a complex field, and the associated single-field potential could be
\begin{equation}
    V(S) = -m^2 SS^\dagger + \lambda(SS^\dagger)^2 - \mu(S^N+S^{\dagger\,N}) ~,    
\end{equation}
where the mass dimension of $\mu$ is $4-N$.\footnote{
As we discussed in Ref.~\cite{Bai:2023cqj}, $V(S)$ is in general non-renormalizable for $N>4$. Moreover, for odd $N$'s, the potential is not bounded from below. Both of these imply the need for UV-completion for the model, which we simply assume that exists and will not affect our phenomenological studies.
} 
Assuming that the symmetry breaking scale is given by $f=f(m^2,\lambda,\mu)$, we can write down the interpolation solution $S(z)=f\exp[i\,\theta(z)]$ between any two adjacent vacua $\langle S\rangle=f\exp[i\,2\pi j/N],f\exp[i\,2\pi (j+1)/N]$ as
\begin{equation}
    \theta(z) = \frac{2\pi j}{N} + \frac{4}{N}\arctan\left[\exp\left(\sqrt{N\mu f^{N-2}}z\right)\right] ~,
\end{equation}
which gives the wall tension
\begin{equation}
    \sigma = 16\mu^{1/2}f^{N/2+1}N^{-3/2} ~.
\end{equation}
If $\mu\sim f^{4-N}$, then $\sigma\sim f^3$ as in the $\mathbb{Z}_2$ case. However, in certain models, $\sigma$ can be associated with other energy scales, such as temperature $T$. For example, in $SU(N)$ Yang-Mills theories, since the $\mathbb{Z}_N$ center symmetry, if spontaneously broken, is not associated with any particular scale in the Lagrangian when $T>\Lambda_{SU(N)}$, the associated domain wall tension will scale as $\sigma\sim T^3$~\cite{Bhattacharya:1990hk}. Another example is to consider the interaction between a scalar order parameter field and another scalar field that is in thermal equilibrium with the primordial plasma~\cite{Ramazanov:2021eya,Babichev:2021uvl}. Such domain walls will ``melt'' away as temperature drops along with the expansion of the universe, as studied in Ref.~\cite{Dankovsky:2024ipq}. As a result, we will explore different parametrizations of the scaling of $\sigma$ and the corresponding evolution patterns. Despite these possible different scenarios, the general evolution of domain walls can be understood in a consistent framework, which we now describe.

At temperature $T=T_{\rm form}=T_c$, domain walls will form and separate the universe into domains of distinct but energetically degenerate vacua via the Kibble-Zurek mechanism~\cite{Kibble:1976sj,Zurek:1985qw}. To minimize the total energy density stored in the wall network, domain walls will begin to expand to reduce the overall curvature. Given the tension of the domain walls $\sigma$, one can discuss the evolution of the characteristic length scale of the domain wall network $L\equiv \sigma/\rho_w$, $\rho_w$ being the energy density of the walls, using the VOS model~\cite{Martins:2016ois}, which is a coupled differential equation system of $L$ and the root mean square velocity of the wall $v$,
\begin{equation}\label{eq:VOS}
\begin{aligned}
    \frac{dL}{dt} &= HL+v^2\frac{L}{\ell_d} + c_w v ~, \\
    \frac{dv}{dt} &= (1-v^2)\left(\frac{k_w}{L}-\frac{v}{\ell_d}\right) ~,
\end{aligned}
\end{equation}
where $\ell_d^{-1}=3H+\ell_f^{-1}+\dot{\sigma}/\sigma$ is the damping factor derived in Eq.~\eqref{eq:l_d:full} and $c_w$ and $k_w$ are the phenomenological parameters given by simulations that describe possible energy losses from the domain wall network. Note that $\dot{\sigma}=0$ in the original VOS model, while the inclusion of a $t$-dependent wall tension $\sigma(t)$ is derived in Appendix~\ref{sec:gVOS}. Simulations show that the VOS model describes the domain wall dynamics associated with general $\mathbb{Z}_N$ symmetries pretty well~\cite{Hiramatsu:2012sc,Kawasaki:2014sqa}, and therefore we will adopt it for our analysis below.

In the radiation-dominated universe, $H$ is given by
\begin{equation}
    H(T) = \sqrt{\frac{\pi^2}{90}g_*(T)}\frac{T^2}{M_{\rm Pl}} ~,
\end{equation}
where $M_{\rm pl}=2.43\times10^{18}$~GeV is the reduced Planck mass and $g_*(T)$ denotes the relativistic degrees of freedom that give the radiation energy density
\begin{equation}
    \rho_R(T) = \frac{\pi^2}{30}g_*(T)T^4 ~.
\end{equation}
Since within the temperature range of our interest ($T>T_{\nu}\simeq0.8$~MeV), $g_*(T)$ will only vary by at most one order of magnitude (see Ref.~\cite{Husdal:2016haj} for example), which implies an $\mathcal{O}(1)$ variation in most of the considered observables, we will fix $g_*(T)=g_*(100~{\rm MeV})\approx 10$ as a constant for convenience.

If the domain wall network enters the scaling regime (see below discussion), its energy density will scale as $\rho_w\propto t^{-1}$, while the radiation energy density scales as $\rho_R\propto t^{-2}$. Eventually, domain walls will dominate the total energy density and contradict astrophysical and cosmological observations such as BBN. To avoid this, we introduce a bias potential $V_{\rm bias}$ so that the walls will begin to collapse at time $t_{\rm ann}$ when $\rho_w(t_{\rm ann})=V_{\rm bias}$, \ie, when the wall tension force $p_T=\rho_w$ balances the bias potential, for which we require $T_{\rm ann}\gtrsim T_{\rm BBN}\approx3$~MeV~\cite{Bai:2021ibt,Bringmann:2023opz}. In the following, we discuss the scaling expansion, frustrating expansion, and collapse of the domain wall network.

\subsection{Scaling expansion}\label{subsec:DW:scaling}

In the situation where friction can be neglected and $\dot{\sigma}=0$, \ie, $\ell_d^{-1}=3H$, Eq.~\eqref{eq:VOS} reduces to
\begin{equation}
\begin{aligned}
    \frac{dL}{dt} &\simeq HL(1+3v^2) + c_wv ~, \\
    \frac{dv}{dt} &\simeq (1-v^2)\left(\frac{k_w}{L}-3Hv\right) ~.
\end{aligned}
\end{equation}
Consequently, the domain wall network will enter the ``scaling regime'' where the expansion behavior is simply described by 
\begin{equation}\label{eq:Lv_sc}
    L_{\rm sc}(t) = L_0 t ,~ v_{\rm sc}(t) = v_0 ~~~[\text{Scaling Expansion}] ~,
\end{equation}
with $L_0=2\sqrt{k_w(k_w+c_w)/3}$ and $v_0=\sqrt{k_w/[3(k_w+c_w)]}$. In Ref.~\cite{Martins:2016ois}, it is shown through simulations for $\mathbb{Z}_2$ domain walls that scaling expansion begins at $T_{\rm scal}\approx T_{\rm form}/30$ with $L_0=1.2$, $v_0=0.42$, implying $k_w=0.66$ and $c_w=0.81$. Similar scaling behavior is also found for general $\mathbb{Z}_N$ domain walls.
In the scaling regime, the energy density of $\mathbb{Z}_N$ domain walls is given by $\rho_w=\mathcal{A}\,\sigma/t$, $\mathcal{A}\approx0.4N$~\cite{Hiramatsu:2012sc,Kawasaki:2014sqa}. One natural interpretation of the $N$-factor is that within one Hubble length, the universe is expected to be on average divided into $N$ segments, each corresponding to one degenerate vacuum, due to the Kibble-Zurek mechanism~\cite{Kibble:1976sj,Zurek:1985qw}. As a result, one expects the universe to be divided into $\sim N^3$ distinct domains per Hubble volume during the scaling regime.

Now, if $\sigma=\sigma(T)\propto T^{x_\sigma} \propto t^{-{x_\sigma}/2}$, one gets $L_0=2\sqrt{k_w(k_w+c_w)/(3-{x_\sigma})}$ and $v_0=\sqrt{k_w/[(3-{x_\sigma})(k_w+c_w)]}$. For $0<{x_\sigma}<3$, a scaling solution with $x_\sigma$-dependent length and velocity scales still exists; however, there is no physical scaling solution for ${x_\sigma}\geq3$, which could also be seen from the fact that $3H+\dot{\sigma}/\sigma=(3-x_\sigma)/2t\leq0$, meaning that there is not only no longer an effective damping term, but, for $x_\sigma>3$, an unphysical ``boosting'' term will be induced instead. This could also be the reason why Ref.~\cite{Dankovsky:2024ipq} cannot consistently get a scaling solution for melting domain walls with $\sigma\propto T^3$.

\subsection{Frustrated expansion}\label{subsec:DW:frustrated}

If the domain walls interact strongly enough with the primordial plasma, they could experience sufficient friction and undergo frustrated expansion. For example, Ref.~\cite{Blasi:2022ayo} studies the behavior of axion-like-particle (ALP) domain walls under significant friction and finds that their typical length scale and velocity are indeed highly suppressed before the Hubble term dominates the damping effect ($H\gg\ell_f^{-1}$). To discuss scenarios of domain wall tension $\sigma$ with different energy dependence, we generalize the framework of Ref.~\cite{Hook:2026grn}, which studies the frustrated behavior of axion strings but can be extended to the study of domain walls, to analyze asymptotic solutions to the VOS model for a domain wall network.

In the rest frame of a planar domain wall moving with velocity $v$, the friction force per area that it experiences is given by~\cite{Blasi:2022ayo}
\begin{equation}
    F_f = -\frac{\sigma}{\ell_f}\gamma(v)\,v ,~\quad  \gamma(v)=\frac{1}{\sqrt{1-v^2}} ~,
\end{equation}
where $F_f$ can also be understood as the net pressure $\Delta P\equiv\vert F_f\vert$ (force per area) acting on the wall. Since the source of $\Delta P$ is usually the asymmetric momentum transfer of plasma particles to the moving wall when they scatter with it from both sides, one typically expects $\Delta P\propto v$ in the non-relativistic limit, assuming that friction is sufficiently strong to suppress the wall acceleration~\cite{Blasi:2022ayo}. Consequently, we have in the non-relativistic limit
\begin{equation}
    \ell_f = \frac{\sigma\gamma(v)\,v}{\Delta P} \approx \frac{\sigma v}{\Delta P} ~.
\end{equation}
We parametrize the friction length as
\begin{equation}
    \ell_f(T) = \frac{T^{x_\sigma}f^{3-{x_\sigma}}}{\beta\,T^{4-{x_{P}}}f^{x_{P}}} \equiv \frac{T^{{x_f}-4}f^{3-{x_f}}}{\beta} = \frac{t^{(4-{x_f})/2}}{M_{\rm Pl}^{1/2}t_c^{(3-{x_f})/2}}\left(\frac{2}{\beta^2}\sqrt{\frac{\pi^2g_*}{90}}\right)^{1/2} ~,
\end{equation}
where the domain wall tension $\sigma\sim T^{x_\sigma}f^{3-{x_\sigma}}$ with $0\leq {x_\sigma}\leq3$, $\beta\sim \Delta P/(v\,T^{4-{x_{P}}}f^{x_{P}})$ with $0\leq {x_{P}}\leq4$ the drag coefficient determined by the microscopic details of the wall-plasma interaction, and ${x_f}\equiv {x_\sigma}+{x_{P}}$ with $0\leq {x_f}\leq 7$.\footnote{We do not consider non-integer values of the $x$'s. Although such floating values of $x$'s might be generated through dimensional transmutation or the anomalous dimensions in renormalization group running, it is unclear to us what kind of model is needed to achieve this type of scaling for $\sigma$ and $\ell_f$.}
For convenience, we have also defined
\begin{equation}\label{eq:Tc_tc}
    T_c = T_{\rm form} = f ,~\quad  t_c = \sqrt{\frac{90}{4\pi^2g_*}}\frac{M_{\rm Pl}}{T_c^2} ~.
\end{equation}
Since $\ell_f^{-1}\propto 1/t^{(4-{x_f})/2}$, it decreases with $t$ faster than $H(t)\propto 1/t$ for ${x_f}<2$. As we discuss later, this is why there is in principle no scaling solution for ${x_f}>2$, while for ${x_f}=2$ there is no transient stage (called the ``Kibble regime'') between the initially frustrated stage (called the ``stretching regime'') and the final scaling regime. Moreover, ${x_f}=3$ is the critical point where $\ell_f\propto(t/M_{\rm Pl})^{1/2}$ is completely independent of $f$, making $\ell_d$ only dependent on $T$ and thus the entire VOS model completely independent of the symmetry breaking scale $f$. In this situation, friction will dominate for the entire time, and the evolution of the wall will be completely determined by the plasma properties up to the initial correlation length set by $f$.

In practice, $\Delta P$ could also scale with other energy scales and even contain a Boltzmann suppression factor in the low-$T$ regime. For example, Refs.~\cite{Blasi:2022ayo,Hassan:2024bvb} show that for an ALP domain wall of mass $m_a$ that couples to fermions of mass $m_f$ or to photons through a Chern-Simons interaction, $\Delta P$ can also scale with $m_a$ and/or $m_f$ under different kinetic circumstances; when $m_{a,f}\gg T$, it will even be Boltzmann-suppressed. In the following discussion, we assume that $T$ is always sufficiently high so that $\Delta P$ is never Boltzmann-suppressed unless otherwise specified. We also parametrize the additional scale-dependence through $\beta f^{x_P}$, \eg, if $\Delta P\propto m^{x_m}$, then $\beta f^{x_P}=(m/f)^{x_m}f^{x_P}=m^{x_m}f^{x_P-x_m}$, so that $\beta$ can be treated as a $T$-independent constant in the analysis below. Under these assumptions, the expansion described by Eq.~\eqref{eq:VOS} can be asymptotically divided into three stages: the stretching regime, the Kibble regime~\cite{Kibble:1981gv}, and the scaling regime~\cite{Hook:2026grn} (see Figure~\ref{fig:Lt_vt}). Nevertheless, it is possible that $\beta\ll1$, such as discussed in the scenario in Ref.~\cite{Blasi:2022ayo} where $T\gg m_{a,f}$. In this case, the asymptotic division of the solution could potentially be invalidated, which we will comment on below.

\subsubsection{Stretching regime}\label{subsubsec:DW:frustrated:stretching}

Immediately after the formation of the walls at $t=t_c$, domain walls are mainly damped by friction ($\ell_f^{-1}\gg H,\dot{\sigma}/\sigma$) and have very low velocities $v\ll1$. This becomes exact when $x_\sigma=3$, $3H+\dot{\sigma}/\sigma=0$ and hence $\ell_d^{-1}=\ell_f^{-1}$. In this regime, the acceleration of the network is slow compared to the curvature-driven acceleration, $\vert dv/dt\vert\ll 1/L$. As a result, the velocity is set by a quasi-static balance between the curvature term and friction, which leads to the simple relation
\begin{equation}
    0 \simeq \frac{k_w}{L} - \frac{v}{\ell_f} \to v(t) \simeq k_w\frac{\ell_f(t)}{L(t)} ~.
\end{equation}
At the same time, the $v$-dependent energy loss terms $v^2L/\ell_d$ and $c_wv$ are much smaller than the Hubble stretching term $HL$, and thus one has
\begin{equation}\label{eq:L_st}
    \frac{dL}{dt} \simeq HL \to L_{\rm st}(t) \simeq L(t_c)\left(\frac{t}{t_c}\right)^{1/2} ~,
\end{equation}
where $L(t_c)\simeq H^{-1}(t_c)/N$, \ie, $N^3$ roughly parametrizes the total number of domains within a Hubble volume at $t=t_c$. Correspondingly,
\begin{equation}\label{eq:v_st}
    v_{\rm st}(t) \simeq \frac{1}{L(t_c)M_{\rm Pl}^{1/2}}\frac{t^{(3-{x_f})/2}}{t_c^{(2-{x_f})/2}}\sqrt{\frac{2}{\beta^2}\left(\frac{\pi^2g_*}{90}\right)^{1/2}} ~.
\end{equation}
In the stretching regime, the growth of the domain wall length scale is highly suppressed, leading to the inclusion of exponentially more domains than $N^3$ within a Hubble patch, each of which has a significantly smaller size than the Hubble volume. On the other hand, the velocity will increase from a very small initial value until the energy loss terms become comparable to the Hubble stretching term, thus ending the stretching regime and starting the Kibble regime for ${x_f}<3$. When ${x_f}=3$, $v_{\rm st}(t)={\rm constant}$, and thus the domain wall network never leaves the stretching regime. This will manifest again when we discuss the $v(t)$ profile in the Kibble regime. Finally, for ${x_f}>3$, there is in principle no asymptotic stretching solution unless $\beta$ contains additional $T$-dependence, such as by including a Boltzmann suppression factor which will cause $v_{\rm st}\sim \exp(m/T)t^{-({x_f}-3)/2}$ to grow exponentially in $t$. This situation can happen when, for example, $T\ll m_{a,f}$ so that $\Delta P\propto m_a^2m_f^2\exp(-m_a/T)$ for ALP-fermion interactions~\cite{Blasi:2022ayo} or $m_a^3T\exp(-m_a/T)$ for ALP-photon interactions~\cite{Hassan:2024bvb}, and effectively this means that friction is too weak to brake the domain walls and one should just directly adopt the scaling solution. In contrast, in the absence of such $T$-dependence for $\beta$ when ${x_f}>3$, the domain walls will never expand and are effectively ``killed'' by friction. Note that the asymptotic stretching solution could be invalid when $\beta\ll1$, such as in the scenario presented in Ref.~\cite{Blasi:2022ayo} where $m_{a,f}\ll T$. In this case, $\beta\sim(m_a^2m_f^2)/f^4\ll1$, and thus $v_{\rm st}$ might not be low enough to justify dropping the $v$-dependent terms, depending on other factors in the system. Alternatively, a working example is the scenario we consider in Section~\ref{subsec:formation:model} where $f>T>m_f$ and thus $\beta\sim m_f^2/f^2$, assuming that the domain wall mass $\sim f$ and the walls couple to fermions in a manner similar to the ALP domain walls.

\subsubsection{Kibble regime}\label{subsubsec:DW:frustrated:Kibble}

Once the wall velocity becomes sufficiently high, we can no longer ignore the $\mathcal{O}(v)$ energy loss term, but transiently we can still ignore the $\mathcal{O}(v^2)$ term. In the meantime, we still expect $\ell_f^{-1}\gg H,\dot{\sigma}/\sigma$ and $\vert dv/dt\vert\ll v/\ell_f$ due to the strong friction, and thus
\begin{equation}
    0 \simeq \frac{k_w}{L} - \frac{v}{\ell_f} \to v(t) \simeq k_w\frac{\ell_f(t)}{L(t)} ~.
\end{equation}
Correspondingly, the evolution of the length scale is governed by
\begin{equation}
    \frac{dL}{dt} \simeq HL + c_wv = HL + c_wk_w\frac{\ell_f}{L} ~,
\end{equation}
where we made the substitution $v=k_w\ell_f/L$. Using a power-law ansatz for $L(t)$, one gets
\begin{equation}\label{eq:L_K}
    L_{\rm K}(t) \simeq \frac{t^{(6-{x_f})/4}}{M_{\rm Pl}^{1/4}t_c^{(3-{x_f})/4}}\sqrt{\frac{4}{4-{x_f}}\frac{c_wk_w}{\beta}\left(\frac{4\pi^2g_*}{90}\right)^{1/4}} ~,
\end{equation}
and
\begin{equation}\label{eq:v_K}
    v_{\rm K}(t) \simeq \frac{t^{(2-{x_f})/4}}{M_{\rm Pl}^{1/4}t_c^{(3-{x_f})/4}}\sqrt{\frac{4-{x_f}}{4}\frac{k_w}{\beta c_w}\left(\frac{4\pi^2g_*}{90}\right)^{1/4}} ~.
\end{equation}
As can be seen, the wall length scale grows faster than the horizon for ${x_f}<2$ and the same with the horizon for ${x_f}=2$, while the velocity could either increase for ${x_f}<2$, remain constant for ${x_f}=2$, or unphysically decrease for ${x_f}>2$ (setting ${x_f}=1$ will reproduce the evolution pattern of axion strings presented in Ref.~\cite{Hook:2026grn}).
For ${x_f}=2$, notice that both $L_{\rm K}(t)\propto  t$ and $v_{\rm K}(t)={\rm constant}$ behave identically to the scaling behavior described in Eq.~\eqref{eq:Lv_sc}. This means that for ${x_f}=2$, there is no asymptotic Kibble regime in between the stretching regime and the scaling regime, and the $v^2L/\ell_f$ term can never grow strong enough to affect the relevant dynamics.
For ${x_f}=3$, recall that $v_{\rm st}(t)$ is a constant [see Eq.~\eqref{eq:v_st}], which agrees with our observation here that these walls cannot be described by the Kibble solution and will always stay in the stretching regime.
Finally, for ${x_f}>3$, the story again relies on the additional $T$-dependence of $\beta$, which we discussed at the end of Appendix~\ref{subsubsec:DW:frustrated:stretching}.

For ${x_f}<2$, we can obtain the asymptotic transition time $t_{\rm K}$ from the stretching regime to the Kibble regime by solving the equation
\begin{equation}
    L_{\rm st}(t_{\rm K}) = L_{\rm K}(t_{\rm K}) ~,
\end{equation}
which gives
\begin{equation}\label{eq:t_K}
    t_{\rm K} = \frac{t_c^{({x_f}-5)/({x_f}-4)}}{M_{\rm Pl}^{1/({x_f}-4)}}\left[\frac{N}{2}\sqrt{\frac{4}{4-{x_f}}\frac{c_wk_w}{\beta}\left(\frac{4\pi^2g_*}{90}\right)^{1/4}}\right]^{4/({x_f}-4)} ~~~[{x_f}<2] ~.
\end{equation}
On the other hand, the asymptotic transition time $t_{\rm sc}$ from the Kibble regime to the scaling regime is given by
\begin{equation}\label{eq:t_sc}
    3H(t_{\rm sc})+\frac{\dot{\sigma}(t_{\rm sc})}{\sigma(t_{\rm sc})}\sim\ell_f^{-1}(t_{\rm sc}) \to t_{\rm sc} = \frac{t_c^{({x_f}-3)/({x_f}-2)}}{M_{\rm Pl}^{1/({x_f}-2)}}\left[\frac{(3-{x_\sigma})^2}{2\beta^2}\sqrt{\frac{\pi^2g_*}{90}}\right]^{1/({x_f}-2)} ~~~[{x_f}<2] ~.
\end{equation}
As for ${x_f}=2$, there is only one transition from the stretching regime to the scaling regime at $t=t_{\rm K}=t_{\rm sc}$ as defined in Eq.~\eqref{eq:t_K}.

\subsubsection{Scaling regime}\label{subsubsec:DW:frustrated:scaling}

After the velocity grows sufficiently high at $t=t_{\rm sc}$ (for ${x_f}<2$, see Eq.~\eqref{eq:t_sc}; for ${x_f}=2$, see Eq.~\eqref{eq:t_K} and the last paragraph in Appendix~\ref{subsubsec:DW:frustrated:Kibble}) such that friction becomes negligible compared to the $H$ and $\dot{\sigma}/\sigma$ terms, the domain wall network will enter the scaling regime as discussed in Appendix~\ref{subsec:DW:scaling}. Note that this only happens for ${x_f}\leq 2$ as discussed previously, which means that as long as friction is sufficiently strong, wall networks with $\ell_f\propto T^{x_f}f^{3-{x_f}}$, ${x_f}>2$ will never adopt the scaling behavior, while for ${x_f}=2$, the $v^2L/\ell_f$ term is always negligible, and the effective scaling behavior of the wall network is described by Eqs.~\eqref{eq:L_K} and \eqref{eq:v_K} instead.

\subsection{Collapse}\label{subsec:DW:collapse}

If a bias potential $V_{\rm bias}$ is generated, potentially along with the QCD phase transition at $T=T_{\rm QCD}\approx150$~MeV, then the domain wall network collapses at $T=T_{\rm ann}\lesssim T_{\rm QCD}$, \ie, when the tension force is caught up by the bias pressure [$\rho_w(t_{\rm ann})=V_{\rm bias}$]. Because of the curvature-driven acceleration, which grows inversely proportional to the domain wall length scale, it is usually expected that the collapse will finish instantaneously (or at least last much shorter than a Hubble time). However, there are two reasons why this might not always be the case: (1) When the vacuum structure of the spontaneously broken symmetry manifests a more-than-two-fold degeneracy, junctions--the intersections of multiple vacua at a single point or along a line segment--will inevitably form and effectively reduce the overall curvature of the walls~\cite{PinaAvelino:2006ia,Avelino:2006xf,Avelino:2008ve,Battye:2011ff}. (2) If the domain walls interact strongly enough with the plasma, then the induced friction could significantly reduce the collapse velocity. In this study, we only consider the second case.

To model the collapse of individual domain walls, we turn to the VOS model for a single defect which has been studied in Ref.~\cite{Avelino:2019wqd}. Denoting the comoving radius of a single wall by $q$, the VOS model expressed in terms of comoving time $\tau$ is given by
\begin{equation}
\begin{aligned}
    \frac{dq}{d\tau} &= -v ~, \\
    \frac{dv}{d\tau} &= (1-v^2)\left(\frac{s}{q}-\frac{a}{\ell_d}v\right) ~,
\end{aligned}
\end{equation}
where we express the damping term given in Eq.~\eqref{eq:l_d:full} in terms of $\tau$ as
\begin{equation}
    \ell_d^{-1} = 3\frac{\mathcal{H}}{a} + \ell_f^{-1} + \frac{1}{a}\frac{\sigma^\prime(\tau)}{\sigma(\tau)} ~.
\end{equation}
Using the physical length $R=a\,q$ and changing the time variable from $\tau$ to $t$, we have
\begin{equation}
\begin{aligned}
    \frac{dR}{dt} &= HR - v ~, \\
    \frac{dv}{dt} &= (1-v^2)\left(\frac{s}{R}-\frac{v}{\ell_d}\right) ~,
\end{aligned}
\end{equation}
where, this time, the damping term is given in $t$ as in Eq.~\eqref{eq:l_d:full}. Note that $s$ is the curvature parameter. In particular, $s=2$ for a spherical wall, $s=1$ for a cylindrical wall, and $s=0$ for a planar wall. Now, to properly describe the dynamics of the collapse, we need to further include the bias potential $V_{\rm bias}$, which leads to
\begin{equation}
\begin{aligned}
    \frac{dR}{dt} &= HR - v ~, \\
    \frac{dv}{dt} &= (1-v^2)\left(\frac{s}{R}+\frac{V_{\rm bias}}{\sigma}-\frac{v}{\ell_d}\right) ~.
\end{aligned}
\end{equation}

Assuming that the damping effect is strong enough so that the ``terminal velocity'' $v_t\ll 1$ is achieved instantaneously, one has
\begin{equation}
    v_t(t) \approx \ell_d(t)\left(\frac{s}{R}+\frac{V_{\rm bias}}{\sigma}\right) ~.
\end{equation}
Note that the terminal velocity defined in this context is $t$-dependent, which is different from the usual constant terminal velocity defined in other mechanical systems. Substituting this into the $dR/dt$ equation, one obtains
\begin{equation}\label{eq:dRdt_vt}
    \frac{dR}{dt} \approx HR - \ell_d(t)\left(\frac{s}{R}+\frac{V_{\rm bias}}{\sigma}\right) ~.
\end{equation}
We examine two cases in the following. 

First, assuming that friction is negligible, the domain wall size is then predicted by the scaling behavior as $L(t)\sim t$. Using the $\mathbb{Z}_N$ model, one has $t_{\rm ann}=\mathcal{A}\,\sigma(T_{\rm ann})/V_{\rm bias}$, which leads to
\begin{equation}
\begin{aligned}
    v_t(t) &= \left(3H+\frac{\dot{\sigma}}{\sigma}\right)^{-1}\left(\frac{s}{R}+\frac{V_{\rm bias}}{\sigma}\right) \\
    &= \frac{2t}{3-{x_\sigma}}\left[\frac{s}{R} + \frac{V_{\rm bias}}{\sigma(T_{\rm ann})}\left(\frac{T_{\rm ann}}{T}\right)^{x_\sigma}\right] = \frac{2t}{3-{x_\sigma}}\left[\frac{s}{R} + \frac{\mathcal{A}}{t_{\rm ann}}\left(\frac{t}{t_{\rm ann}}\right)^{{x_\sigma}/2}\right] ~~~[\text{Frictionless}] ~,
\end{aligned}
\end{equation}
where we have taken $\sigma(T)=T^{x_\sigma}f^{3-{x_\sigma}}$. Given $T_{\rm ann}$ and $V_{\rm bias}$, one can then determine
\begin{equation}
    f = \left[\sqrt{\frac{90}{4\pi^2g_*}}\frac{M_{\rm Pl}V_{\rm bias}}{\mathcal{A}\,T_{\rm ann}^{2+{x_\sigma}}}\right]^{1/(3-{x_\sigma})} ~~~[\text{Frictionless}] ~.
\end{equation}
Notice that
\begin{equation}
    v_t(t) \geq \frac{2t}{3-{x_\sigma}}\frac{\mathcal{A}}{t_{\rm ann}}\left(\frac{t}{t_{\rm ann}}\right)^{{x_\sigma}/2} \geq \frac{2\mathcal{A}}{3} ~~~[\text{Frictionless}] ~,
\end{equation}
where, naively plugging in $\mathcal{A}=0.4N$, one has $v_t(t)\geq 0.8N/3\geq 0.53$ for $N\geq2$. As a result, a non-relativistic terminal velocity can never be reached in a frictionless scenario. Alternatively, we can solve for $v(t)$ in the non-relativistic limit assuming $s=0$ and $v(t_{\rm ann})=0$, which gives
\begin{equation}
    v(t) = 0.16N\left[
        \left(\frac{t}{t_{\rm ann}}\right)^{({x_\sigma}+2)/2}-\left(\frac{t}{t_{\rm ann}}\right)^{({x_\sigma}-3)/2}
    \right] ~~~[\text{Frictionless and }s=0] ~.
\end{equation}
Therefore, even for the minimal case with ${x_\sigma}=0$, $v(t)$ is still growing linearly in $t$ and can easily reach the speed of light within a Hubble time, making it unlikely for the frictionless collapse to remain non-relativistic.

Next, we turn to the friction-dominated case, \ie, $\ell_d^{-1}\approx \ell_f^{-1}$ (note that this becomes exact when $x_\sigma=3$). Parametrizing $\ell_f=T^{{x_f}-4}f^{3-{x_f}}/\beta$ where ${x_f}=x_\sigma+x_P$ as in Appendix~\ref{subsec:DW:frustrated}, we have
\begin{equation}
    v_t(t) = \ell_f\left(\frac{s}{R}+\frac{V_{\rm bias}}{\sigma}\right) ~~~[\text{Friction-dominant}] ~.
\end{equation}
Suppose the wall expansion has always remained in the stretching regime (see Appendix~\ref{subsubsec:DW:frustrated:stretching}) before the collapse begins, one has $L(t)\sim (t/t_c)^{1/2}H^{-1}(t_c)=2t^{1/2}t_c^{1/2}$ where $t_c$ are defined in Eq.~\eqref{eq:Tc_tc}. Consequently, one can derive 
\begin{equation}
    t_{\rm ann}=\mathcal{A}^2\,\sigma^2(T_{\rm ann})/(4V_{\rm bias}^2t_c)=0.04\,N^2\sigma^2(T_{\rm ann})/(V_{\rm bias}^2\,t_c) ~,
\end{equation}
and thus 
\begin{equation}
    f = \left(\frac{5}{N}\sqrt{\frac{90}{4\pi^2g_*}}\frac{M_{\rm Pl}V_{\rm bias}}{T_{\rm ann}^{1+x_\sigma}}\right)^{1/(4-x_\sigma)} ~~~[\text{Friction-dominant}] ~.
\end{equation}
To make sure that $t_K\gtrsim t_{\rm ann}$ for $x_f\leq2$ [see Eq.~\eqref{eq:t_K}] so that the collapse begins during the stretching regime (this will not be a concern if $x_f=3$, while we do not consider $x_f>3$, as discussed in Section~\ref{subsubsec:DW:frustrated:stretching}), one needs to require
\begin{equation}
    f
    \lesssim
    \left[\frac{5}{N}\frac{V_{\rm bias}}{T_{\rm ann}^{x_\sigma}}\left(\frac{N}{2}\sqrt{\frac{4}{4-{x_f}}\frac{c_wk_w}{\beta}}\right)^{\frac{2}{{x_f}-4}}\left(\sqrt{\frac{90}{4\pi^2g_*}}M_{\rm Pl}\right)^{\frac{x_f-5}{x_f-4}}\right]^{\left[\frac{{x_f}-5}{{x_f}-4}+(4-x_\sigma)\right]^{-1}} ~~~[{x_f}\leq 2] ~,
\end{equation}
and correspondingly
\begin{equation}
    V_{\rm bias} \lesssim \frac{N}{5}\left[\frac{4-x_f}{N^2}\frac{\beta}{c_wk_w}\right]^{\frac{x_\sigma-4}{x_f-5}}\left(\sqrt{\frac{90}{4\pi^2g_*}}M_{\rm Pl}\right)^{\frac{x_\sigma-x_f+1}{x_f-5}}T_{\rm ann}^{-\frac{x_\sigma-5x_f+21}{x_f-5}} ~~~[{x_f}\leq 2] ~.
\end{equation}
As long as this condition is satisfied, we can derive
\begin{equation}\label{eq:v_t_friction}
\begin{aligned}
    v_t(t) &= \frac{T^{x_f-4}f^{3-x_f}}{\beta}\left[\frac{s}{R}+\frac{V_{\rm bias}}{\sigma(T_{\rm ann})}\left(\frac{T_{\rm ann}}{T}\right)^{x_\sigma}\right] \\
    &= \left(\frac{4\pi^2g_*}{90}\right)^{1/2}\frac{T_c}{\beta M_{\rm Pl}}\left[\frac{s}{R}+\frac{0.2Nt^{x_\sigma/2}}{t_c^{1/2}t_{\rm ann}^{(1+x_\sigma)/2}}\right]\frac{t^{(4-x_f)/2}}{t_c^{(2-x_f)/2}}~~~[\text{Friction-dominant}] ~.
\end{aligned}
\end{equation}
Again, we examine the lower bound
\begin{equation}
    v_t(t) \geq  \left(\frac{4\pi^2g_*}{90}\right)^{1/2}\frac{0.2N}{\beta}\frac{T_c}{M_{\rm Pl}}\left(\frac{T_c}{T_{\rm ann}}\right)^{3-x_f} ~~~[\text{Friction-dominant}] ~,
\end{equation}
which is Planck-suppressed. Plugging Eq.~\eqref{eq:v_t_friction} into Eq.~\eqref{eq:dRdt_vt}, one gets
\begin{equation}
    \frac{dR}{dt} \approx \frac{R}{2t} - \left(\frac{4\pi^2g_*}{90}\right)^{1/2}\frac{T_c}{\beta M_{\rm Pl}}\left[\frac{s}{R}+\frac{0.2Nt^{x_\sigma/2}}{t_c^{1/2}t_{\rm ann}^{(1+x_\sigma)/2}}\right]\frac{t^{(4-x_f)/2}}{t_c^{(2-x_f)/2}} ~~~[\text{Friction-dominant}] ~,
\end{equation}
subject to the initial condition $R(t_{\rm ann})=2\,t_{\rm ann}^{1/2}t_c^{1/2}/N$. Notice that when $HR-v_t(t) > 0$, even though the wall velocity is pointing inward, the domain wall size will still temporarily increase along with Hubble expansion until the latter becomes caught up by the curvature force.

\section{Gravitational waves from frustrated domain walls}\label{sec:GW_details}

In this appendix, we provide a detailed derivation of the GW spectrum generated by the collapse of frustrated domain walls, focusing on the stretching regime discussed in Appendix~\ref{sec:DW}. We first begin with Einstein's quadrupole formula~\cite{Einstein:1918btx}, which states that the gravitational radiation power is roughly scaling as $P_{\rm GW}\sim G(d^3Q/dt^3)^2$, $Q$ being the transverse-traceless part of the quadrupole moment of matter and $G=1/(8\pi M_{\rm Pl}^2)$ the Newton constant. For a domain wall, $Q(t)\sim \sigma(t)L(t)^4$~\cite{Gleiser:1998na}. The energy density of GW's released per unit Hubble time is the given by $\rho_{\rm GW}\sim P_{\rm GW}H^{-1}/L^3$, where the factor $L^{-3}$ can be thought of as the number density of GW sources. Since our focus is the frustrated domain walls that never leave the stretching regime, we have
\begin{equation}
    Q(t) \sim \frac{16}{N^4}\left(\sqrt{\frac{90}{4\pi^2g_*}}M_{\rm Pl}\right)^{3/2}t_c^{(x_\sigma+1)/2}t^{(4-x_\sigma)/2} ~,
\end{equation}
\begin{equation}
    P_{\rm GW}(t) \sim \left[(4-x_\sigma)(2-x_\sigma)x_\sigma\right]^2\frac{M_{\rm Pl}}{2\pi N^8}\left(\sqrt{\frac{90}{4\pi^2g_*}}\right)^{3}t_c^{x_\sigma+1}t^{-x_\sigma-2} ~,
\end{equation}
\begin{equation}
\begin{aligned}
    \rho_{\rm GW}(t) &\sim \left[(4-x_\sigma)(2-x_\sigma)x_\sigma\right]^2\frac{M_{\rm Pl}}{8\pi N^5}\left(\sqrt{\frac{90}{4\pi^2g_*}}\right)^{3}t_c^{x_\sigma-1/2}t^{-x_\sigma-5/2} \\
    &= \frac{\left[(4-x_\sigma)(2-x_\sigma)x_\sigma\right]^2}{8\pi N^5M_{\rm Pl}^2}\frac{T^{2x_\sigma+5}}{f^{2x_\sigma-1}} ~.
\end{aligned}
\end{equation}
Notice that $\rho_{\rm GW}=0$ for $x_\sigma=0,2,4$, which arises from the $d^3Q/dt^3$ term that comes from the quadrupolar nature of GW's. Since we only consider $x_\sigma=1,3\leq 3$, the prefactor is the same for both cases, $\left[(4-x_\sigma)(2-x_\sigma)x_\sigma\right]^2=9$. We remark that this estimation only applies to the ideal domain wall geometry considered so far, while in reality there might be multiple length scales involved to make $d^3Q/dt^3\neq0$ for even $x_\sigma$'s. We first analyze the peak amplitude, which is generated at $T_{\rm ann}$ and can be obtained from
\begin{equation}
    \rho_{\rm GW}(T_{\rm ann}) \sim \frac{9}{8\pi}\left(\frac{1}{5}\sqrt{\frac{4\pi^2g_*}{90}}\right)^{\frac{2x_\sigma-1}{4-x_\sigma}}\frac{T_{\rm ann}^{\frac{4x_\sigma+19}{4-x_\sigma}}}{N^{\frac{21-7x_\sigma}{4-x_\sigma}}M_{\rm Pl}^{\frac{7}{4-x_\sigma}}V_{\rm bias}^{\frac{2x_\sigma-1}{4-x_\sigma}}} ~.
\end{equation}
Consequently, the observed peak frequency at the present time $t_0$ is roughly given by~\cite{Saikawa:2017hiv}
\begin{equation}
    f_{\rm peak} \simeq \left(\frac{R(t_{\rm ann})}{R(t_0)}\right)\frac{1}{2\pi R(t_{\rm ann})} = \left(\frac{g_{*s0}}{g_{*s}(T_{\rm ann})}\right)^{1/3}\frac{1}{2\pi R(t_{\rm ann})} ~,
\end{equation}
where $g_{*s}(T)$ is the relativistic entropy degrees of freedom at temperature $T$ and $g_{*s0}=g_{*s}(t_0)$, with the peak amplitude
\begin{equation}
    \Omega_{\rm GW}h^2(t_0)\Big\vert_{\rm peak} = \frac{\rho_{\rm GW}(t_0)h^2}{\rho_c(t_0)}\Big\vert_{\rm peak} = \Omega_{\rm rad} h^2\left(\frac{g_{*}(T_{\rm ann})}{g_{*0}}\right)\left(\frac{g_{*s0}}{g_{*s}(T_{\rm ann})}\right)^{4/3}\Omega_{\rm GW}(T_{\rm ann})\Big\vert_{\rm peak} ~,
\end{equation}
where we take $g_{*s}(T_{\rm ann})\approx g_{*}(T_{\rm ann})$ and we have $g_{*s0}=g_{*s}(t_0)=3.91$ and $g_{*0}=g_*(t_0)=3.36$. Here, $\Omega_{\rm rad}h^2=4.15\times10^{-5}$ is the density parameter of radiations at the present time and $h=H_0/100~{\rm km\cdot sec^{-1}Mpc^{-1}}$ is the reduced Hubble parameter. Moreover,
\begin{equation}
    \Omega_{\rm GW}(T_{\rm ann})\Big\vert_{\rm peak} = \frac{1}{\rho_c(T_{\rm ann})}\left(\frac{d\rho_{\rm GW}(T_{\rm ann})}{d\ln k}\right)\Big\vert_{\rm peak} ~,
\end{equation}
where $k$ is the comoving wavenumber and $\rho_c=3H^2/(8\pi G)=3H^2M_{\rm Pl}^2$ is the critical energy density. For a scaling domain wall, simulations give that~\cite{Hiramatsu:2013qaa}
\begin{equation}
    \left(\frac{d\rho_{\rm GW}(T_{\rm ann})}{d\ln k}\right)\Big\vert_{\rm peak} = \tilde{\epsilon}_{\rm GW}\rho_{\rm GW}\Big\vert_{\rm peak} ,~\quad  \tilde{\epsilon}_{\rm GW}=0.7\pm0.4 ~.
\end{equation}
We assume that this also applies to the stretching domain wall and simply take $\tilde{\epsilon}_{\rm GW}=0.7$. Finally, we obtain
\begin{equation}
    \Omega_{\rm GW}h^2(t_0)\Big\vert_{\rm peak} \simeq 1.18\times10^{-5}\left(\sqrt{\frac{90}{4\pi^2g_*}}\right)^{\frac{9-4x_\sigma}{4-x_\sigma}}\frac{5^{\frac{1-2x_\sigma}{4-x_\sigma}}T_{\rm ann}^{\frac{3+8x_\sigma}{4-x_\sigma}}}{N^{\frac{21-7x_\sigma}{4-x_\sigma}}M_{\rm Pl}^{\frac{7}{4-x_\sigma}}V_{\rm bias}^{\frac{2x_\sigma-1}{4-x_\sigma}}} ~~~[x_\sigma=1,3] ~,
\end{equation}
with
\begin{equation}
    f_{\rm peak} \simeq 5.82\times10^{-2} \left(\sqrt{\frac{4\pi^2g_*}{90}}\right)^{\frac{3-x_\sigma}{4-x_\sigma}}\frac{5^{\frac{1}{4-x_\sigma}}N^{\frac{3-x_\sigma}{4-x_\sigma}}V_{\rm bias}^{\frac{1}{4-x_\sigma}}}{M_{\rm Pl}^{\frac{3-x_\sigma}{4-x_\sigma}}T_{\rm ann}^{\frac{2x_\sigma-3}{4-x_\sigma}}} ~~~[x_\sigma=1,3] ~.
\end{equation}
Numerically, they are
\begin{equation}\label{eq:GW_peak_amp}
    \Omega_{\rm GW}h^2(t_0)\Big\vert_{\rm peak} \simeq \begin{cases}
        4.90\times10^{-54}\left(\frac{7}{N}\right)^{14/3}\left(\frac{T_{\rm ann}}{50\,{\rm MeV}}\right)^{11/3}\left(\frac{(1\,{\rm MeV})^4}{V_{\rm bias}}\right)^{1/3} &,~ x_\sigma=1 \\[2ex]
        5.17\times10^{-112}\left(\frac{T_{\rm ann}}{50\,{\rm MeV}}\right)^{27}\left(\frac{(1\,{\rm MeV})^4}{V_{\rm bias}}\right)^{5} &,~ x_\sigma=3
    \end{cases} ~,
\end{equation}
and
\begin{equation}
    f_{\rm peak} \simeq \begin{cases}
        2.94\times10^{6}\,{\rm Hz}\left(\frac{N}{7}\right)^{2/3}\left(\frac{T_{\rm ann}}{50\,{\rm MeV}}\right)^{1/3}\left(\frac{V_{\rm bias}}{(1\,{\rm MeV})^4}\right)^{1/3} &,~ x_\sigma=1 \\[2ex]
        5.63\times10^{14}\,{\rm Hz}\left(\frac{V_{\rm bias}}{(1\,{\rm MeV})^4}\right)\left(\frac{50\,{\rm MeV}}{T_{\rm ann}}\right)^{3} &,~ x_\sigma=3
    \end{cases} ~.
\end{equation}

\end{appendix}
\setlength{\bibsep}{3pt}
\providecommand{\href}[2]{#2}\begingroup\raggedright\endgroup


\begin{thebibliography}{10}

\bibitem{Planck:2018vyg}
{\bf Planck} Collaboration, N.~Aghanim et~al., {\it {Planck 2018 results. VI.
  Cosmological parameters}},  {\em Astron. Astrophys.} {\bf 641} (2020) A6,
  [\href{http://arxiv.org/abs/1807.06209}{{\tt arXiv:1807.06209}}]. [Erratum:
  Astron.Astrophys. 652, C4 (2021)].

\bibitem{Petraki:2013wwa}
K.~Petraki and R.~R. Volkas, {\it {Review of asymmetric dark matter}},  {\em
  Int. J. Mod. Phys. A} {\bf 28} (2013) 1330028,
  [\href{http://arxiv.org/abs/1305.4939}{{\tt arXiv:1305.4939}}].

\bibitem{Foot:2004pa}
R.~Foot, {\it {Mirror matter-type dark matter}},  {\em Int. J. Mod. Phys. D}
  {\bf 13} (2004) 2161--2192,
  [\href{http://arxiv.org/abs/astro-ph/0407623}{{\tt astro-ph/0407623}}].

\bibitem{Bai:2013xga}
Y.~Bai and P.~Schwaller, {\it {Scale of dark QCD}},  {\em Phys. Rev. D} {\bf
  89} (2014), no.~6 063522, [\href{http://arxiv.org/abs/1306.4676}{{\tt
  arXiv:1306.4676}}].

\bibitem{Ritter:2024sqv}
A.~C. Ritter and R.~R. Volkas, {\it {Explaining the cosmological dark matter
  coincidence in asymmetric dark QCD}},  {\em Phys. Rev. D} {\bf 110} (2024),
  no.~1 015032, [\href{http://arxiv.org/abs/2404.05999}{{\tt
  arXiv:2404.05999}}].

\bibitem{Brzeminski:2023wza}
D.~Brzeminski and A.~Hook, {\it {A Dynamical Explanation of the Dark
  Matter-Baryon Coincidence}},  {\em Phys. Rev. Lett.} {\bf 132} (2024), no.~20
  201001, [\href{http://arxiv.org/abs/2310.07777}{{\tt arXiv:2310.07777}}].

\bibitem{Banerjee:2024xhn}
A.~Banerjee, D.~Brzeminski, and A.~Hook, {\it {Predicting the Dark Matter --
  Baryon Abundance Ratio}},  \href{http://arxiv.org/abs/2410.22412}{{\tt
  arXiv:2410.22412}}.

\bibitem{Kibble:1976sj}
T.~W.~B. Kibble, {\it {Topology of Cosmic Domains and Strings}},  {\em J. Phys.
  A} {\bf 9} (1976) 1387--1398.

\bibitem{Zurek:1985qw}
W.~H. Zurek, {\it {Cosmological Experiments in Superfluid Helium?}},  {\em
  Nature} {\bf 317} (1985) 505--508.

\bibitem{Witten:1984rs}
E.~Witten, {\it {Cosmic Separation of Phases}},  {\em Phys. Rev. D} {\bf 30}
  (1984) 272--285.

\bibitem{Aoki:2006we}
Y.~Aoki, G.~Endrodi, Z.~Fodor, S.~D. Katz, and K.~K. Szabo, {\it {The Order of
  the quantum chromodynamics transition predicted by the standard model of
  particle physics}},  {\em Nature} {\bf 443} (2006) 675--678,
  [\href{http://arxiv.org/abs/hep-lat/0611014}{{\tt hep-lat/0611014}}].

\bibitem{Bhattacharya:2014ara}
T.~Bhattacharya et~al., {\it {QCD Phase Transition with Chiral Quarks and
  Physical Quark Masses}},  {\em Phys. Rev. Lett.} {\bf 113} (2014), no.~8
  082001, [\href{http://arxiv.org/abs/1402.5175}{{\tt arXiv:1402.5175}}].

\bibitem{Bai:2024amm}
Y.~Bai and T.-K. Chen, {\it {Approaching stable quark matter}},  {\em Eur.
  Phys. J. C} {\bf 86} (2026), no.~2 147,
  [\href{http://arxiv.org/abs/2410.19678}{{\tt arXiv:2410.19678}}].

\bibitem{Bai:2025zpm}
Y.~Bai and T.-K. Chen, {\it {Excluding Stable Quark Matter: Insights from the
  QCD Vacuum Energy}},  \href{http://arxiv.org/abs/2502.20241}{{\tt
  arXiv:2502.20241}}.

\bibitem{Holdom:2017gdc}
B.~Holdom, J.~Ren, and C.~Zhang, {\it {Quark matter may not be strange}},  {\em
  Phys. Rev. Lett.} {\bf 120} (2018), no.~22 222001,
  [\href{http://arxiv.org/abs/1707.06610}{{\tt arXiv:1707.06610}}].

\bibitem{Zhitnitsky:2002qa}
A.~R. Zhitnitsky, {\it {'Nonbaryonic' dark matter as baryonic color
  superconductor}},  {\em JCAP} {\bf 10} (2003) 010,
  [\href{http://arxiv.org/abs/hep-ph/0202161}{{\tt hep-ph/0202161}}].

\bibitem{Bai:2018vik}
Y.~Bai and A.~J. Long, {\it {Six Flavor Quark Matter}},  {\em JHEP} {\bf 06}
  (2018) 072, [\href{http://arxiv.org/abs/1804.10249}{{\tt arXiv:1804.10249}}].

\bibitem{Bai:2021ibt}
Y.~Bai and M.~Korwar, {\it {Cosmological constraints on first-order phase
  transitions}},  {\em Phys. Rev. D} {\bf 105} (2022), no.~9 095015,
  [\href{http://arxiv.org/abs/2109.14765}{{\tt arXiv:2109.14765}}].

\bibitem{Bringmann:2023opz}
T.~Bringmann, P.~F. Depta, T.~Konstandin, K.~Schmidt-Hoberg, and C.~Tasillo,
  {\it {Does NANOGrav observe a dark sector phase transition?}},  {\em JCAP}
  {\bf 11} (2023) 053, [\href{http://arxiv.org/abs/2306.09411}{{\tt
  arXiv:2306.09411}}].

\bibitem{Bai:2023cqj}
Y.~Bai, T.-K. Chen, and M.~Korwar, {\it {QCD-collapsed domain walls: QCD phase
  transition and gravitational wave spectroscopy}},  {\em JHEP} {\bf 12} (2023)
  194, [\href{http://arxiv.org/abs/2306.17160}{{\tt arXiv:2306.17160}}].

\bibitem{Lee:2020tmi}
D.~Lee, U.-G. Mei{\ss}ner, K.~A. Olive, M.~Shifman, and T.~Vonk, {\it
  {{\ensuremath{\theta}} -dependence of light nuclei and nucleosynthesis}},
  {\em Phys. Rev. Res.} {\bf 2} (2020), no.~3 033392,
  [\href{http://arxiv.org/abs/2006.12321}{{\tt arXiv:2006.12321}}].

\bibitem{Martins:2016ois}
C.~J. A.~P. Martins, I.~Y. Rybak, A.~Avgoustidis, and E.~P.~S. Shellard, {\it
  {Extending the velocity-dependent one-scale model for domain walls}},  {\em
  Phys. Rev. D} {\bf 93} (2016), no.~4 043534,
  [\href{http://arxiv.org/abs/1602.01322}{{\tt arXiv:1602.01322}}].

\bibitem{Hiramatsu:2012sc}
T.~Hiramatsu, M.~Kawasaki, K.~Saikawa, and T.~Sekiguchi, {\it {Axion cosmology
  with long-lived domain walls}},  {\em JCAP} {\bf 01} (2013) 001,
  [\href{http://arxiv.org/abs/1207.3166}{{\tt arXiv:1207.3166}}].

\bibitem{Kawasaki:2014sqa}
M.~Kawasaki, K.~Saikawa, and T.~Sekiguchi, {\it {Axion dark matter from
  topological defects}},  {\em Phys. Rev. D} {\bf 91} (2015), no.~6 065014,
  [\href{http://arxiv.org/abs/1412.0789}{{\tt arXiv:1412.0789}}].

\bibitem{Hook:2026grn}
A.~Hook, R.~Mondal, and S.~Mukherjee, {\it {Putting the Brakes on Axion
  Strings: Friction and Its Impact on the QCD Axion Abundance}},
  \href{http://arxiv.org/abs/2603.00237}{{\tt arXiv:2603.00237}}.

\bibitem{Bhattacharya:1990hk}
T.~Bhattacharya, A.~Gocksch, C.~Korthals~Altes, and R.~D. Pisarski, {\it
  {Interface tension in an SU(N) gauge theory at high temperature}},  {\em
  Phys. Rev. Lett.} {\bf 66} (1991) 998--1000.

\bibitem{Kibble:1981gv}
T.~W.~B. Kibble, {\it {Phase Transitions in the Early Universe}},  {\em Acta
  Phys. Polon. B} {\bf 13} (1982) 723.

\bibitem{Kolb:1990vq}
E.~W. Kolb and M.~S. Turner, {\em {The Early Universe}}, vol.~69.
\newblock Taylor and Francis, 5, 2019.

\bibitem{Blasi:2022ayo}
S.~Blasi, A.~Mariotti, A.~Rase, A.~Sevrin, and K.~Turbang, {\it {Friction on
  ALP domain walls and gravitational waves}},  {\em JCAP} {\bf 04} (2023) 008,
  [\href{http://arxiv.org/abs/2210.14246}{{\tt arXiv:2210.14246}}].

\bibitem{Bagherian:2025puf}
H.~Bagherian, M.~Ekhterachian, and S.~Stelzl, {\it {The bearable inhomogeneity
  of the baryon asymmetry}},  {\em JHEP} {\bf 01} (2026) 068,
  [\href{http://arxiv.org/abs/2505.15904}{{\tt arXiv:2505.15904}}].

\bibitem{Alcock:1985vc}
C.~Alcock and E.~Farhi, {\it {The Evaporation of Strange Matter in the Early
  Universe}},  {\em Phys. Rev. D} {\bf 32} (1985) 1273.

\bibitem{ParticleDataGroup:2024cfk}
{\bf Particle Data Group} Collaboration, S.~Navas et~al., {\it {Review of
  particle physics}},  {\em Phys. Rev. D} {\bf 110} (2024), no.~3 030001.

\bibitem{DiVecchia:1980yfw}
P.~Di~Vecchia and G.~Veneziano, {\it {Chiral Dynamics in the Large n Limit}},
  {\em Nucl. Phys. B} {\bf 171} (1980) 253--272.

\bibitem{Fodor:2016bgu}
Z.~Fodor, C.~Hoelbling, S.~Krieg, L.~Lellouch, T.~Lippert, A.~Portelli,
  A.~Sastre, K.~K. Szabo, and L.~Varnhorst, {\it {Up and down quark masses and
  corrections to Dashen's theorem from lattice QCD and quenched QED}},  {\em
  Phys. Rev. Lett.} {\bf 117} (2016), no.~8 082001,
  [\href{http://arxiv.org/abs/1604.07112}{{\tt arXiv:1604.07112}}].

\bibitem{Hoferichter:2015hva}
M.~Hoferichter, J.~Ruiz~de Elvira, B.~Kubis, and U.-G. Mei{\ss}ner, {\it
  {Roy{\textendash}Steiner-equation analysis of pion{\textendash}nucleon
  scattering}},  {\em Phys. Rept.} {\bf 625} (2016) 1--88,
  [\href{http://arxiv.org/abs/1510.06039}{{\tt arXiv:1510.06039}}].

\bibitem{Hoferichter:2015tha}
M.~Hoferichter, J.~Ruiz~de Elvira, B.~Kubis, and U.-G. Mei{\ss}ner, {\it
  {Matching pion-nucleon Roy-Steiner equations to chiral perturbation theory}},
   {\em Phys. Rev. Lett.} {\bf 115} (2015), no.~19 192301,
  [\href{http://arxiv.org/abs/1507.07552}{{\tt arXiv:1507.07552}}].

\bibitem{Bodeker:2017cim}
D.~Bodeker and G.~D. Moore, {\it {Electroweak Bubble Wall Speed Limit}},  {\em
  JCAP} {\bf 05} (2017) 025, [\href{http://arxiv.org/abs/1703.08215}{{\tt
  arXiv:1703.08215}}].

\bibitem{DiClemente:2024lzi}
F.~Di~Clemente, M.~Casolino, A.~Drago, M.~Lattanzi, and C.~Ratti, {\it {Strange
  quark matter as dark matter: 40~yr later, a reappraisal}},  {\em Mon. Not.
  Roy. Astron. Soc.} {\bf 537} (2025), no.~2 1056--1069,
  [\href{http://arxiv.org/abs/2404.12094}{{\tt arXiv:2404.12094}}].

\bibitem{Bramante:2026wzh}
J.~Bramante, {\it {Very Heavy and Composite Dark Matter: Theory and
  Experimental Searches}},  \href{http://arxiv.org/abs/2602.23708}{{\tt
  arXiv:2602.23708}}.

\bibitem{DeRujula:1984axn}
A.~De~Rujula and S.~L. Glashow, {\it {Nuclearites: A Novel Form of Cosmic
  Radiation}},  {\em Nature} {\bf 312} (1984) 734--737.

\bibitem{Herrin:2005kb}
E.~T. Herrin, D.~C. Rosenbaum, and V.~L. Teplitz, {\it {Seismic search for
  strange quark nuggets}},  {\em Phys. Rev. D} {\bf 73} (2006) 043511,
  [\href{http://arxiv.org/abs/astro-ph/0505584}{{\tt astro-ph/0505584}}].

\bibitem{SinghSidhu:2019tbr}
J.~Singh~Sidhu and G.~D. Starkman, {\it {Reconsidering astrophysical
  constraints on macroscopic dark matter}},  {\em Phys. Rev. D} {\bf 101}
  (2020), no.~8 083503, [\href{http://arxiv.org/abs/1912.04053}{{\tt
  arXiv:1912.04053}}].

\bibitem{Sugiyama:2026kpv}
S.~Sugiyama, M.~Takada, N.~Yasuda, and N.~Tominaga, {\it {Microlensing
  constraints on Primordial Black Hole abundance with Subaru Hyper Suprime-Cam
  observations of Andromeda}},  \href{http://arxiv.org/abs/2602.05840}{{\tt
  arXiv:2602.05840}}.

\bibitem{Bai:2018bej}
Y.~Bai and N.~Orlofsky, {\it {Microlensing of X-ray Pulsars: a Method to Detect
  Primordial Black Hole Dark Matter}},  {\em Phys. Rev. D} {\bf 99} (2019),
  no.~12 123019, [\href{http://arxiv.org/abs/1812.01427}{{\tt
  arXiv:1812.01427}}].

\bibitem{Tamta:2024pow}
M.~Tamta, N.~Raj, and P.~Sharma, {\it {Entering the window of primordial black
  hole dark matter with x-ray microlensing}},  {\em Phys. Rev. D} {\bf 111}
  (2025), no.~4 043043, [\href{http://arxiv.org/abs/2405.20365}{{\tt
  arXiv:2405.20365}}].

\bibitem{Jiang:2025xln}
S.~Jiang, A.~Yang, and F.~P. Huang, {\it {Macroscopic Dark Matter under siege:
  from White Dwarf Data to Gravitational Wave Detection}},
  \href{http://arxiv.org/abs/2511.23263}{{\tt arXiv:2511.23263}}.

\bibitem{Paczynski:1985jf}
B.~Paczynski, {\it {Gravitational microlensing by the galactic halo}},  {\em
  Astrophys. J.} {\bf 304} (1986) 1--5.

\bibitem{Griest:1990vu}
K.~Griest, {\it {Galactic Microlensing as a Method of Detecting Massive Compact
  Halo Objects}},  {\em Astrophys. J.} {\bf 366} (1991) 412--421.

\bibitem{EROS-2:2006ryy}
{\bf EROS-2} Collaboration, P.~Tisserand et~al., {\it {Limits on the Macho
  Content of the Galactic Halo from the EROS-2 Survey of the Magellanic
  Clouds}},  {\em Astron. Astrophys.} {\bf 469} (2007) 387--404,
  [\href{http://arxiv.org/abs/astro-ph/0607207}{{\tt astro-ph/0607207}}].

\bibitem{Wyrzykowski:2010bh}
L.~Wyrzykowski et~al., {\it {The OGLE View of Microlensing towards the
  Magellanic Clouds. II. OGLE-II SMC data}},  {\em Mon. Not. Roy. Astron. Soc.}
  {\bf 407} (2010) 189--200, [\href{http://arxiv.org/abs/1004.5247}{{\tt
  arXiv:1004.5247}}].

\bibitem{Wyrzykowski:2015ppa}
{\L}.~Wyrzykowski et~al., {\it {Black Hole, Neutron Star and White Dwarf
  Candidates from Microlensing with OGLE-III}},  {\em Mon. Not. Roy. Astron.
  Soc.} {\bf 458} (2016), no.~3 3012--3026,
  [\href{http://arxiv.org/abs/1509.04899}{{\tt arXiv:1509.04899}}].

\bibitem{2015AcA....65....1U}
A.~{Udalski}, M.~K. {Szyma{\'n}ski}, and G.~{Szyma{\'n}ski}, {\it {OGLE-IV:
  Fourth Phase of the Optical Gravitational Lensing Experiment}},  {\em Acta
  Astron.} {\bf 65} (2015), no.~1 1--38,
  [\href{http://arxiv.org/abs/1504.05966}{{\tt arXiv:1504.05966}}].

\bibitem{Mroz:2024wia}
P.~Mr{\'o}z et~al., {\it {Limits on Planetary-mass Primordial Black Holes from
  the OGLE High-cadence Survey of the Magellanic Clouds}},  {\em Astrophys. J.
  Lett.} {\bf 976} (2024), no.~1 L19,
  [\href{http://arxiv.org/abs/2410.06251}{{\tt arXiv:2410.06251}}].

\bibitem{Bond:2001kt}
I.~A. Bond et~al., {\it {Real-time difference imaging analysis of moa galactic
  bulge observations during 2000}},  {\em Mon. Not. Roy. Astron. Soc.} {\bf
  327} (2001) 868, [\href{http://arxiv.org/abs/astro-ph/0102181}{{\tt
  astro-ph/0102181}}].

\bibitem{2018PASJ...70S...1M}
S.~{Miyazaki}, Y.~{Komiyama}, S.~{Kawanomoto}, Y.~{Doi}, H.~{Furusawa},
  T.~{Hamana}, Y.~{Hayashi}, H.~{Ikeda}, Y.~{Kamata}, H.~{Karoji}, M.~{Koike},
  T.~{Kurakami}, S.~{Miyama}, T.~{Morokuma}, F.~{Nakata}, K.~{Namikawa},
  H.~{Nakaya}, K.~{Nariai}, Y.~{Obuchi}, Y.~{Oishi}, N.~{Okada}, Y.~{Okura},
  P.~{Tait}, T.~{Takata}, Y.~{Tanaka}, M.~{Tanaka}, T.~{Terai}, D.~{Tomono},
  F.~{Uraguchi}, T.~{Usuda}, Y.~{Utsumi}, Y.~{Yamada}, H.~{Yamanoi},
  H.~{Aihara}, H.~{Fujimori}, S.~{Mineo}, H.~{Miyatake}, M.~{Oguri},
  T.~{Uchida}, M.~M. {Tanaka}, N.~{Yasuda}, M.~{Takada}, H.~{Murayama}, A.~J.
  {Nishizawa}, N.~{Sugiyama}, M.~{Chiba}, T.~{Futamase}, S.-Y. {Wang}, H.-Y.
  {Chen}, P.~T.~P. {Ho}, E.~J.~Y. {Liaw}, C.-F. {Chiu}, C.-L. {Ho}, T.-C.
  {Lai}, Y.-C. {Lee}, D.-Z. {Jeng}, S.~{Iwamura}, R.~{Armstrong},
  S.~{Bickerton}, J.~{Bosch}, J.~E. {Gunn}, R.~H. {Lupton}, C.~{Loomis},
  P.~{Price}, S.~{Smith}, M.~A. {Strauss}, E.~L. {Turner}, H.~{Suzuki},
  Y.~{Miyazaki}, M.~{Muramatsu}, K.~{Yamamoto}, M.~{Endo}, Y.~{Ezaki},
  N.~{Ito}, N.~{Kawaguchi}, S.~{Sofuku}, T.~{Taniike}, K.~{Akutsu}, N.~{Dojo},
  K.~{Kasumi}, T.~{Matsuda}, K.~{Imoto}, Y.~{Miwa}, M.~{Suzuki}, K.~{Takeshi},
  and H.~{Yokota}, {\it {Hyper Suprime-Cam: System design and verification of
  image quality}},  {\em Publ. Astron. Soc. Jap.} {\bf 70} (2018) S1.

\bibitem{2018PASJ...70S...2K}
Y.~{Komiyama}, Y.~{Obuchi}, H.~{Nakaya}, Y.~{Kamata}, S.~{Kawanomoto},
  Y.~{Utsumi}, S.~{Miyazaki}, F.~{Uraguchi}, H.~{Furusawa}, T.~{Morokuma},
  T.~{Uchida}, H.~{Miyatake}, S.~{Mineo}, H.~{Fujimori}, H.~{Aihara},
  H.~{Karoji}, J.~E. {Gunn}, and S.-Y. {Wang}, {\it {Hyper Suprime-Cam: Camera
  dewar design}},  {\em Publ. Astron. Soc. Jap.} {\bf 70} (2018) S2.

\bibitem{2018PASJ...70...66K}
S.~{Kawanomoto}, F.~{Uraguchi}, Y.~{Komiyama}, S.~{Miyazaki}, H.~{Furusawa},
  F.~{Finet}, T.~{Hattori}, S.-Y. {Wang}, N.~{Yasuda}, and N.~{Suzuki}, {\it
  {Hyper Suprime-Cam: Filters}},  {\em Publ. Astron. Soc. Jap.} {\bf 70}
  (2018), no.~4 66.

\bibitem{2018PASJ...70S...3F}
H.~{Furusawa}, M.~{Koike}, T.~{Takata}, Y.~{Okura}, H.~{Miyatake}, R.~H.
  {Lupton}, S.~{Bickerton}, P.~A. {Price}, J.~{Bosch}, N.~{Yasuda}, S.~{Mineo},
  Y.~{Yamada}, S.~{Miyazaki}, F.~{Nakata}, S.~{Koshida}, Y.~{Komiyama},
  Y.~{Utsumi}, S.~{Kawanomoto}, E.~{Jeschke}, J.~{Noumaru}, K.~{Schubert},
  I.~{Iwata}, F.~{Finet}, T.~{Fujiyoshi}, A.~{Tajitsu}, T.~{Terai}, and C.-H.
  {Lee}, {\it {The on-site quality-assurance system for Hyper Suprime-Cam:
  OSQAH}},  {\em Publ. Astron. Soc. Jap.} {\bf 70} (2018) S3.

\bibitem{eXTP:2016rzs}
{\bf eXTP} Collaboration, S.~N. Zhang et~al., {\it {eXTP -- enhanced X-ray
  Timing and Polarimetry Mission}},  {\em Proc. SPIE Int. Soc. Opt. Eng.} {\bf
  9905} (2016) 99051Q, [\href{http://arxiv.org/abs/1607.08823}{{\tt
  arXiv:1607.08823}}].

\bibitem{STROBE-X}
{\bf STROBE-X Science Working Group} Collaboration, P.~S. Ray et~al., {\it
  {STROBE-X: X-ray Timing and Spectroscopy on Dynamical Timescales from
  Microseconds to Years}},  \href{http://arxiv.org/abs/1903.03035}{{\tt
  arXiv:1903.03035}}.

\bibitem{Graham:2015apa}
P.~W. Graham, S.~Rajendran, and J.~Varela, {\it {Dark Matter Triggers of
  Supernovae}},  {\em Phys. Rev. D} {\bf 92} (2015), no.~6 063007,
  [\href{http://arxiv.org/abs/1505.04444}{{\tt arXiv:1505.04444}}].

\bibitem{Graham:2018efk}
P.~W. Graham, R.~Janish, V.~Narayan, S.~Rajendran, and P.~Riggins, {\it {White
  Dwarfs as Dark Matter Detectors}},  {\em Phys. Rev. D} {\bf 98} (2018),
  no.~11 115027, [\href{http://arxiv.org/abs/1805.07381}{{\tt
  arXiv:1805.07381}}].

\bibitem{Madsen:1988zgf}
J.~Madsen, {\it {Astrophysical Limits on the Flux of Quark Nuggets}},  {\em
  Phys. Rev. Lett.} {\bf 61} (1988) 2909--2912.

\bibitem{Watts:2006hk}
A.~L. Watts and S.~Reddy, {\it {Magnetar oscillations pose challenges for
  strange stars}},  {\em Mon. Not. Roy. Astron. Soc.} {\bf 379} (2007) L63,
  [\href{http://arxiv.org/abs/astro-ph/0609364}{{\tt astro-ph/0609364}}].

\bibitem{Seto:2004zu}
N.~Seto and A.~Cooray, {\it {Search for small-mass black hole dark matter with
  space-based gravitational wave detectors}},  {\em Phys. Rev. D} {\bf 70}
  (2004) 063512, [\href{http://arxiv.org/abs/astro-ph/0405216}{{\tt
  astro-ph/0405216}}].

\bibitem{Adams:2004pk}
A.~W. Adams and J.~S. Bloom, {\it {Direct detection of dark matter with
  space-based laser interferometers}},
  \href{http://arxiv.org/abs/astro-ph/0405266}{{\tt astro-ph/0405266}}.

\bibitem{LISA:2017pwj}
{\bf LISA} Collaboration, P.~Amaro-Seoane et~al., {\it {Laser Interferometer
  Space Antenna}},  \href{http://arxiv.org/abs/1702.00786}{{\tt
  arXiv:1702.00786}}.

\bibitem{TianQin:2015yph}
{\bf TianQin} Collaboration, J.~Luo et~al., {\it {TianQin: a space-borne
  gravitational wave detector}},  {\em Class. Quant. Grav.} {\bf 33} (2016),
  no.~3 035010, [\href{http://arxiv.org/abs/1512.02076}{{\tt
  arXiv:1512.02076}}].

\bibitem{Hu:2017mde}
W.-R. Hu and Y.-L. Wu, {\it {The Taiji Program in Space for gravitational wave
  physics and the nature of gravity}},  {\em Natl. Sci. Rev.} {\bf 4} (2017),
  no.~5 685--686.

\bibitem{Crowder:2005nr}
J.~Crowder and N.~J. Cornish, {\it {Beyond LISA: Exploring future gravitational
  wave missions}},  {\em Phys. Rev. D} {\bf 72} (2005) 083005,
  [\href{http://arxiv.org/abs/gr-qc/0506015}{{\tt gr-qc/0506015}}].

\bibitem{Dewdney:2009tmd}
P.~E. Dewdney, P.~J. Hall, R.~T. Schilizzi, and T.~J. L.~W. Lazio, {\it {The
  Square Kilometre Array}},  8, 2009.

\bibitem{Manchester:2013ndt}
R.~N. Manchester, {\it {The International Pulsar Timing Array}},  {\em Class.
  Quant. Grav.} {\bf 30} (2013) 224010,
  [\href{http://arxiv.org/abs/1309.7392}{{\tt arXiv:1309.7392}}].

\bibitem{Kramer:2013kea}
{\bf EPTA} Collaboration, M.~Kramer and D.~J. Champion, {\it {The European
  Pulsar Timing Array and the Large European Array for Pulsars}},  {\em Class.
  Quant. Grav.} {\bf 30} (2013) 224009.

\bibitem{Brazier:2019mmu}
A.~Brazier et~al., {\it {The NANOGrav Program for Gravitational Waves and
  Fundamental Physics}},  \href{http://arxiv.org/abs/1908.05356}{{\tt
  arXiv:1908.05356}}.

\bibitem{Burdin:2014xma}
S.~Burdin, M.~Fairbairn, P.~Mermod, D.~Milstead, J.~Pinfold, T.~Sloan, and
  W.~Taylor, {\it {Non-collider searches for stable massive particles}},  {\em
  Phys. Rept.} {\bf 582} (2015) 1--52,
  [\href{http://arxiv.org/abs/1410.1374}{{\tt arXiv:1410.1374}}].

\bibitem{Cyncynates:2016rij}
D.~Cyncynates, J.~Chiel, J.~Sidhu, and G.~D. Starkman, {\it {Reconsidering
  seismological constraints on the available parameter space of macroscopic
  dark matter}},  {\em Phys. Rev. D} {\bf 95} (2017), no.~6 063006,
  [\href{http://arxiv.org/abs/1610.09680}{{\tt arXiv:1610.09680}}]. [Addendum:
  Phys.Rev.D 95, 129903 (2017)].

\bibitem{Marsquake}
B.~Fernando, I.~J. Daubar, C.~Charalambous, P.~M. Grindrod, A.~Stott,
  A.~Al~Ateqi, D.~Atri, S.~Ceylan, J.~Clinton, M.~Fillingim, E.~Hauber, J.~R.
  Hill, T.~Kawamura, J.~Liu, A.~Lucas, R.~Lorenz, L.~Ojha, C.~Perrin,
  S.~Piqueux, S.~Stähler, D.~Tirsch, C.~Wilson, N.~Wójcicka, D.~Giardini,
  P.~Lognonné, and W.~B. Banerdt, {\it A tectonic origin for the largest
  marsquake observed by insight},  {\em Geophysical Research Letters} {\bf 50}
  (2023), no.~20 e2023GL103619,
  [\href{http://arxiv.org/abs/https://agupubs.onlinelibrary.wiley.com/doi/pdf/10.1029/2023GL103619}{{\tt
  https://agupubs.onlinelibrary.wiley.com/doi/pdf/10.1029/2023GL103619}}].

\bibitem{Theodosopoulos:2026ehn}
D.~P. Theodosopoulos, K.~Freese, C.~Kelso, and P.~Stengel, {\it {Projected
  Sensitivity of Paleo-Detectors to Dark Matter Effective Interactions with
  Nuclei}},  \href{http://arxiv.org/abs/2603.13629}{{\tt arXiv:2603.13629}}.

\bibitem{Jacobs:2014yca}
D.~M. Jacobs, G.~D. Starkman, and B.~W. Lynn, {\it {Macro Dark Matter}},  {\em
  Mon. Not. Roy. Astron. Soc.} {\bf 450} (2015), no.~4 3418--3430,
  [\href{http://arxiv.org/abs/1410.2236}{{\tt arXiv:1410.2236}}].

\bibitem{Price:1986ky}
P.~B. Price and M.~H. Salamon, {\it {Search for Supermassive Magnetic Monopoles
  Using Mica Crystals}},  {\em Phys. Rev. Lett.} {\bf 56} (1986) 1226--1229.

\bibitem{Burns:2023sgx}
A.-K. Burns, T.~M.~P. Tait, and M.~Valli, {\it {PRyMordial: the first three
  minutes, within and beyond the standard model}},  {\em Eur. Phys. J. C} {\bf
  84} (2024), no.~1 86, [\href{http://arxiv.org/abs/2307.07061}{{\tt
  arXiv:2307.07061}}].

\bibitem{Atreya:2014sca}
A.~Atreya, A.~Sarkar, and A.~M. Srivastava, {\it {Reviving quark nuggets as a
  candidate for dark matter}},  {\em Phys. Rev. D} {\bf 90} (2014), no.~4
  045010, [\href{http://arxiv.org/abs/1405.6492}{{\tt arXiv:1405.6492}}].

\bibitem{SinghSidhu:2019cpq}
J.~Singh~Sidhu and G.~Starkman, {\it {Macroscopic Dark Matter Constraints from
  Bolide Camera Networks}},  {\em Phys. Rev. D} {\bf 100} (2019), no.~12
  123008, [\href{http://arxiv.org/abs/1908.00557}{{\tt arXiv:1908.00557}}].

\bibitem{1986AJ.....92..595H}
J.~G. {Hills}, {\it {Limitations on the masses of objects constituting the
  missing mass in the Galactic disk and the Galactic halo}},  {\em Astron. J.}
  {\bf 92} (1986) 595--599.

\bibitem{2017ExA....43..237H}
R.~M. {Howie}, J.~{Paxman}, P.~A. {Bland}, M.~C. {Towner}, M.~{Cupak}, E.~K.
  {Sansom}, and H.~A.~R. {Devillepoix}, {\it {How to build a continental scale
  fireball camera network}},  {\em Exp. Astron.} {\bf 43} (2017), no.~3
  237--266.

\bibitem{Ramazanov:2021eya}
S.~Ramazanov, E.~Babichev, D.~Gorbunov, and A.~Vikman, {\it {Beyond freeze-in:
  Dark matter via inverse phase transition and gravitational wave signal}},
  {\em Phys. Rev. D} {\bf 105} (2022), no.~6 063530,
  [\href{http://arxiv.org/abs/2104.13722}{{\tt arXiv:2104.13722}}].

\bibitem{Babichev:2021uvl}
E.~Babichev, D.~Gorbunov, S.~Ramazanov, and A.~Vikman, {\it {Gravitational
  shine of dark domain walls}},  {\em JCAP} {\bf 04} (2022), no.~04 028,
  [\href{http://arxiv.org/abs/2112.12608}{{\tt arXiv:2112.12608}}].

\bibitem{Dankovsky:2024ipq}
I.~Dankovsky, S.~Ramazanov, E.~Babichev, D.~Gorbunov, and A.~Vikman, {\it
  {Numerical analysis of melting domain walls and their gravitational waves}},
  {\em JCAP} {\bf 02} (2025) 064, [\href{http://arxiv.org/abs/2410.21971}{{\tt
  arXiv:2410.21971}}].

\bibitem{Husdal:2016haj}
L.~Husdal, {\it {On Effective Degrees of Freedom in the Early Universe}},  {\em
  Galaxies} {\bf 4} (2016), no.~4 78,
  [\href{http://arxiv.org/abs/1609.04979}{{\tt arXiv:1609.04979}}].

\bibitem{Hassan:2024bvb}
S.~Hassan, G.~R. Kane, J.~March-Russell, and G.~Obied, {\it {Chern-Simons
  induced thermal friction on axion domain walls}},  {\em JHEP} {\bf 03} (2025)
  022, [\href{http://arxiv.org/abs/2410.19906}{{\tt arXiv:2410.19906}}].

\bibitem{PinaAvelino:2006ia}
P.~Pina~Avelino, C.~J. A.~P. Martins, J.~Menezes, R.~Menezes, and J.~C. R.~E.
  Oliveira, {\it {Frustrated expectations: defect networks and dark energy}},
  {\em Phys. Rev. D} {\bf 73} (2006) 123519,
  [\href{http://arxiv.org/abs/astro-ph/0602540}{{\tt astro-ph/0602540}}].

\bibitem{Avelino:2006xf}
P.~P. Avelino, C.~J. A.~P. Martins, J.~Menezes, R.~Menezes, and J.~C. R.~E.
  Oliveira, {\it {Scaling of cosmological domain wall networks with
  junctions}},  {\em Phys. Lett. B} {\bf 647} (2007) 63--66,
  [\href{http://arxiv.org/abs/astro-ph/0612444}{{\tt astro-ph/0612444}}].

\bibitem{Avelino:2008ve}
P.~P. Avelino, C.~J. A.~P. Martins, J.~Menezes, R.~Menezes, and J.~C. R.~E.
  Oliveira, {\it {Dynamics of domain wall networks with junctions}},  {\em
  Phys. Rev. D} {\bf 78} (2008) 103508,
  [\href{http://arxiv.org/abs/0807.4442}{{\tt arXiv:0807.4442}}].

\bibitem{Battye:2011ff}
R.~A. Battye, J.~A. Pearson, and A.~Moss, {\it {X-type and Y-type junction
  stability in domain wall networks}},  {\em Phys. Rev. D} {\bf 84} (2011)
  125032, [\href{http://arxiv.org/abs/1107.1325}{{\tt arXiv:1107.1325}}].

\bibitem{Avelino:2019wqd}
P.~P. Avelino, {\it {Parameter-free velocity-dependent one-scale model for
  domain walls}},  {\em Phys. Rev. D} {\bf 101} (2020), no.~2 023514,
  [\href{http://arxiv.org/abs/1910.07011}{{\tt arXiv:1910.07011}}].

\bibitem{Einstein:1918btx}
A.~Einstein, {\it {{\"U}ber Gravitationswellen}},  {\em Sitzungsber. Preuss.
  Akad. Wiss. Berlin (Math. Phys. )} {\bf 1918} (1918) 154--167.

\bibitem{Gleiser:1998na}
M.~Gleiser and R.~Roberts, {\it {Gravitational waves from collapsing vacuum
  domains}},  {\em Phys. Rev. Lett.} {\bf 81} (1998) 5497--5500,
  [\href{http://arxiv.org/abs/astro-ph/9807260}{{\tt astro-ph/9807260}}].

\bibitem{Saikawa:2017hiv}
K.~Saikawa, {\it {A review of gravitational waves from cosmic domain walls}},
  {\em Universe} {\bf 3} (2017), no.~2 40,
  [\href{http://arxiv.org/abs/1703.02576}{{\tt arXiv:1703.02576}}].

\bibitem{Hiramatsu:2013qaa}
T.~Hiramatsu, M.~Kawasaki, and K.~Saikawa, {\it {On the estimation of
  gravitational wave spectrum from cosmic domain walls}},  {\em JCAP} {\bf 02}
  (2014) 031, [\href{http://arxiv.org/abs/1309.5001}{{\tt arXiv:1309.5001}}].

\end{thebibliography}
\end{document}